\documentstyle{article}
\newcommand{\resetcounters}{
	\setcounter{equation}{0}
        \setcounter{figure}{0}
        \setcounter{table}{0}}

\begin{document}

\bibliography{plain}

\title{QUANTUM ENVIRONMENTS: SPINS BATHS, OSCILLATOR BATHS, and applications to
QUANTUM MAGNETISM}
\author{P.C.E. STAMP \\ Physics Department, and Canadian Institute for Advanced Research,\\
University of British Columbia, 6224 
Agricultural Rd., Vancouver, B.C., Canada V6T 1Z1\\
and \\ L.C.M.I./ Max Planck Institute, Ave. des Martyrs, Grenoble 38042, France \\ **** \\
email: stamp@physics.ubc.ca}
\maketitle

\begin{abstract} The low-energy physics of systems coupled to their surroundings is understood 
by truncating to effective Hamiltonians; these tend to reduce to a few canonical forms, involving
coupling to "baths' of oscillators or spins. The method for doing this is demonstrated using
examples from magnetism, superconductivity, and measurement theory, as is the way one then solves for 
the low-energy dynamics. Finally, detailed application is given to the exciting recent Quantum
Relaxation and tunneling work in nanomagnets.
\end{abstract}

\section{Introduction}
\resetcounters

These lecture notes are taken from lectures given in April 1997 at the Institute for 
Nuclear Theory in Seattle (University of Washington).  Their main point was to explain how 
condensed matter theorists deal with problems in which some degree/degrees of freedom of 
interest, are coupled to a background ``quantum environment". In almost all cases the 
interesting
degrees of freedom are showing quantum behaviour, and often they are mesocopic or even 
macroscopic.

The tactic adopted in the lectures was to explain the general framework within which the 
theory operates, and then to demonstrate its operation with the help of examples.  These 
examples are taken from various fields (including superconductivity, the Kondo problem, and 
one-dimensional systems), but the emphasisis is on the very active new field of ``Quantum 
Nanomagnetism".  Great scientific interest has been generated (as well as some rather misleading
articles in the more popular press)
by recent experiments in crystals of magnetic macromolecules 
\cite{pau,nov,bar,fri,tho,her,san}, which show 
resonant tunneling of their magnetisation. 
Other experiments which have also generated considerable press include 
coherence experiments on the ferritin biological macromolecule \cite{aws,gid}, and tunneling 
experiments on domain walls in magnetic wires \cite{gio,wer}. I thus spent some time
explaining how the theoretical methods may be applied to these real systems.  For another very 
important example of large-scale quantum behaviour, see the lectures of A.J. Leggett in this 
volume, who discusses macroscopic quantum phenomena in superconducting systems.

The notes are divided into 3 parts.  The first part (Chapter 2) deals with the essential 
theoretical step in which one sets 
up a low-energy description of the system of interest.  The result of this is an ``effective 
Hamiltonian", which claims to accurately describe both the quantum 
degrees of freedom of interest, and the quantum 
environment (and their coupling), below a certain energy cut-off.  We end up with  
2 kinds of quantum environment.  One is the oscillator bath model, which is very old -  in 
its modern form it goes back to papers of Feynman et al. \cite{fey}, at the beginning of the 
1960's.  Except in rather 
unusual cases, it well describes an environment of {\it delocalized} 
modes (eg. phonons, photons, conduction electrons, magnons, etc).  It was introduced in this 
form into the discussion of tunneling by Caldeira and Leggett \cite{cal} in 1981.  The second 
environment is the ``spin bath" model, which has been developed recently by myself and 
Prokof'ev \cite{pro,sta,prok,prok2,prok3} (see also refs. \cite{shi,sach,castro,sta2}), 
stimulated 
particularly by problems in quantum nanomagnetism (although it can be applied outside this 
domain \cite{sach,sta2,nag,yam}; to illustrate this, brief space is 
given to superconducting 
SQUIDs and spin 
chains).  It describes environmental modes like paramagnetic or nuclear spins, or other similar 
cases in which each environmental mode has only a few (often only two) levels of interest,  
\underline{\rm and} where they are very weakly coupled.
The main point of this chapter is 
that readers may see clearly how the ``truncation" to an effective Hamiltonian works, and
to carry it 
out explicitly on some model examples in a tutorial way.

Once one has an effective Hamiltonian, the next task (chapter 3) 
is to solve for the dynamics of the 
degrees of freedom we are interested in, by ``integrating out" (averaging over) the 
environmental modes. Doing this for an oscillator bath is standard, so I don't spend too much 
time on it, except to delineate the main ideas and results.  More time is spent on the spin 
bath, where some new tricks are needed, and where the results are often radically different 
from the oscillator bath. Technical details are kept to a minimum, and ideas are 
emphasized. Again, it is then shown how these results apply to 
some real systems.

In both Chapters 2 and 3, I try to show how a very wide variety of physical problems tends to 
reduce to just a very few ``canonical models" at low energies. Many readers will not be 
surprised by this- it is of course the idea underlying the whole philosophy of ``universality
classes" and the renormalisation group.

Finally in the 3rd part of these lectures (chapter 4) I turn to some real down-to-earth physics, 
and discuss recent advances in our understanding of quantum magnets, particularly at the 
nano/mesoscopic level, where all sorts of interesting tunneling phenomena occur.  These 
include simple dissipative tunneling and quantum relaxation; one also gets resonant tunneling 
relaxation and there is evidence for quantum coherence at the mesoscopic level.  Although not 
all experiments are properly understood at present, it is becoming very clear that much of 
the physics is controlled by the combined effect of nuclear spins and magnetic dipolar 
fields on the tunneling entities; moreover, it is {\it essential} to take proper account 
of the nuclear dynamics, which justifies  {\it a posteriori} the interest in the spin bath 
model.  There is a rich experimental literature here, on magnetic molecules 
and particles, and on magnetic wires, as well as related experiments on spin chains and on the 
``colossal resistance" Mn- perovskite materials.

>From this summary the reader will see that I neglect discussion of other currently-used 
models of quantum environments such as the ``Landau-Zener" model \cite{shi, whe,bul}.  
This is unfortunate, because it is much easier to connect this model with studies in quantum 
chaos, then the ``bath" models I have described.  My only excuses are (i) lack of space, and 
(ii) lack of experience - condensed matter physicists are typically forced 
to deal with 
real solids, liquids, or glasses, in which statistical averages over environmental modes
are unavoidable.

\section{EFFECTIVE HAMILTONIANS:  SPIN BATHS AND OSCILLATOR BATHS}

In this first lecture I discuss, from both the physical and mathematical points of view, 
a rather extraordinary simplifying feature of almost all physical problems, which is 
fundamental to our understanding of complex systems having many degrees of freedom.  It 
is simply this - that provided we are only interested in a certain {\it energy scale} 
(or a limited range of energies), we may {\it throw away} most of the degrees of 
freedom - they do not play any direct role in the physics in the energy range of interest.  
Of course this can't be done willy-nilly; it must be done carefully, and those degrees of 
freedom which have been dumped will still play an {\it indirect} role in the physics.

Readers will of course be familiar with this idea in many forms - one's 
first encounter with it, as a student, is often in the form of the ``coarse-graining" 
assumptions inherent in statistical mechanics or thermodynamics.  In this lecture we 
begin by showing how one actually carries out the truncation of the Hamiltonian of a 
complex system down to a low-energy ``effective Hamiltonian".  This is done explicitly for 
one real physical example, in some detail, so that readers can see what is involved.

Another important idea, also rather extraordinary, is that if one carries out such 
truncations for many different physical systems, the same kinds of low-energy effective 
Hamiltonian recur again and again.  These ``universality classes" thus become very important, 
and the corresponding effective Hamiltonians become worthy of study in their own right.  
In this lecture we study 2 such Hamiltonians. One is the ``oscillator bath" model, in which 
some simple central system is coupled to an environment of harmonic oscillators.  The 
simplest such central system is a single 2-level system; the model is then the famous 
``spin-boson" model.  We will also look at a model in which {\it two} central spin systems 
couple to the oscillator bath
(the ``PISCES" model, where ``PISCES" is an acronym for 
``Pair of Interacting Spins Coupled to an Environmental Sea"); and I mention a number of other
such "canonical" models.

The other model effective Hamiltonian we shall study is the ``spin-bath" model, where 
now the environment itself is represented by a set of 2-level systems.  The simplest 
model of this kind is the ``central spin" model, where the central system is also 
represented by a 2-level system. In the same way as for the oscillator bath environments, one
can also develop a number of canonical spin bath models. Typically the spin bath models arise
for environments of {\it localised} modes, whereas the oscillator baths represent delocalised
modes.

I will discuss various examples in this lecture, to illustrate how these 
simple archetypal models really do represent the low-energy physics of a huge variety of 
real systems. What I will {\it not} do here is discuss the dynamics of these models- 
this is saved for the next lecture (section 3).

\subsection{Truncation to Low Energies}

The idea of a low-energy ``truncated" effective Hamiltonian (or Lagrangian) goes back a long 
way - in classical mechanics to hydrodynamics, and in quantum theory to the old spin 
Hamiltonians.  In its modern form (partially inspired by Landau's treatment of turbulence 
) it is often discussed in the renormalization group (RNG) framework \cite{wil2}. 
For the purposes of these lectures, we can formulate the problem as follows.

Typically one is presented with a reasonably accurately known ``high-energy" or ``bare" 
Hamiltonian (or Lagrangian) for a quantum system, valid below some ``ultraviolet" upper 
energy cut-off energy $E_c$, and having the form
\begin{equation}
\tilde{H}_{\mbox{\scriptsize Bare}} = \tilde{H}_o (\tilde{P},
\tilde{Q} ) + \tilde{H}_{\mbox{\scriptsize int}} (\tilde{P},
\tilde{Q}; \tilde{p}, \tilde{q}) +\tilde{H}_{\mbox{\scriptsize
env}} (\tilde{p}, \tilde{q}) \; ~~~(E <E_c)\;,
\label{1.1}
\end{equation}
where $\tilde{Q}$ is an $\tilde{M}$-dimensional coordinate describing that part of the 
system we are interested in (with $\tilde{P}$ the corresponding conjugate momentum), and 
$(\tilde{p}, \tilde{q})$ are $\tilde{N}$-dimensional coordinates describing all other degrees 
of freedom which may couple to $(\tilde{P}, \tilde{Q} )$. Conventionally one refers to 
$(\tilde{p}, \tilde{q})$  as environmental coordinates.

What is important about $\tilde{H}_{\mbox{\scriptsize Bare}}$ is that its form is known well 
(it can of course be regarded as a low-energy form of some other even higher-energy 
Hamiltonian, in a chain extending ultimately back to quarks and leptons). If, however, one is 
only interested in physics below a much lower energy scale $\Omega_o$, then the question is - 
how can we find a new effective Hamiltonian, having the form
\begin{equation}
H_{\mbox{\scriptsize eff}} = H_o (P,Q ) + H_{\mbox{\scriptsize int}}
(P,Q; p, q) +H_{\mbox{\scriptsize env}} (p, q) \; ~~~(E <\Omega_o)\;,
\label{1.2}
\end{equation}
in the truncated Hilbert space of energies below $\Omega_o$? In this 
$H_{\mbox{\scriptsize eff}}$, $P$ and $Q$ are generalised $M$-dimensional coordinates of 
interest, and $p,q$ are $N$-dimensional environmental coordinates coupled to them. Since 
we have truncated the total Hilbert space, we have in general that $M<\tilde{M}$ and 
$N<\tilde{N}$. This truncation is illustrated in Fig. 1.

\vspace{3in}

FIG. 1: The truncation Procedure (schematic). In (a) we see how the Hilbert space of the high-
energy Hamiltonian (with UV cut-off $E_c$) is truncated down to low energies (UV cut-off $\Omega_o$),
both for the system and the environment. 
In (b) the corresponding flow in the coupling constant space of effective Hamiltonians (here shown
for 2 couplings $\alpha_1$ and $\alpha_2$).

\vspace{5mm}

At first glance it is not at all obvious why anyone would want to make this
truncation, because its inevitable effect is to generate various couplings
between the low-energy modes which were not there before. However there are 2 very good 
reasons for truncating. First, in spite of the new couplings
appearing at low energy, it almost always turns out that $H_{\mbox{\scriptsize eff}}$ is easier
 to handle than $\tilde{H}_{\mbox{\scriptsize Bare}}$ (particularly in predicting the 
dynamics of the low-energy variables $P,Q$ of interest), simply because there are fewer 
variables to deal with. Second, and much more fundamental, 
the truncation pushes the new $H_{\mbox{\scriptsize eff}}$ towards some low-energy 
``fixed point" Hamiltonian (Fig. 1 again). Moreover, many
different physical systems may ``flow" to the same fixed point. This is particularly true for 
that part of $H_{\mbox{\scriptsize eff}}$
which describes the environment, and this allows theorists to speak of 
``universality classes" of quantum environment, each describing many different physical 
systems. Each of these different physical systems will then flow towards the same fixed point, 
and they will all be described by the same form for $H_{\mbox{\scriptsize env}} (p, q)$, 
albeit with different values for the couplings. In fact the low-energy couplings parametrize 
the position of a given system, with a given UV cut-off, in the space of available couplings 
(``effective Hamiltonian" space). As one varies the UV cut-off, the couplings change and any 
given system moves in the coupling space; but all systems in a given universality class move
towards one fixed point as the UV cutoff $\Omega_0$ is reduced.

In this context one understands the various coupling terms in $H_{eff}$, as a parametrisation 
of the {\it path taken} by $H_{eff}$, as it approaches the fixed point.

>From this point of view it is not so surprising that the description of quantum environments 
reduces to the discussion of just a few ``universality classes".  I emphasize this just because 
physicists are often quite surprised that one can discuss such a wide variety of, eg., 
tunneling problems with such a restricted class of models.  Of course, in asserting that 
these fixed points exist, I have simply swapped one mystery for another - instead of 
scratching our heads over the way in which so many physical systems resemble each other 
in low energies, we are now left wondering {\it why} there must be a flow of 
$H_{eff}$ towards one or other fixed point (ie., why fixed points?). This is often 
how physics proceeds, by substituting one conundrum for another (albeit a more 
precisely formulated one!).  

Let us leave this question hanging for now, and simply note here the enormous 
{\it pragmatic} interest of this result.  We shall see, in the 2 universality 
classes of interest, that we can parametrise very simply the form of the low-energy 
physics.  In the case of oscillator bath models, this will be in terms of a 
function $J(\omega)$, where $\omega$ is a frequency - this is the famous Caldeira-Leggett 
spectral function \cite{cal}, which convolves the coupling to the bath with the bath density
of states.  In the case of spin bath models, one has a density of states 
function $W(\epsilon)$ for the spin bath (usually Gaussian), and parameters which describe
both the coupling to the spin bath, and spin diffusion within the spin bath.  
At low temperatures and low applied fields, we will only care about the low-energy 
behaviour of $\omega$ or $\epsilon$.
This will allow us in some cases to make rather sweeping statements about whole classes 
of physical systems (as well as calculating their detailed dynamics, for which see the 
next lecture, in Ch. 3).  To take an example, we will be able to make remarks about quantum 
measurements which apply to large classes of measuring devices and measured systems, and 
yet which are still {\it realistic} (in the sense that they apply to real 
physical systems, instead of just to some idle toy model).

Having said this, I emphasize that a lot of the physics is in the details (a fact often 
ignored in the quantum measurement literature!).  Thus, rather than giving some (necessarily 
rather abstract) attempt at a general discussion of the derivation and validity of these 2 
models, I will instead show how it is done for some particular real systems.  After this, 
some at least of the general features will be evident (as well as the pitfalls in a too 
general approach!).

\subsection{Spin Baths and Nanomagnets- a worked example}

Let's begin our formal approach to the truncation procedure, by picking a particular 
example of a physical system which reduces to a spin bath model at low energies.  We 
shall look at a ``nanomagnet", by which we mean a magnetically-ordered particle or 
molecule which is sufficiently small that all the electronic spins causing the magnetism 
are aligned along one axis (note that a {\it macroscopic} magnet does 
{\it not} satisfy this criterion - it is often full of domains, with spins pointing 
in various directions; and spin waves or other spin fluctuations can also cause the spin 
axis to wander around a macroscopic sample).  The size of such monodomain particles does 
not often exceed 2000\AA; yet a particle this large can contain as many as 10$^8$ 
electronic spins!  Because all microscopic spins are aligned, the total system then behaves 
like a ``giant spin", denoted by $\vec{S}$.

As we shall see, the most important environment to which $\vec{S}$ is coupled is the set of 
{\it nuclear} spins inside it (and often to nuclear and paramagnetic impurity spins 
outside it as well).  This is how the spin bath arises. Our aim in this section is to 
truncate this physical system to low energies.  Recall that the reason for going through 
this is so you can see how it works on a realistic example.  You don't have to care about 
nanomagnets to follow the exercise - the point is to understand the truncation procedure, 
and see how to apply it elsewhere.

\vspace{7mm}

{\bf 2.2(a) HIGH ENERGY HAMILTONIAN}

Suppose we start by looking for a high-energy effective Hamiltonian, analogous to (2.1), but 
for a magnetic system which is coupled to a set of weakly-coupled 
spin-$1/2$ spins (ie., a "spin bath", instead of an "oscillator bath").
If the magnetic degrees of freedom are described by some continuous variable $Q$, and the 
oscillator coordinates $(\tilde{p}, \tilde{q})$ are replaced by a set of $N$ spin-$1/2$ 
variables $\{ \hat{\vec{\sigma}}_k \}$, i.e.,
two-level systems, we get a high-energy  
Hamiltonian of general form
\begin{equation}
H = H_o (P,Q ) + H_{\mbox{\scriptsize int}} (P,Q;\{ \hat{\vec{\sigma}} \} ) 
+H_{\mbox{\scriptsize env}} (\{ \hat{\vec{\sigma}} \}) \;;
\label{1.3}
\end{equation}
\begin{equation}
H_{\mbox{\scriptsize int}} (P,Q;\{ \hat{\vec{\sigma}} \} ) = \sum_{k=1}^N 
\bigg[ F_{\parallel}(P,Q) \hat{\sigma}_k^z + \big( F_{\perp}(P,Q) \hat{\sigma}_k^- +h.c. 
\big) \bigg] \;;
\label{1.4}
\end{equation}
\begin{equation}
H_{\mbox{\scriptsize env}} (\{ \hat{\vec{\sigma}} \}) =\sum_{k=1}^N
\vec{h}_k . \hat{\vec{\sigma}}_k + {1 \over 2} \sum_{k=1}^N\sum_{k'=1}^N
V_{kk'}^{\alpha \beta} \hat{\sigma}_k^\alpha \hat{\sigma}_{k'}^\beta \;,
\label{1.5}
\end{equation}
for energy scales $E<E_c$. In this Hamiltonian, we have a "central
system", moving in a space described by the continuous coordinate $Q$, which couples 
simultaneously to the environmental spin variables $\{ \hat{\vec{\sigma}}_k \}$.  An example 
of such a Hamiltonian, which has been worked out in detail recently, is that 
for a magnetic soliton, 
like a domain wall, at position $Q$, 
coupled by hyperfine interactions to a set of $N$ spin-$1/2$ nuclear 
spins \cite{dube}; this example will be discussed in section 2.3(b). In the case of nuclear spins 
$V_{kk'}^{\alpha \beta}$ describes the extremely weak internuclear 
dipolar coupling; typically $\vert V_{kk'}^{\alpha \beta} \vert \le 10^{-7} ~K$; 
and $\vec{h}_k$ is any external field that might influence these nuclei.

However in the case of nanomagnetic
grains or magnetic macromolecules things simplify a great deal, because the continuous
coordinate $Q$ can be replaced by a "giant spin" vector moving in the compact space
of the surface of a sphere; we get a high-energy Hamiltonian
of form
\begin{equation}
H(\vec{S};\{ \hat{\vec{\sigma}} \} ) = H_o (\vec{S} ) + {1 \over S}
\sum_{k=1}^N \omega_k \vec{S} \cdot \hat{\vec{\sigma}}_k +H_{\mbox{\scriptsize env}} 
(\{ \hat{\vec{\sigma}} \}) \;;
\label{1.6}
\end{equation}
where $H_o(\vec{S})$ is a ``giant spin" Hamiltonian, describing a
quantum rotator with spin quantum number  $S=\vert \vec{S} \vert
\gg 1$, and $H_{\mbox{\scriptsize env}} (\{ \hat{\vec{\sigma}} \})$
is the same as in (\ref{1.5}).

The assumption here is that we have a ``monodomain" magnetic particle or macromolecule, 
in which the very strong exchange interactions lock the microscopic electronic spins 
$\vec{s}_j$ into either a giant ferromagnetic (FM) moment, with $\vec{S}$ = ${\sum_j} 
\vec{s}_j$ (summed over ionic spin sites) or a giant antiferromagnetic (AFM) N\'{e}el vector, 
with $\vec{N} \equiv \vec{S} = {\sum_j} (-1)^j \vec{s}_j$ 
(for a simple staggered N\'{e}el order).  
With this assumption, the usual hyperfine coupling to the nuclear spins $\vec{I}_k$ 
can be written
\begin{equation}
H_{Hyp} \ = \ \sum_{k=1}^{N} \omega_k \mbox{\b{s}}_k \cdot 
\mbox{\b{I}}_k \ \ \rightarrow \frac{1}{S} \sum_{k=1}^{N} 
\omega_k \mbox{\b{S}} \cdot \mbox{\b{I}}_k
\label{1.7}
\end{equation}
(or a more complicated tensor interaction if we wish).  In these lectures I will assume that 
$\mid \vec{I}_k \mid = I = \frac{1}{2}$, and write $\vec{I}_k \rightarrow 
\sigma_k$, i.e. the nuclear spins will be described by spin-$\frac{1}{2}$ Pauli 
matrices.  
In many cases even if $I \neq \frac{1}{2}$, the low-energy nuclear spin dynamics is well 
described by a 2-level system; the effect of the other levels appears in the field 
$\vec{h}_k$ in (5).

The ``giant spin" Hamiltonian $H_o(\vec{S})$ itself can either be written as a continuous 
function of the vector $\vec{S}$ (assuming that $S$ is constant), or in terms of the usual spin
operators for $\vec{S}$, for which we have the general form
\begin{equation}
H_o(\mbox{$\b{S}$}) \ = \ \sum_{l=1}^{S} \frac{^{\parallel}K_l}{S^{l-1}} \ 
{\mbox{$\hat{S}$}}_{z}^{l} \ + \ \frac{1}{2} \sum_{r=1}^{S} \frac{^{\perp}K_r}{S^{r-1}} \ 
(\hat{S}^{r}_{+} \ + \ \hat{S}^{r}_{-})
\label{1.8}
\end{equation}
This form of the Hamiltonian separates the longitudinal part $^{\parallel}{\cal H} (\hat{S}_z)$
(which conserves $S_z$) from the transverse part $^{\perp}{\cal H}(\hat{S}_{+}, \hat{S}_{-})$.
In the symmetric case (all $l$ even) a pioneering semi-classical study of (1.8) was carried out 
by van Hemmen et al. \cite{van,sch}.  A very simple example of the giant spin Hamiltonian in 
the symmetric case is the biaxial form: 
\begin{equation}
H_o (S) \ = \ \frac{1}{S} \left[^{\parallel}K_2 \hat{S}^{2}_{z} \ + \ 
^{\perp}K_2 \hat{S}^{2}_{y}\right]\
\label{1.9}
\end{equation}
in which tunneling between the classical minima (at $S_z$ = $\pm S$) is accomplished by the 
symmetry-breaking transverse $^{\perp}K_2$ term, if $^{\parallel}K_2$ is negative.

\vspace{3in}

FIG. 2: The excited states of a nanomagnet (schematic): The $2S + 1$ states of the giant spin
appear at low energies. All states that break the constraint $\sum_{j} \vec{s}_j = \vec{S}$, with
$\vert \vec{S} \vert = s$, are referred to as internal magnon states. As $S \rightarrow 0$, the 
energy at the bottom of the magnon band tends to the interspin exchange $J$, but it falls as $S$
increases, because the magnons can spread out. When it reaches the top of the giant spin manifold,
the two manifolds mix strongly in the overlap region shown.

\vspace{5mm}

I refer here to the giant spin Hamiltonian for a nanomagnet as a "high-energy"
Hamiltonian, but of course it is obvious that Hamiltonians like (8) or (9) are 
themselves the result of the truncation of an even higher energy (more "microscopic") electronic 
spin Hamiltonian like, eg.,
\begin{equation}
H_o(\{ \mbox{$\b{s}$}_j\}) \ = \ \frac{1}{2} \sum_{<ij>} J^{\alpha \beta}_{ij} 
\hat{s}^{\alpha}_{i} \hat{s}^{\beta}_{j} \ + \ \frac{1}{2} \sum_{j}K^{\alpha \beta}_{j} 
\hat{s}^{\alpha}_j \hat{s}^{\beta}_j
\label{1.10}
\end{equation}
If $\mid s_j\mid = s$ (so there are $L = S/s$ electronic spins), this Hamiltonian acts in a huge 
Hilbert space of dimension $(2s+1)^L$.  I would like to strongly emphasize here a 
practical point of some importance (which will be familiar to anyone working in 
mesoscopic or nanoscopic physics). Even if we knew all the couplings 
$J^{\alpha \beta}_{ij}$, $K^{\alpha \beta}_{j}$ (which is hardly likely given the internal
complexity of most nanomagnets), truncation of $H_o(\b{S})$ at low energies 
is {\it practically impossible} if $L > 5-10$, even with supercomputers.  Instead one attempts to 
{\it measure} the parameters in (8) or (9), thereafter treating them as 
``fundamental" (similar situations are encountered in most problems involving strong 
interactions, ranging from QCD to Fermi liquid theory).  Moreover, 
even this is not easy - 
experiments such as ESR (Electron Spin Resonance) 
can only parametrize terms with small $\ell$ or $r$ in (8).  
Unfortunately the tunneling matrix element $\Delta_S$, between levels$\mid S_z >$ = 
$\pm \mid S>$ of the longitudinal part of (8) is just as much influenced by higher 
transverse couplings as by lower ones (a point to be discussed in more detail in the 
last section). Thus in any practical situation, we will probably {\it not} know all
of the important terms in our giant spin Hamiltonian.

In any case it is clear that $H_o(\vec{S})$ is only strictly meaningful if the higher states in 
$H_o(\{\vec{S}_j\})$ are not excited; these ``internal magnon" states (of which there are a 
huge number) break the constraint $\sum_j \vec{s}_j = S$ 
(or $\sum_j (-1)^{j} \vec{s}_j = \vec{N}$).  
What this implies is shown schematically in Fig. 2; as the 
nanomagnet gets bigger, internal magnon states (where internal spin flips overcome the powerful
 $J_{ij}$ exchange) can lower their energy by spreading over the whole sample.  Once these 
start impinging on the $(2S+1)$ - dimensional giant spin manifold, we may be in trouble. 
Note that the giant spin manifold is spread over $\sim SK$ 
in energy 
(where $K$ is the dominant anisotropy term in $H_o(\vec{S})$); values of $K$ are typically in the 
range 0.1-10 $K$.

\vspace{7mm}

{\bf 2.2(b) TRUNCATION to LOW ENERGIES (QUANTUM REGIME)}

Suppose however that one 
is interested in energies or temperatures where only the 2 lowest levels 
of $H_o(\vec{S})$ are involved. This arises in particular when one is interested in 
quantum phenomena, since the dynamics of the system go over into a temperature-independent
quantum regime once the higher levels can no longer be thermally activated.
We are then at liberty, in the absence of the nuclear spins, 
to truncate $H_o(\vec{S})$ to a simple 2-level Hamiltonian $H_o(
\vec{\tau})$,
 which is typically written as
\begin{equation}
H_o({\vec{\tau}}) \ = \ \Delta_o \hat{\tau}_x \ + \ \xi \hat{\tau}_x \
\label{1.11}
\end{equation}
where $\vec{\tau}$ is a Pauli spin.  It is very important, in the context of macroscopic 
quantum phenomena, that $H_o(\vec{\tau})$ is meaningful below a UV cut-off energy $\Omega_o 
\sim 
K$ {\it which is independent of} $S$ (at least until the monodomain assumption fails, 
and we return to (10)). This means that we enter the quantum regime at temperatures which
are often as high as several degrees Kelvin, even for very large spins.

Our essential task 
now becomes clear - we need to continue our truncation of (6) down to low T.  
For the simple giant spin Hamiltonian (no nuclear spins),
the passage from (8) to (11) has been discussed by many authors for $S \gg 1$.  Korenblit 
and Shender \cite{kor} used perturbation expansions (valid if $^{\perp}{\cal H}$/
$^{\parallel}{\cal H} \ll 1$); 
van 
Hemmen and Sut\H{o} \cite{van} used WKB methods (valid for arbitrary $^{\perp}{\cal H}$/
$^{\parallel}{\cal H}$, 
provided $S$ is large enough), and Enz and Schilling introduced instanton methods \cite{enz} 
(in many cases, formally equivalent to the WKB method, but not so general - one requires a 
semiclassical path or paths). Quite a lot can (and has been) said about this kind of 
truncation, but the problem is now solved, and you can read about it in the literature.
What interests us here is the fact that such calculations are basically {\it irrelevant}
to the real world, because of the strong coupling to the nuclear spins. This is a much more 
complicated problem, involving many more degrees of freedom- yet such problems are quite
generic in this game, since there is always going to be some environment which 
relaxes the system.

In what follows, I will describe how the truncation can be done using instanton methods. This
is certainly not the only way one could imagine it being done, and in fact it would be very
interesting to see it done otherwise. Since this course is pedagogical, I will not try to to the
reduction for a general Hamiltonian; instead it is done for the simple biaxial Hamiltonian given 
above.

\vspace{4mm}

{\bf (i) \underline{Free Giant Spins}}:
To warm up, lets briefly recap how the instanton analysis works for the simple {\it isolated} biaxial 
Hamiltonian (1.9) (cf. ref \cite{enz,van2,los,tup}).  We start by choosing a basis in the 
truncated 
(2-level) space such that the eigenstates of $\hat{\tau}_z$ correspond to the 2 semiclassical 
minimum states of $H_o(\vec{S})$, defined by coherent state vectors $\vert \vec{n}_1 \rangle 
$ and $\vert \vec{n}_2 \rangle $, such that $\vec{n}_1 \vert \vec{S} \vert \vec{n}_1 
\rangle = S \vec{n}_1$ and $\vec{n}_2 \vert \vec{S} \vert \vec{n}_2 \rangle = S 
\vec{n}_2$; the eigenstates of $H_{eff}^o(\vec{\tau})$ are then linear combinations of 
$\vert \vec{n}_1 \rangle $ and $\vert \vec{n}_2 \rangle $, which we can determine once we have 
found the four matrix elements $\langle \vec{n}_{\alpha} \vert H^o_{eff} \vert \vec{n}_{\beta } 
\rangle $ with $\alpha ,\beta =1,2 $.

Formally one can do this as follows, for the free spin. Consider the path
integral expression for the transition amplitude $\Gamma^o_{\alpha \beta } (t)$, during the 
time $t$; this is given by \cite{}:
\begin{equation}
\Gamma^o_{ \alpha \beta }(t) = \langle \vec{n}_{\alpha} \vert  e^{-iH_o(\vec{S})t} \vert 
\vec{n}_{\beta } \rangle = \int_{\vec{n}(\tau =0 )= \vec{n}_{\beta } }^{\vec{n} (\tau=t) 
=\vec{n}_{\alpha}} {\cal D} \vec{n} (\tau ) \exp \left\{ -\int_0^t d \tau {\cal L}_o (\tau ) 
\right\} \;,
\label{1.12}
\end{equation}
where the free spin  Euclidean Lagrangian is
\begin{equation}
{\cal L}_o = - iS \dot{\theta } \varphi \sin \theta + H_o(\vec{n} )\;.
\label{1.13}
\end{equation}
Here $\theta$ and $\varphi$ are the usual polar and azimuthal angles for the unit vector 
$\vec{n}(\tau )$.

Now in the semiclassical approximation there are two fundamental time
scales in the paths $\vec{n} (\tau )$ in (\ref{1.12}); these are $\Omega_o^{-1}$, the time 
required for the instanton traversal to be made between states, and $\Delta_o^{-1}$, the 
typical time elapsing between instantons. By definition, an effective Hamiltonian is supposed 
to reproduce the slow dynamics of the system in the truncated Hilbert space of
the two lowest levels, i.e., for long time scales an evolution operator is approximated as
\begin{equation}
\Gamma^o_{ \alpha \beta }(t) \approx \left( e^{-iH_{eff} t} \right)_{ \alpha \beta }\;.
\label{1.14}
\end{equation}
Since $\Delta_o$ is exponentially smaller than $\Omega_o$, and the nondiagonal elements are 
$ \sim \Delta_o $, we can write
\begin{eqnarray}
\Gamma^o_{\alpha \beta  }(t) &= &\langle \vec{n}_{\alpha} \vert   e^{-iH_{eff} t} \vert 
\vec{n}_{\beta } \rangle \nonumber \\  & \approx &  \delta _{\alpha \beta } -it  \langle 
\vec{n}_{\alpha} \vert H^o_{eff} \vert \vec{n}_{\beta } \rangle  \;;  \;\;\;\; (\Omega_o^{-1} 
\ll t \ll \Delta_o^{-1}) \;;
\label{1.15}
\end{eqnarray}
Then we immediately find the matrix elements of $H^o_{eff} (\vec{\tau })$
for $\alpha \ne \beta $ as
\begin{equation}
\left( H^o_{eff}(\vec{\tau }) \right)_{\alpha \beta  } = {i \over t}
\Gamma^o_{\alpha \beta  }(t)\;; \;\;\;\; (\Omega_o^{-1} \ll t \ll \Delta_o^{-1})\;.
\label{1.16}
\end{equation}
As a concrete example, consider the easy-axis/easy-plane Hamiltonian
(\ref{1.9}), where
\begin{equation}
H_o(\vec{n})= S K_{\parallel} \bigg[ \sin ^2\theta + { K_{\perp} \over K_{\parallel} }\: 
\sin ^2\theta  \sin ^2 \varphi \bigg] \;,
\label{1.17}
\end{equation}
The two lowest states  are $\vec{n}_{1}=\hat{\vec{z}}$ and  $\vec{n}_{2 }=-\hat{\vec{z}}$; 
henceforth we write these states as $\vert \Uparrow \rangle $ and  $\vert \Downarrow \rangle $. 
In the usual case where
$K_{\perp} / K_{\parallel} \gg 1$ (so that the tunneling amplitude is
appreciable) one has only small oscillations of $\varphi $ about the
semiclassical trajectories $\varphi =0$ or $\pi$, and by eliminating $\varphi$ one has
\begin{equation}
{\cal L}_o (\theta) = {S \over 4 K_{\perp}} \dot{\theta}^2 + S K_{\parallel} \sin ^2\theta \;,
\label{1.18}
\end{equation}
giving a classical equation of motion $\dot{\theta } = \Omega_o \sin \theta$, and instanton 
solution \cite{Chud88,Sta92}, going from $\vert \Uparrow \rangle $ to $\vert 
\Downarrow \rangle$
\begin{equation}
\sin \theta (\tau ) =1/ \cosh (\Omega_o \tau ) 
\label{1.19}
\end{equation}
(with the instanton centered at $\tau =0$);
in this system the "bounce" or "small oscillation" frequency is
\begin{equation}
\Omega_o = 2 (K_{\parallel}K_{\perp})^{1/2}  \;.
\label{1.20}
\end{equation}
The frequency $\Omega_o$ then sets the ultraviolet cut-off for the Hilbert space of $H^o_{eff} 
(\vec{\tau })$, and one finds, by substituting the semiclassical solution into (\ref{1.12}) and 
evaluating the determinant over the quadratic fluctuations around the semiclassical solution 
\cite{Callan} (the zero mode contribution gives a factor $it$, in the usual way), that from 
(\ref{1.16}) we get \cite{van,enz,van2,los,Chud88,Sta92,Ios}:
\begin{equation}
\hat{H}^o_{eff} (\vec{\tau }) = \frac{\Delta_o ( S )}{2} \hat{\tau}_{x}  \;,
\label{1.21}
\end{equation}
\begin{equation}
\Delta_o ( S ) = -\sum_{\eta =\pm }\sqrt{{2 \over \pi} Re A_o^{(\eta )}}
\Omega_o \exp \{ -A_o^{(\eta )} \} \equiv 2\Delta_o \cos \pi S \;,
\label{1.22}
\end{equation}
\begin{equation}
A_o^{(\eta )} = 2S ( K_{\parallel}/K_{\perp} )^{1/2} + i\eta \pi S  \;,
\label{1.23}
\end{equation}
where the action $A_o^{(\eta )}$ is that for the transition between the two
semiclassical minima, either clockwise ( $\eta =+$) or anticlockwise ($\eta =-$); the phase 
$\eta \pi S $ is the Kramers/Haldane phase, coming from the linear in time derivatives kinetic 
term in (\ref{1.13}).
Without this phase, we would simply have a splitting $\vert \Delta_o \vert  = 
\sqrt{2  Re A_o/\pi } \Omega_o \exp \{ -A_o \}$ with $ A_o = 2S 
( K_{\parallel}/K_{\perp} )^{1/2}$.

For this symmetric problem (where the 2 semiclassical states $ \vert \Uparrow \rangle $ and
$\vert \Downarrow \rangle $ are degenerate), the {\it diagonal} matrix elements in $\hat{H}^o_{eff}$
are zero (in fact if we computed them directly, again by evaluating the derminant, we would find
a value $\Omega_o$ for each- but this is just the energy of the 2 
states measured from the bottom of each well, and it makes sense to redefine the energy zero
to make these diagonal elements zero).
If instead of considering the symmetric problem, 
I had also added an external bias field, causing
an energy difference $\xi$ in the absence of tunneling, then the diagonal elements would simply 
be $\pm \xi/2$, and we would end up with an effective Hamiltonian 
\begin{equation}
H^o_{eff} =  \frac{\Delta_{0}}{2} \hat{\tau}_{x} + \frac{\xi}{2}
\hat{\tau}_{z}
\label{2whamiltonian}
\end{equation}
with eigenvalues
\begin{equation}
E_{\pm} = \pm |E| =
\pm \frac{1}{2} \sqrt{ \Delta^{2} + \epsilon^{2} }
\end{equation}
and eigenfunctions
\begin{equation}
\psi_{\pm} = A_{\pm} \, \left[ \, (E_{\pm}+\epsilon)| \Uparrow \rangle -
\Delta_{0} | \Downarrow \rangle \, \right] \, ,
\end{equation}
\begin{equation}
A_{\pm} = \left[ \frac{1}{(E_{\pm}+\epsilon)^{2}+(\Delta_{0})^{2}} \right]^{1/2}
\end{equation}
Suppose this 2-level system starts off at $t = 0$ in the state $\vert \Uparrow \rangle$. Then after 
a time $t$, the system is described by a density matrix in the same basis, given by
\begin{equation}
\rho(t) = \left(
\begin{array}{cc}
1-\frac{\Delta_{0}^{2}}{E^{2}} \sin^{2} Et & -i \frac{\Delta_{0}}{E} \sin Et \\
i \frac{\Delta_{0}}{E} \sin Et & \frac{\Delta_{0}^{2}}{E^{2}} \sin^{2} Et
\end{array} \right)
\label{2wbiasdensityt}
\end{equation}
In the equivalent 2-well problem, the
initial ``wave-packet'' $\vert \Uparrow \rangle $
partially oscillates between the 2 wells- the diagonal elements give the
occupation probability of the wells, and the off-diagonal elements
describe oscillatory quantum interference between
them, which is suppressed by
the bias; when $\xi \gg \Delta_{0}$, the system 
stays in one well, in the absence of any coupling to
the environment, even if this means it is in an excited (high-energy) state.
To remove this "blocking" of transitions to the lower energy state (and to give 
irreversible {\it relaxation}, as opposed to just oscillations), we 
need some kind of {\it dynamic environment}.

Notice in passing the limitations of this instanton derivation - it would not work in the 
absence of a few well-defined semiclassical paths.  Thus if we replaced the transverse term 
$^{\perp}K_2 S^{2}_{y}$ by the rotationally invariant $^{\perp}K_2(S^{2}_{x} + 
S^{2}_{y})/2$, there would be no favoured semiclassical path, and we would be forced back onto 
the WKB or perturbative analyses.
However when we can use the instanton formalism, it is easily adaptable to include the spin 
bath as well; we now turn to this task.

\vspace{4mm}

{\bf (ii) \underline{Including the Spin Bath}}:
I will not give all the details here (for which see Tupitsyn et al. \cite{tup}),
just the main ideas.
First, the basic physics.  Before we couple the giant spin to the spin bath, the spin bath 
spectrum, containing 2$^N$ lines, is almost completely degenerate - only the tiny internuclear 
coupling splits these lines.  However the hyperfine coupling is very large (it ranges from 1.4 
mK, in Ni, to nearly 0.5 K, in Ho, {\it per nuclear spin}, and it drastically alters 
the environmental spectrum.  The nuclear levels, for a given giant spin state, now find 
themselves spread in a Gaussian multiplet of half-width $E_o \sim \omega_o N^{\frac{1}{2}}$ 
in energy (where $\omega_o$ is a typical hyperfine coupling) around the giant spin state; 
$E_o$ is many orders of magnitude greater than the width $T^{-1}_{2} \sim 10^4 - 10^5$ Hz of 
the nuclear multiplet before coupling.  In this sense the nuclear spin bath degrees of freedom 
are \underline{\rm slaved} to the giant spin.  Notice also that if the giant spin was formerly 
able to tunnel (because of near resonance between states $\mid S_z>$ and $\mid - S_z>$), this is 
unlikely now, because an extra {\it internal} bias field $\epsilon = \tau_z \Sigma_k \omega_k
\sigma^{z}_k$ acts on \b{$\tau$} (indeed on $\vec{S}$ itself), 
and typically $\epsilon \gg \Delta_o$, pushing the 
giant spin way off resonance. Thus the hyperfine coupling to even a single 
nuclear spin drastically alters the giant spin dynamics as well. before doing anything formal, lets
see qualitatively what must happen.

  (i) In the absence of any nuclear dynamics this ``degeneracy blocking" would only allow a tiny 
fraction of giant spins to make any tunneling transitions at all \cite{pro,prok,prok2,gar} 
(Fig. 3). We will see in Chapter 4 that in the low temperature limit, it has been recently
discovered that such giant spins can have very fast dynamics, which is (to 
me at any rate!) a very convincing
demonstration of the role that the nuclear spin {\it dynamics} must play in the relaxation 
(since at these temperatures this is the only environment left with any dynamics at all!). This
justifies {\it a posteriori} the theoretical effort 
that has been put into understanding these effects!

(ii) A second effect of the spin bath is ``topological decoherence" \cite{pro,sta,prok}.  Each 
transition of the giant spin causes a time-dependent perturbation (via $H_{hyp}$) on the 
nuclear bath, which can cause transitions in the bath - moreover, this perturbation causes a 
{\it phase change} in the nuclear bath state.  Since this phase change varies from 
one transition of $\vec{S}$ to the next, the net effect is to {\it randomize} the phase 
of $\vec{S}$, ie., to cause phase decoherence.  From the measurement point of view the 
nuclear spins are measuring time-varying fields due to $\vec{S}$ (ie., inhomogeneous in 
time rather than in space, as in the usual Stern-Gerlach experiment); this is a kind of 
``reverse Stern-Gerlach measuring apparatus" \cite{sta2}.

(iii) A final effect of the nuclear spins on the tunneling is that of "orthogonality blocking". Suppose 
that, semiclassically speaking, all nuclear spins are parallel/antiparallel to the net field
acting on them (due to $\vec{S}$ and/or an external field), {\it before} $\vec{S}$ tunnels. What
happens to them afterwards? The answer is that if they are still parallel/antiparallel to the 
new field, nothing at all. But in general the new field is not parallel/antiparallel to the old one,
and so the bath spins must now precess (or quantum mechanically, make transitions) in the new
field. The "mismatch" between old and new nuclear eigenstates causes a severe suppression of
the tunneling of $\vec{S}$ (as well as decohering it).

Well, this all sounds very nice; but
how do we deal formally with this physics?  We wish to truncate to a low energy 
Hamiltonian $H_{eff} (\vec{\tau}, \{ \vec{\sigma}_k \})$, valid for energies 
$ \ll \Omega_o$ (cf. Fig. 1), where $\Omega_o \sim K$.  First, let's cheat and look at the final 
answer!  One finds for the general problem of a giant spin coupled to a spin bath that
\cite{pro,prok,tup}
\begin{eqnarray}
H_{\mbox{\scriptsize eff}} & = & \left\{ 2 \Delta_o \hat{\tau }_ -  \cos
\bigg[ \pi S - i \sum_k (\alpha_k \vec{n} \cdot \hat{\vec{\sigma }}_k  +
 \beta_o \vec{n} \cdot \vec{H}_o)  \bigg] + H.c. \right\} \nonumber \\
& + & {{\hat \tau }_z \over 2} \sum_{k=1}^N \omega_k^{\parallel} \:
{\vec l}_k \cdot {\hat {\vec \sigma }}_k  + {1 \over 2} \sum_{k=1}^N
\omega_k^{\perp}\: {\vec m}_k \cdot {\hat {\vec \sigma }}_k
+\sum_{k=1}^N \sum_{k'=1}^N V_{kk'}^{\alpha \beta } \hat{\sigma}_k^\alpha
\hat{\sigma}_{k'}^\beta ~ \;.
\label{1.24}
\end{eqnarray}
I have also added an external field $\vec{H}_o$. This is the {\it general} form; we shall see 
below that for the biaxial nanospin, with the simple contact hyperfine interaction in (7), it 
simplifies somewhat. 
This effective Hamiltonian looks forbidding (although we shall see it is not so difficult to 
use); let us now look at each term in turn.

The first thing to notice is the separation into a diagonal term (in $\hat{\tau}_z$) and 
a non-diagonal one (in $\hat{\tau}_+, \hat{\tau}_-$). The non-diagonal term 
operates during transitions 
of $\vec{S}$, and it also causes 
transitions in the nuclear bath, because of the 
time-dependent field $\omega_k \vec{S}/S$ acting on 
each $\hat{\vec{\sigma}}_k$ during a transition.  The diagonal term operates when $\vec{S}$ is in 
one of its two quiescent states $\vec{S}_1$ and $\vec{S}_2$.  Defining as before the basis states 
$\mid \vec{n}_1 > \equiv \mid \Uparrow >$ and $\mid \vec{n}_2 > 
\equiv \mid \Downarrow >$, we have corresponding fields $\vec{\gamma}_{k}^{(1)}$ 
and $\vec{\gamma}_{k}^{(2)}$ acting on $\vec{\sigma}_k$. We define the {\it sum} and 
the {\it difference} terms as
\begin{eqnarray}
\omega_k^{\parallel}{\vec l}_k & =& {\vec \gamma }_k^{(1)} - {\vec \gamma }_k^{(2)} \nonumber \\
 \omega_k^{\perp} {\vec m}_k & =& {\vec \gamma }_k^{(1)} + {\vec \gamma }_k^{(2)}\;.
\label{1.25}
\end{eqnarray}
where the ${\vec l}_k$ and ${\vec m}_k$ are unit vectors. Then the truncated diagonal 
interaction  takes the form (we project on states
$\vert \Uparrow \rangle $ and $\vert \Downarrow \rangle $ using
standard $(1+\hat{\tau}_z)/2$ and $(1-\hat{\tau}_z)/2$ operators)
\begin{equation}
H^D_{eff} = \sum_{k=1}^N \bigg\{ {\vec \gamma}_k^{(1)} {1+\hat{\tau}_z \over 2} + 
{\vec \gamma}_k^{(2)} {1-\hat{\tau}_z \over 2} \bigg\} \cdot {\hat {\vec \sigma }}_k \equiv   
{1 \over 2} \bigg\{{\hat \tau }_z \sum_{k=1}^N \omega_k^{\parallel} \: {\vec l}_k \cdot 
{\hat {\vec \sigma }}_k  + \sum_{k=1}^N \omega_k^{\perp}\: {\vec m}_k \cdot {\hat 
{\vec \sigma }}_k \bigg\}\;,
\label{1.26}
\end{equation}
i.e., one term which changes when $\vec{S}_1 \to \vec{S}_2$, and one
which does not. Usually $\omega_k^{\parallel} \gg \omega_k^{\perp}$ ,
and $\omega_k^{\parallel} \sim \omega_o$.  Thus we get the diagonal term in (\ref{1.24}). For the 
biaxial system with contact hyperfine interaction, the diagonal term is trivial; we get
\begin{equation}
\omega_k^{\parallel} = \omega_k \;; \;\;\; \omega_k^{\perp} = 0 \;; \;\;\; (biaxial \; system)
\end{equation}
Notice that it is this diagonal term which is responsible for the degeneracy blocking just
mentioned- if we look at the {\it combined} density of states for the giant spin plus the
nuclear levels, we see that there will be a huge multiplet of $2^N$ levels around
each giant spin level. We will see later (section 3.4(a)) under what circumstances this takes
the Gaussian form mentioned above.

Turning now to the non-diagonal term, we notice that unless $\omega_k\ll \Omega_o$, the 
hyperfine coupling itself can mediate interactions between the 2 lowest giant spin levels, 
and the higher levels.  In practise $\omega_k/\Omega_o \ll 1$ almost always and we have a 
small parameter. Then instead of the bare transition matrix $\Gamma_{\alpha \beta}^{o}$ in 
(\ref{1.14}), we calculate
\begin{eqnarray}
&&\Gamma_{\alpha \beta }( \{ \sigma_k^{(\alpha )}, \sigma_k^{(\beta )} \}; t) = \nonumber \\ 
\prod_{k=1}^N \int_{\vec{\sigma}_k^{(\alpha )}}^{\vec{\sigma}_k^{(\beta )}} {\cal D} 
\vec{\sigma}_k(\tau )& &  \int_{\vec{n}_\alpha }^{\vec{n}_\beta } {\cal D} \vec{n}(\tau )  
exp \left\{   -\int d\tau \big[ {\cal L}_o(\tau ) +  \sum_{k=1}^N{\cal L}_k^o(\tau ) +\delta 
{\cal L}_{\sigma}(\tau ) \big]  \right\} \;.
\label{1.27}
\end{eqnarray}
where ${\cal L}^{o}_k (\tau) = -(i/2) \dot \Theta_k \varphi_k \sin 
\Theta_k$ is the nuclear spin Lagrangian, and $\delta {\cal L}_{\sigma}(\tau) = 
\Sigma_{k = 1} \vec{\gamma}_k (\tau) \cdot \vec{\sigma}_k (\tau)$,
 with $\vec{\gamma}_k (\tau) = \omega_k \vec{S}(\tau)/2S$ the time-dependent 
field, from $\vec{S}$, acting on $\vec{\sigma}_k$.  During the transition, 
$\vec{S}(\tau)$ varies over a timescale $\Omega^{-1}_{o}$, and so we get a time-dependent
 perturbation $\sim \omega_k/\Omega_{o}$ (ie., a sudden perturbation).

To calculate $\Gamma_{\alpha \beta}$ one first finds the new instanton trajectory by minimizing 
${\cal L}_o (\tau) + \delta {\cal L}(\tau)$, and then calculates the transition amplitude from 
$\mid \vec{n}_{\alpha}>$ to $\mid \vec{n}_{\beta}>$; for details see Tupitsyn et al. ~\cite{tup}.  
For the simple biaxial Hamiltonians (1.9), one gets for the transition element 
$\Gamma_{\Uparrow\Downarrow}(t)$ after a time $t$ (the "non-diagonal" amplitude):
\begin{equation}
\hat{ \Gamma}_{ \Downarrow \Uparrow }(t) = it \sum_{\eta =\pm } \sqrt{{2 \over \pi } Re A_o } 
\Omega_o \exp \left\{ -A_o^{(\eta )} - \eta \sum_k (\alpha_k \vec{n} \cdot \hat{\vec{\sigma}}_k 
+ \beta_o \vec{n}_o \cdot \vec{H}_o) \right\}  \;; \;\;\; (\Omega_o^{-1} \ll t \ll 
\Delta_o^{-1} ) 
\label{1.28}
\end{equation}
where the vectors $\alpha_k \vec{n}$ and $\beta_o \vec{n}_o$ have to be determined in terms of the 
parameters in the original high-energy Hamiltonian, in the course of minimizing the action. For the
case of the biaxial giant spin Hamiltonian with contact hyperfine interactions, one easily gets
\begin{equation}
\alpha_k \vec{n} = {\pi \omega_k \over 2 \Omega_o} \big(\hat{\vec{x}},~
i \sqrt{K_{\parallel} / K_{\perp }}~ \hat{\vec{y}} \big)  \;;
\label{1.29}
\end{equation}
\begin{equation}
\beta_o \vec{n}_o = {\pi \gamma_e S \over \Omega_o} \big(\hat{\vec{x}},~
i \sqrt{K_{\parallel} / K_{\perp }} ~\hat{\vec{y}} \big)  \;,
\label{1.30}
\end{equation}

>From this result the non-diagonal term in (\ref{1.24}) follows immediately. The dimensionless parameter
$\alpha_{k}$ is particularly interesting, since it parameterizes the inelastic effect, on 
the nuclear spin $\underline{\sigma}_k$, of transitions made by $\vec{S}$ 
(and it leads to topological decoherence ~\cite{pro}).  Note that it is {\it complex}, and that 
for $\omega_k/\Omega_o \ll 1$, we have $\vert \alpha_k \vert \sim \pi\omega_k/2\Omega_o$.
  Since the {\it probability} that $\vec{\sigma}_k$ will flip is 
$\frac{1}{2}|\alpha_k|^2$, this result is what we would expect (it follows from 
time-dependent perturbation theory in the sudden approximation).  The only subtlety is that we 
cannot apply perturbation theory directly, since (at least in the instanton formalism), the 
tunneling takes place in {\it imaginary} time $\tau$ (not real time $t$); this is 
why the more detailed treatment of Tupitsyn et al. \cite{tup} is necessary. Roughly speaking, 
the real part of $\alpha$ adds an extra phase to the existing giant spin Berry phase (leading
to topological decoherence), and the imaginary part is a renormalisation of the tunneling
action caused by the nuclear spins. 

This completes the discussion of how this particular spin bath effective Hamiltonian is 
derived - I have given details so you can see some of what is involved.  In Chapter 3 we will 
see how to extract the dynamics of the giant spin (and we will then see the quantitative 
realisation of the concepts of topological decoherence, degeneracy blocking, and orthogonality
blocking, discussed above).  But first we look at a few other examples
of systems coupled to a spin bath.

\subsection{Some Other Systems Coupled To  Spin Baths}

The previous section attempted to explain the details of a particular truncation procedure in 
some pedagogical detail.  However there are other important physical examples of systems coupled 
to spin baths, which can be profitably studied if you wish to understand the truncation 
procedure better.  I shall briefly recount how two such systems work, and then mention some 
other examples which can be found in the literature.  The first example involves a Josephson 
superconducting ring, coupled to nuclear and paramagnetic spins; and the second involves a 
magnetic soliton (a domain wall) coupled to nuclear spins.

\vspace{7mm} 

{\bf 2.3(a) \underline{MQC in SUPERCONDUCTORS, and the SPIN BATH}}

Let us start with a system which, in the {\it absence} of the spin bath, is thought 
to truncate at low energies to a 2-level system, just like the giant spin system.  I am thinking 
here of a superconducting ``ring" with a single Josephson ``weak link"; a typical geometry is 
shown in Fig.3.  

\vspace{3in}

FIG. 3: The geometry chosen for the analysis of the Josephson ring. Topologically, the system is a torus,
with the weak link allowing the passage of a fluxon between the inside and the outside of the ring,
if the current through the weak link exceeds $I_c$. The geometry is discussed in the text.

\vspace{5mm}

Following the work of a number of authors, it is understood that the flux 
$\Phi$ passing through the ring moves in an adiabatic (``high-energy") potential $V(\Phi)$ of form
\begin{equation}
V(\Phi) = \frac{1}{2L}(\Phi - \Phi_e)^2 - 2\pi E_{J} \cos(2\pi \Phi/\Phi_o)
\end{equation}
where $\Phi_o$ is the flux quantum $h/2e$, $L$ is the ring inductance, $\Phi_e$ is the external 
imposed flux, $\Phi$ the total flux (both of these through the ring) and $E_{J} = I_c \Phi_o/2\pi$ is 
the weak link ``Josephson coupling energy", with $I_c$ the critical current through the ring.  
This ``RF SQUID", and the possibility of macroscopic tunneling of $\Phi$ through the Josephson 
cosine potential, has been discussed in great detail in the literature \cite{cal,AJL}(and see also 
Leggett's chapter in this volume).  There is also a kinetic term $T = \frac{1}{2} C \dot{\Phi}^2$ 
in the system Hamiltonian; and as a result at low energies it finds itself near the bottom of a 
potential well, able both to oscillate (small oscillation frequency 
$\Omega_o \sim 2\pi(E_j/\pi C)^{1/2}/\Phi_o)$ or to tunnel to the nearest potential well.  
If these 2 wells are the 2 lowest in energy (with others at energies higher by $\Omega_o$ 
or more) then when $kT < \Omega_o/2\pi$, we can again model the system by a 
2-level system.

There is also a resistive coupling to normal electrons and Bogoliubov quasiparticles, which 
can be understood in terms of a coupling to a bath of oscillators (see next section).  What 
we are interested in here is - what will be the coupling of the flux $\Phi$ to any spins in 
the ring?  These may be either nuclear (of which there is a very large number) or  
paramagnetic spins (coming from magnetic impurities).  We shall see in this and the next 
chapter that this question is of fundamental importance for the search for ``Macroscopic 
Quantum Coherence" in RF SQUID rings.  In what follows I give a qualitative discussion only, 
based on a detailed calculation by myself and Prokof'ev. \cite{prok3}

Suppose to start with we consider the geometry shown in Fig. 3. We assume a cube of dimension
 1x1x1$\:$cm ($L=1\:cm$)
 of type-I superconducting material, with London penetration
 depth $ \lambda_L=5\times 10^{-6}\:cm $, and a hole in the
 center of radius $R =0.2\: cm$,
 and surface area $ S=\pi R^2 = 0.1\:cm^2 $.
 The magnetic field inside the hole corresponding to a 
 half-flux quantum is  
 \begin{equation}
 B_o={\pi \hbar c \over e \pi R^2}=2\times 10^{-6}\: G\;.
 \label{1.SQ}
 \end{equation}
There is a slit in the cube, 
bridged by the wire-shaped junction, 
of length $l= 10^{-4}\:cm $ and diameter  
$d=2\times 10^{-5}\: cm$. The current density in the junction is enhanced over that elsewhere
on the system surface by a factor
$L/d = 5\times 10^4$, and correspondingly the magnetic field in the
junction is as high as
$B_{jun} \sim B_o L/d = 10^{-1}\: G$.

There are both nuclear spins and paramagnetic impurities in the spin bath. Consider first the 
nuclear spins; assuming all nuclei have spins, we find that in the bulk of the ring, within 
a penetration depth of the surface, there is a number
\begin{equation}
N_{ring} =(L\times 2\pi R \times \lambda_L ) \times 10^{23}
\approx 5 \times 10^{17} \;,
\label{3.SQ}
\end{equation}
of nuclear spins coupling to the ring current; and in the junction itself, a number
\begin{equation}
N_{jun}=(l\times \pi d \times \lambda_L ) \times 10^{23}
\approx 3 \times 10^{9} \;.
\label{4.SQ}
\end{equation}
coupling to the junction current.
In the "high-temperature" limit the average nuclear polarisation is $M = \sqrt{N}$, and so the 
typical coupling between the junction current and the nuclear spins is 
$\Gamma_{ring} \sim  \omega_{ring} M = 2\times 10^{-13}K \times 7\times 10^{8}
\approx 10^{-4}K$, where $\omega_{ring} = \mu_n B_o$, and $\mu_n$ is the nuclear Bohr magneton.
For the junction nuclei one has a coupling
$\Gamma_{jun} \sim  \omega_{jun}  \sqrt{N_{jun}}
\approx 5\times 10^{-4}K$, where $\omega_{jun} = \mu_n B_{jun}$, which is actually similar, 
because there is a linear increase in the magnetic field (since $B_{jun} \sim 1/d$),
but a quadratic suppression in the number of spins (ie., $N_{jun} \sim ld$). Notice the 
magnetic coupling is $\ll kT$, justifying the high-$T$ assumption. Notice also we have
taken no account of a possible coupling to substrate spins (the ring is in superfluid He-4!), which
would have a much larger coupling again to the current. We have also assumed perfect screening
from external fields.

This coupling gives a longitudinal bias energy $\xi \sim 10^{-4}-10^{-3} K$, 
acting on the tunneling flux
coordinate $\Phi$, which is rather bigger than any presently realistic numbers for the
SQUID tunneling matrix element $\Delta_o$; we are again in a situation of strong degeneracy blocking.
However, it also turns out that we have {\it very} strong orthogonality blocking! This is because 
each nuclear spin feels the dipolar fields from the other nuclear spins; in general there will be 
some component perpendicular to the field from the SQUID, of strength $\sim 1G$, and 
associated energy $\omega_k^{\perp} \sim V_{NN} \sim 10^{-7} K$, where $V_{NN}$ is the nearest 
neighbour internuclear dipolar coupling energy. Physically, when the SQUID flips, the field on each 
nuclear spin hardly changes its direction, being dominated by the more slowly varying 
(but much stronger) nuclear dipolar field.

We may summarize this analysis of the nuclear spin effects on the flux dynamics in the form of the 
effective Hamiltonian
\begin{equation}
H{\mbox{\scriptsize eff}}= \Delta_o \hat{\tau}_x + {1 \over 2}
\sum_{k=1}^{N} ( \omega^{\perp}_k\hat{\sigma}_k^z + 
\hat{\tau}_z
\omega^{\parallel}_k \hat{\sigma}_k^z ) \;
\label{cases.3}
\end{equation}
where $\omega_k^{\parallel} = \omega_{ring}$ or $\omega_{jun}$ (with values $\omega_{ring} \sim 
2 \times 10^{-13} K$ and $\omega_{jun} \sim 10^{-8} K$ respectively), whilst $\omega_k^{\perp}
\sim 10^{-7} K$ as above. Thus  $\omega^{\parallel}_k/\omega^{\perp}_k \ll 1$, which is the 
opposite limit considered to that for the giant spin! Notice further that both these couplings
are much less than $\Delta_o$ ($\xi$ is bigger than $\Delta_o$ only because there are so many
nuclei involved). Thus the nuclear bath is no longer slaved to the central system at all! In the 
next chapter we will show that this allows us to map this problem to that of an oscillator
bath, coupled to $\Phi$.

Consider now 
the effect of paramagnetic impurities, with concentration $n_{pm}$, a single-spin coupling 
$\omega_{pm} \sim 10^3 \times \omega_{ring}$ to the current, and hence a total coupling energy
$\Gamma_{pm} \sim 10^{6}n_{pm} \Gamma_{ring} \sim n_{pm}\times 100~K$. This is obviously bigger 
still, unless the superconductor is very pure indeed! The dynamics of these will come either from
the Kondo effect (in the superconductor), 
or from flip-flop processes between these impurities. The former go at a rate $\sim T_K$, the Kondo 
temperature (which varies over many orders of magnitude depending on the impurity; the latter would
go at a rate $(T_2^{-1})_{pm} \sim 10^{9} n_{pm}$ Hz, except that in pure samples these
flip-flop processes will themselves be blocked by the local dipolar coupling between the impurity 
abd nearby nuclear spins (of strength $\sim 10^{-4} K$); thus this will happen once 
$n_{pm} \ll 10^{-3}$.

Just as for the Giant spins, to properly understand the flux dynamics we must then consider the 
nuclear bath dynamics, which can in principle relieve this blocking. We will deal with this in
the next chapter.

\vspace{7mm}

{\bf 2.3(b) \underline{MAGNETIC DOMAIN WALL TUNNELING}}

Let us now move to a quite different example, in which at low energies the central system 
does not behave at all like a 2-level system, but instead like a particle moving continuously 
in a one-dimensional potential.  Of course we expect that many physical systems will behave 
like this at low energies, and many of them will couple to spin environments - but the 
following example is the only one which has so far been worked out in detail \cite{dube,staDW}.

The reason that a magnetic domain wall moves like a 1-dimensional particle in many realistic 
situations is that it is a membrane-like soliton whose "flexural" oscillations are
hindered by the strong magnetic dipolar field, which tries to keep the wall flat. As an 
example we take the famous "Bloch wall", which has a Hamiltonian
\begin{eqnarray}
{\cal H} &=& \int d {\bf r}  [J  ({\bf \nabla m})^{2}
- K_{\|} m_{z}^{2} + K_{\bot} m_{x}^{2}] \nonumber \\
&=& \int d {\bf r} [ J  (({\bf \nabla}\theta)^{2} + \sin^{2} \theta
 ({\bf \nabla}\phi)^{2})
 - K_{\|} \cos^{2} \theta + K_{\bot}
 \cos^{2} \phi \sin^{2} \theta ]
 \end{eqnarray}
 representing a ferromagnet with easy axis along the $z$ axis
 and the $z-y$ plane being
 easy. In 3-dimensions, the units of $J$ are $\mbox{J/m}$ and the units of the
 anisotropy constants are in $\mbox{J/m}^{3}$.
 
The domain wall corresponding to this Hamiltonian is perpendicular to the
$x$ axis, with the magnetisation rotating in the $z-y$ plane. We refer to
the wall by its center, located at a position $Q$ along the $x$ axis
The new frame of reference is thus $(x_{1},x_{2},x_{3}) =(z,y,x)$.
This is represented in Fig.4 below.

The components of the magnetisation are given by:
\begin{eqnarray}
\hat{m}^{B}_{1} &=& C \tanh \left( \frac{x_{3}-Q(t)}{\lambda_{B}}
\right) \nonumber \\
\hat{m}^{B}_{2} &=& \chi \left( 1 - \frac{\dot{Q}^{2}(t)}{8 c_{0}^{2}}
\right)
\mbox{sech}  \left( \frac{x_{3} -Q(t)}{\lambda_{B}} \right)
\label{magcomp} \\
\hat{m}^{B}_{3} &=& C \frac{\dot{Q}(t)}{2 c_{0}} \mbox{sech} \left(
\frac{x_{3} -Q(t)}{\lambda_{B}} \right) \nonumber
\end{eqnarray}
 $C = \pm 1$ is the
 "topological charge"
 of the wall and $\chi = \pm 1$ is the "chirality". The
 topological charge corresponds to the
 direction along which the wall moves under the application of an external
 magnetic field in a direction parallel to the easy axis, while the
 chirality refers to the sense of the rotation of the magnetisation
 inside
 the wall.
A static Bloch wall only rotates in the easy plane. However, as soon as it
moves it creates a demagnetising field which causes the spins to precess and
the appearance of a component of the magnetisation out of the plane, directly
proportional to the wall velocity.
The precession of the spins also
causes the appearance of an inertial term, the D\"oring mass, given by
\begin{equation}
M_{w} = \frac{S_{w} M_{0}^{2}}{\gamma_{g}^{2} (J K_{\|})^{1/2} }
\left[ \frac{1}{( 1+ K_{\bot}/K_{\|})^{1/2}-1} \right]^{2}
\label{mdoring}
\end{equation}
where $S_{w}$ is the surface area of the wall.

\vspace{3in}

FIG. 4: A standard Bloch wall in a uniaxial ferromagnet, showing the direction of the 
electronic spins as one passes along a path perpendicular to the wall plane. The chirality is 
defined by the sense in which the spins turn, and the topological charge by the difference in
magnetisation along the easy axis, between the 2 sides of the wall.

\vspace{5mm}

Now from this phenomenological picture of the wall, one can proceed to a treatment which
eliminates the details of the wall profile altogether, and simply describes everything
in terms of the coordinate $Q$ of the wall centre (assuming a wall held flat by the
dipolar or "demagnetisation fields- in realistic situations wall curvature is small and
has little effect on what follows). In this case the wall will have a kinetic energy
(parametrised by the D\"oring mass), and there will also be potential terms. There are 
2 obvious sources for these:

(i) If defects are present in the sample, it will be energetically favourable
for the magnetisation to rotate at this site since there is no
associated energy cost.
We
will assume that the radius $R_{d}$ corresponding to the defect volume
is much smaller than $\lambda_{B}$, the domain wall width.
The wall thus becomes pinned by a
potential of the form \cite{staDW}
\begin{equation}
V(Q) = V_{0} \; \mbox{sech}^{2} (Q/\lambda_{B})
\label{pinpot}
\end{equation}
with $V_{0}$ proportional to the volume of the defect.
We further assume that there
is a very small concentration of defects, so that there is only 1
important pinning center for the wall. This would correspond to an ideal
experimental situation.

(ii) The application of an external  magnetic field ${\bf H}_{e}$ in the
direction
of the easy axis couples to the magnetisation to give a potential term
linear in $Q$.

We can then write a ``bare''
 Hamiltonian (ie., neglecting the environment) for the wall \cite{staDW}
 \begin{equation}
 H_{w} = \frac{1}{2} M_{w} \dot{Q}^{2} - V(Q) - 2 S_{w} \mu_{B} M_{0}
 H_{e} Q
 \end{equation}
 where we put the topological charge $C = 1$ for brevity.

What now of the environment? In the literature you will find extensive discussion of the 
effect of magnons \cite{Sta92,staDW} (ie., spin waves),electrons \cite{Tat}, 
and phonons \cite{dube} on the 
wall dynamics. However these are all oscillator baths. What we are interested in here is 
any spin bath effects. It turns out here that, just as for the nanomagnets, these are far 
more important than any oscillator bath effects. To see why this is we recall again the 
strength of hyperfine interactions to nuclear spins. 
Nuclear hyperfine effects vary enormously between magnetic systems.
The weakest is in Ni, where only $ 1 \%$ of the nuclei have spins,
and the hyperfine coupling is only
$\omega_{0} = 28.35 \, \mbox{MHz}$
($\sim 1.4 \, \mbox{mK}$). On the other hand, in the
case of rare earths, $\omega_{0}$ varies from $1$ to $10 \, \mbox{GHz}$
($0.05$ to $0.5 \, \mbox{K}$).
In this latter case the hyperfine coupling energy to a {\it single}
nucleus may be comparable to the other energy scales in the problem !

To derive an effective interaction Hamiltonian, we consider our system
of ferromagnetically ordered spins to be coupled locally to $N$ nuclear
spins
${\bf I}_k$
at positions ${\bf r}_{k}$ ($ k=1,2,3, ... \, N$), which for a set
of dilute nuclear spins (where only one isotope has a nuclear spin) will
be random. The total Hamiltonian for the coupled system is then
\begin{equation}
H = H_{m} + \sum_{k=1}^{N} \omega_{k} {\bf s}_{k} \cdot
{\bf I}_{k} +
\frac{1}{2} \sum_{k} \sum_{k'} V_{k k'}^{\alpha \beta}
I_{k}^{\alpha} I_{k'}^{\beta}
\end{equation}
where $H_{m}$ is the electronic Hamiltonian for the magnetisation
(Eq. (42)), written in terms of the electronic spins
${\bf s}_k$ at the sites where there happen to be nuclear
spins,
and $\omega_{k}$ is the hyperfine coupling at
${\bf r}_{k}$; $V_{k k'}^{\alpha \beta}$ is the internuclear dipolar
interaction,, with strength
$|V_{k k'}^{\alpha \beta}| \sim 1-100 \, \mbox{kHz}$ ($0.05-0.5
\, \mu\mbox{K}$).
In terms of the continuum magnetisation
${\bf M}({\bf r})$, we have
\begin{equation}
H=H_{m} + \sum_{k=1}^{N} \omega_{k}
\int \frac{d^{3}r}{\gamma_{g}} \delta ({\bf r}-{\bf r}_{k})
[ M_{z}({\bf r}) I_{k}^{z} + ( M_{x}({\bf r}) I_{k}^{x}
+  M_{y}({\bf r}) I_{k}^{y})] +
\frac{1}{2} \sum_{k} \sum_{k'} V_{k k'}^{\alpha \beta}
I_{k}^{\alpha} I_{k'}^{\beta}
\label{inutile}
\end{equation}

I would like to draw your attention to the similarity between this equation and that quoted 
at the very beginning of section 2.2. We have in essence reduced our problem to the coupling
of a single 1-dimensional coordinate to a bath of spins. I will not go into any further detail 
here as to how one can rewrite this Hamiltonian in terms of "degeneracy blocking" diagonal
terms, plus non-diagonal terms operating when the domain wall moves (for which see the 
original paper \cite{dube}). Suffice it to say that the longitudinal term gives a total
"hyperfine potential field" acting on the wall, of diagonal form
\begin{equation}
U(Q) = \frac{ \omega_{0} M_{0}}{\gamma_{g}} \sum_{k=1}^{N} \int
d^{3}r \delta ({\bf r}-{\bf r}_{k})
\left( 1 - C \tanh \left(\frac{x_{3}-Q}{\lambda_{B}}\right)
\right)I_{k}^{w}
\label{pinpotnucspin}
\end{equation}
where the nuclear spins now have their axis of quantisation defined according to the {\it local}
magnetisation orientation; the axis $w$ is along ${\vec M}(r)$, and so the component 
$I_k^w$ is diagonal. This potential has the spatial form of a {\it random walk} (Fig.5); however 
it also fluctuates in time (because of nuclear spin diffusion). There are of course non-diagonal 
terms as well, for which see Dub\'e and Stamp \cite{dube}.

As discussed in detail in the original paper, the effect of this potential on the wall dynamics
can be very large indeed, particularly for rare earth magnets.

\vspace{3in}

FIG. 5: Typical form of the "random walk" potential $U(Q)$ acting on a magnetic domain wall, 
due to the hyperfine coupling to disordered nuclear spins. We 
assume that the temperature is higher than the hyperfine coupling to the nuclei. In reality this 
potential also fluctuates in time, because of nuclear spin diffusion.

\vspace{7mm}

This concludes our detailed discussion of the spin bath models, and where they come from. I
should emphasize that there are other "canonical" spin bath models which can be studied (and which
for the most part have not been, at least not at the time of writing). Two obvious ones are

(i) A model in which a simple oscillator is coupled to a spin bath. There are many obvious 
systems for which such a model applies- two of them are immediately apparent, for we can
simply put either the SQUID flux or the domain wall mentioned above, into a harmonic potential.
This would describe the SQUID trapped in one well, near the bottom, or a domain wall trapped
deep in a pinning potential.

(ii) The "Landau-Zener plus spin bath model", in which a 2-level system, coupled to a spin
bath, is also {\it driven} by some external ($c$- number) field. This problem is of considerable
interest for experiments on quantum nanomagnets (where one applies an AC field to the magnets,
and looks at the absorption), and it is also of theoretical interest, as an example of a 
dissipative landau-Zener model. The simplest such model involves the single passage of the
2-level system through resonance.

The reader can certainly think of more such examples (particularly after referring forward to 
the discussion of canonical oscillator bath models, which have been much more thoroughly
studied- see section 2.4(c)).

\subsection{Coupling to The Oscillator Bath}

Let us now turn to that class of effective Hamiltonians $H_{eff}$ which can be written in terms 
of an environment of oscillators (the ``oscillator bath" models).  These models are not 
only very old (early examples include QED, as well as Tomonaga's model of 1-dimensional 
electrons, and early theories of both spin waves and the electron 
gas \cite{Tomo}), but also of great generality - they apply to almost any problem in which the 
central system is coupled to a set of {\it delocalized modes} (whether these be bosonic 
or fermionic, by the way).  The lengthy history of this model means that I will refrain from 
giving a detailed treatment of truncation to its various low-energy forms - instead, I refer 
the reader to relevant original references or reviews.  The reader is urged to consult these - 
the truncation is not without interesting subtleties, particularly where non-linear interactions 
to the environmental modes are involved.

Many physicists (and even more chemists) are suspicious of the claimed generality for the 
oscillator bath model for quantum environments, despite the justifications of Feynman and 
Vernon ~\cite{fey}, and Caldeira and Leggett \cite{cal}.  The essential claim is that at 
low energies, our general effective Hamiltonian (2) can very often be written as
\begin{eqnarray}
H_{\mbox{\scriptsize eff}}^{\mbox{\scriptsize osc}} & =&  H_o (P,Q ) + \sum_{k=1}^{N} 
\bigg[ F_k(P,Q)x_k +G_k(P,Q)p_k \bigg] \nonumber \\ & +& {1 \over 2} \sum_{k=1}^{N} \left(  
{p_k^2 \over m_k} + m_k \omega_k^2 x_k^2 \right)\;,
\label{1.31}
\end{eqnarray}
where $P,Q$ are the momentum and coordinate of interest, and the $\{x_k\}$ are harmonic 
oscillators with frequency $\omega_k < \Omega_o$, the UV cut-off.  The crucial point is 
that the couplings $F_k$ and $G_k$ are weak ($\sim N^{-\frac{1}{2}}$), and this 
justifies one in stopping at linear coupling.  In most cases the oscillators represent 
delocalized modes of the environment, and the $N^{-\frac{1}{2}}$ factor then just comes from 
normalisation of their wavefunctions. 

Now in fact physicists should 
{\it not} be surprised at 
the generality of (1.31),
since it embodies precisely the same assumptions as conventional response function theory
\cite{lan2} - that 
the effect of the central coordinates $P,Q$, on each environmental mode, is
weak ($\sim N^{-\frac{1}{2}}$) and may be treated 
in 2nd order perturbation theory.  Then, by an
appropriate diagonalisation to normal
modes \cite{fey2, cal}, one
may always write (1.31).
>From this point of view it is clear that this should work for, say electrons - although
electrons are fermions, the low-energy modes of the Fermi liquid are incoherent particle-hole
pairs,
plus collective modes \cite{lan2}. It also works nicely for superconductors \cite{cal,AJL,Eck}
(although in this case one really needs 2 oscillator baths \cite{Eck}), and in fact oscillator 
bath models have been applied to a vast array of electronic excitations, ranging from
Luttinger liquid quasiparticles to magnons.

In what follows I will attempt to convince you that the "oscillator bath' models really are
very generally applicable to environments of delocalised modes. The discussion will be 
fairly informal- readers wanting rigour should go to the literature.

\vspace{7mm}

{\bf 2.4(a) SPIN-BOSON MODELS}

Consider the case where $H_o(P,Q)$ describes 
a system moving in a 2-well potential, at which point, as with the giant spin, one truncates 
down to the 2 lowest levels therein, to get the celebrated ``spin-boson" model \cite{leg}:
\begin{eqnarray}
H_{\mbox{\scriptsize SB}} & =&  \Delta_o \hat{\tau}_x +\xi_H \hat{\tau}_z +
 \sum_{k=1}^{N} \bigg[ (c_k^{\parallel} \hat{\tau}_z )x_k  + (c_k^{\perp} \hat{\tau}_- q_k  
+ H.c. ) \bigg] \nonumber \\ & +& {1 \over 2} \sum_{k=1}^{N} \left(  {p_k^2 \over m_k} + m_k 
\omega_k^2 x_k^2 \right)   \;,
\label{1.32}
\end{eqnarray}
with couplings $c^{\|}_k, c^{\perp}_k \sim  O(N^{-\frac{1}{2}})$, and 
$\vec{\tau}$ again describing the 2-level central tunneling system.  Often the term 
$c^{\perp}_k$ is dropped, since formally $c^{\perp}_k/c^{\|}_k \sim
\Delta_o/\Omega_o$; however sometimes one has $c^{\|}_k = 0$ (eg. for symmetry reasons),
 and $c^{\perp}_k$ survives and is important \cite{kag}.

Now at this point we could go through the same kind of exercise as before, and derive a
spin-boson Hamiltonian at low energy for some model system. This has in fact been done
by many authors for systems like SQUID's, or the Kondo problem, etc. However, to give 
some continuity to the discussion, let's stick to the giant spin model for the moment,
since we have already seen how it truncates to a 2-level system in isolation. To see how
it also couples to oscillators, we will consider a giant spin in an environment of 
phonons and electrons. Readers interested in SQUID's should go to Leggett's lectures; and 
there is of course a huge literature on the spin-boson model.

\vspace{4mm}

{\bf (i) \underline{Giant Spin coupled to Phonons}}: 
We thus go back to our giant spin model (eg. (8)), and ask how one should now incorporate the
coupling of phonons and electrons to $\vec{S}$.  Once we have done this, we can also ask how one 
reduces to a form like (\ref{1.32}), once we truncate down to the lowest 2 levels of $\vec{S}$.

Let us start with phonons - this case is almost trivial, since the phonon representation is 
already an oscillator one.  The coupling between $\vec{S}$ and the phonon variables $b_q$, $b^{+}_{q}$ 
is just the standard magnetoacoustic one, which can either be understood from macroscopic 
considerations \cite{trem,lan3}, or at a microscopic level \cite{abr}.  In either case, one has 
couplings like
\begin{equation}
{\cal H}^{\phi}_2 \sim {\Omega_o \over S} \hat{S}_x\hat{S}_z \left(
{m_e \over M_a } \right)^{1/4} \sum_{\vec{q}} \left(
{\omega_q \over \Theta_D }  \right)^{1/2}
 [b_{\vec{q}}+b_{\vec{q}}^{\dag}] \;,
 \label{1.ph}
\end{equation}
where m$_e$ is the electron mass, $M_a$ the mass of the molecule, and  $\Theta_D \sim c_s a^{-1}$ is 
the Debye temperature (with $a$ the lattice spacing in a molecular crystal of giant spins, 
and $c_s$ the sound velocity).  This interaction describes a non-diagonal process in which a phonon
 is emitted or absorbed with a concomitant change of $\pm 1$ in $S_z$ 
(since $S_x = \frac{1}{2}(S_{+} + S_{-}))$. 
One also has diagonal terms of similar form, in which $S_z S_x$ is replaced by, eg. $S^{2}_{z}$; 
and there are also higher couplings to, eg, pairs of phonons. \cite{trem}.  The bare phonon 
Hamiltonian is just the usual form
\begin{equation}
H_{\phi} \ = \ \sum_{q} \; \omega_q (b^{+}_{q} b_{q} \ + \ \frac{1}{2})
\label{1.34}
\end{equation}
and can be written in the usual oscillator form with 
the transformation $x_q = (2m_q \omega_q)^
{-1/2} [b^{+}_q + b_{-q}]$.
Armed with (1.33) and (1.34), one can discuss the dynamics of a tunneling giant spin coupled to 
phonons, as has been done recently by the Grenoble-Firenze group \cite{vil,pol}.  At low T 
($kT \ll \Omega_o$) one can truncate the giant spin to a 2-level system as before.  The general 
technique for doing this is straightforward \cite{prok2,kag}.

\vspace{4mm}

{\bf (ii) \underline{The "Giant Kondo" Problem}}:
Consider now a much more interesting example of a giant spin coupled to 
electrons.  Many situations can be envisaged, depending on whether the nanomagnet, 
or the matrix/substrate in which it is embedded, or both, are conducting; and a great deal of 
physics revolves around how the electrons move across the boundary between the nanomagnet and
the background matrix.  Some of the this is discussed in ref. \cite{prok2} (but a lot more is not!).

Here we take a simplified example which has the virtue of bringing out the essential physics.  We 
assume that an electronic fluid freely permeates both the conducting nanomagnet
(of volume $V_o$), and the background 
matrix; and we assume that the individual electronic spins couple to the mobile electrons via a Kondo 
exchange, so that at the giant spin level we have
\begin{equation}
H \ = \ H_o(\vec{S}) \ + \ \sum_{\vec{k}, \sigma} \epsilon_k c^{+}_{\vec{k}\sigma} 
c_{\vec{k}\sigma} \ + \ \frac{1}{2} \sum_{i \epsilon V_{o}} J_{i} \vec{\hat{s}}_i. 
\vec{\hat{\sigma}}^{\alpha \beta} \ \sum_{\vec{k} \vec{q}} \ e^{i\vec{q} . 
\vec{r}_j} c^{+}_{\vec{k}+\vec{q},\alpha} c_{\vec{k} \beta}
\end{equation}
where $H_o(\vec{S})$ is the usual Giant spin Hamiltonian, whereas $\vec{s}_i$ is an individual 
electronic 
spin at site $i$, and position $\vec{r}_i$ (cf. eq. (10)); 
$c^{+}_{k \sigma}$ creates a conduction electron in momentum 
state $\mid \vec{k} >$, with spin projection $\sigma = \pm 1$ along $\vec{\hat{z}}$; 
and $\vec{\sigma}
^{\alpha \beta}$ is a Pauli matrix.  I have written this interaction in terms of the individual 
electronic spins in order to show its microscopic origin as an exchange interaction; but of course 
the locking together of all the $\vec{s}_i$ into a giant 
$\vec{S}$ means that we can immediately rewrite the 
interaction as
\begin{equation}
H_{int}^{GK} \ = \ \frac{1}{2} \bar{J} \underline{\hat{S}}. 
\underline{\hat{\sigma}}^{\alpha \beta} \
 \sum_{\vec{k}\vec{q}} 
 F_q \ c^{+}_{\vec{k}+\vec{q}_i \alpha} \ c_{\vec{k}_i \beta}
\label{1.35}
\end{equation}
where $\bar{J}$ is the mean value of the $J_i$, and $F_q = \int(d^{3}r/V_o)\rho(\vec{r})
e^{i\vec{q}.\vec{r}}$ is a "form factor" which integrates the number 
density $\rho(\vec{r})$ of electronic 
spins over the volume $V_o$ of the nanomagnet.

This form of the coupling is very famous in condensed matter physics - it is a "Kondo coupling", 
but instead of being between a single spin-1/2 and conduction electrons, it couples a spin $\vec{S}$ 
(with $2S+1$ levels) to the electrons.  It is an unusual form of what has come to be known as the 
"multi-channel Kondo problem". In the energy regime over which the giant spin 
description is valid, one may use the "Giant Kondo" interaction in (\ref{1.35}). At low $T$, a truncation 
to 2 levels produces the simple "spin-boson" Hamiltonian, with only the 
diagonal coupling $\hat{\tau}_z c_k x_k$, and with a Caldeira-Leggett spectral function
\begin{equation}
J(\omega) \ = \ \pi \alpha_{\kappa} \omega
\end{equation}
\begin{equation}
\alpha_{\kappa} \sim 2g^2 S^2 \int_{V_o} \frac{d^{3}r d^{3}r^{\prime}}{V^{2}_{o}} \left(\frac
{sin k_F \mid \underline{r} - \underline{r}\prime \mid}{k_F \mid r - r\prime \mid} \right)^2 \ 
 \sim \ g^2S^{4/3}
\label{1.37}
\end{equation}
where $k_F$ is the Fermi momentum, $g = \bar{J}N(0)$ is the dimensionless Kondo coupling, and 
$N(0)$ the
 Fermi surface density of states; typically $g \sim  0.1$, so that $\alpha_{\kappa}$ 
 is not necessarily 
small. Consider, eg., a very small FM particle made from $Fe$, embedded in a conducting film
or similar substrate/background matrix. Suppose it is only 15 Angstroms across; such a particle
could still have a spin $S \sim 300$. If the conductor were, eg., $Cu$, then with a $g \sim 0.1$, 
we would have $\alpha_{\kappa} \sim 20$. As we will see in Chapter 3, the dynamics of this
system would then be not only overdamped but {\it frozen}, in the quantum regime.

The reason for this remarkably simple effective Hamiltonian is that the number of microscopic electronic 
spins $\vec{s}_i$ in the nanomagnet $\sim  O(S)$, whereas the number of interaction channels 
(essentially angular momentum channels for electrons scattering off $\vec{S}$) is $\sim S^{2/3}$ 
(proportional to the cross-sectional area of the nanomagnet), which means 
that the effective coupling 
{\it per microscopic electronic spin} is weak ($\sim S^{-1/3}$).  A renormalization 
group analysis of the problem substantiates this result \cite{prok2}.

We see that, as previously advertised, a coupling to fermionic electrons has now been reduced 
to an effective coupling to bosonic oscillators.  Why this happens can already be guessed from 
the bilinear form of the fermionic 
terms in (\ref{1.35}).  To an incoherent particle-hole superposition 
$\sum_{k,\sigma} c^{+}_{k+q, \sigma} c_{k \sigma}$ 
we associate a bosonic operator $b^{+}_q$.  If we treat 
the coupling perturbatively (and we have just seen we can) then it is obvious that using the same 
transformation between $b^{+}_q, b_q$, and $x_q$ 
as we need for the phonons, we can go immediately 
to the oscillator bath representation.  This perturbative argument
is of course just a special case of
 the general arguments given by Feynman and Vernon \cite{fey} and Caldeira \& Leggett \cite{cal}.  In 
the case of a stronger coupling (as in the single spin Kondo problem) the argument can actually be 
made in a modified form \cite{leg} (and likewise in the context of the quantum diffusion of defects 
in solids \cite{kag1}).

>From these examples it will perhaps be clear that a very large number of physical systems can be 
mapped, at low T, to a spin-boson model.  Essentially all one requires is that the central system 
of interest may be truncated to a 2-level system, and that it be coupled to a set of extended modes.  
We also need to have some good reason for ignoring any couplings to a spin bath (as we shall discuss 
below, it is not so easy to ignore environmental spins, so that such reasons may be hard to find!).  
Thus, in the sense described at the beginning of this section, 
the spin-boson Hamiltonian constitutes 
an important example of a ``universality class" of low-energy Hamiltonians.

\vspace{7mm}

{\bf 2.4(b) TWO SPINS- The "PISCES" MODEL}

I think it is worthwhile dwelling a bit on another such universality class, which also involves an 
oscillator bath environment.  This one involves not just one but \underline{two} spins, each 
coupled to an oscillator bath.  The effective Hamiltonian is then in general of the form
\begin{equation}
H = H ( \mbox{\boldmath $\tau$}_{1}, \mbox{\boldmath $\tau$}_{2} ) +
H_{osc} (\{ {\bf x_{k}} \} ) + H_{int} (\mbox{\boldmath $\tau$}_{1},
\mbox{\boldmath $\tau$}_{2} ; \{ {\bf x_{k}} \} )
\label{htotal}
\end{equation}
where $\mbox{\boldmath $\tau$}_{1}$ and $\mbox{\boldmath $\tau$}_{2}$
are 2 Pauli spins and the ${\bf x_{k}}$, with ${\bf k} = 1,2,...N$,
are the oscillator coordinates. 
This model has been studied in great detail by M. Dub\'{e} and myself \cite{PIS1}; it is interesting 
in a wide variety of physical situations, particularly at the microscopic level (examples include, 
eg., coupled Anderson or Kondo spins, coupled tunneling defects, coupled chromophores 
in biological molecules, 2-level atoms coupled through the EM field, or nucleons coupled 
via the meson field).  Perhaps even more interesting are the applications at the macroscopic 
level, where $\vec{\tau}_1$ and $\vec{\tau}_2$ represent the 2 lowest levels of a pair of mesoscopic 
or macroscopic quantum systems.  Some concrete examples include 2 coupled SQUID's, or 2 coupled 
nanomagnets \cite{PIS2}.  I should add that the initial development of the model was really directed 
towards the understanding of {\it quantum measurements}.  In this case one of the spins 
represents a measured system, and the other the low-energy subspace of the measuring apparatus 
degrees of freedom - we shall see in a minute how this can happen.

We shall be interested here in situations where 
$\vec{\tau}_1$ and $\vec{\tau}_2$ represent mesoscopic and/or macroscopic systems. In this case, our 
low-energy Hamiltonian will be as above, with the restricted form 
\begin{equation}
H_{0} =  - \frac{1}{2} (\Delta_{1} \hat{\tau}_{1}^{x} +
			 \Delta_{2} \hat{\tau}_{2}^{x})
+ \frac{1}{2} K_{zz} \hat{\tau}_{1}^{z} \hat{\tau}_{2}^{z}
\end{equation}
\begin{equation}
H_{osc} = \frac{1}{2} \sum_{{\bf k}=1}^{N} m_{{\bf k}}
			 ( \dot{{\bf x}}_{{\bf k}}^{2}
			 + \omega_{{\bf k}}^{2} {\bf x_{k}}^{2})
\end{equation}
\begin{equation}
H_{int} = \frac{1}{2} \sum_{{\bf k}=1}^{N}
			 ( c_{{\bf k}}^{(1)} e^{i {\bf k \cdot R}_{1}}
			 \hat{\tau}_{1}^{z} + c_{{\bf k}}^{(2)} e^{i {\bf k \cdot R}_{2}}
			 \hat{\tau}_{2}^{z})
			 {\bf x_{k}}
			 \label{hintpisces}
\end{equation}
			 where the N oscillator bath modes $\{ {\bf x_{k}} \}$ are
			 assumed for simplicity (but without real loss of generality)
			 to be momentum eigenstates, and 
			 $\vec{R}_1$ and $\vec{R}_2$ are the "positions"
of the two spins. 
			 The two spins are coupled through a {\it direct}
			 interaction term $K_{zz}$ having ferro- or antiferromagnetic form,
			 depending on whether $K_{zz}$ is negative or positive. 
The direct interaction itself
comes from integrating out very high energy modes (with frequencies even greater than the UV
cut-off $\Omega_0$ implicit in (\ref{hintpisces}); for frequencies much lower than $\Omega_0$
it can be considered to be static. 
			 In the absence of the coupling to the environmental sea, the
			 spins have their levels split by ``tunneling'' matrix elements
			 $\Delta_{1}$ and $\Delta_{2}$. We work in the basis in which
			 $ | \uparrow \rangle$ and $ | \downarrow \rangle$, eigenstates of
			 $ \hat{ {\bf \tau}}^{z}$, are degenerate until split by the off-diagonal
tunneling.

We refer to Hamiltonians like this as ``PISCES" Hamiltonians, where PISCES is an abbreviation for 
``Pair of Interacting Spins Coupled to an Environmental Sea".
$H_{PISCES}$ as written above is clearly not the most general
model of this kind.
A much wider range of direct couplings is possible; instead of
$K_{zz} \hat{\tau}_{1}^{z} \hat{\tau}_{2}^{z}$ we could use
\begin{equation}
H_{int}^{dir} = \frac{1}{2} \sum_{\mu \nu}
K_{\mu \nu} \hat{\tau}_{1}^{\mu} \hat{\tau}_{2}^{\nu}
\label{direct}
\end{equation}
We could also use more complicated indirect couplings
(ie., couplings to the bath)
like $\frac{1}{2}  \hat{\tau}^{\mu}_{\alpha} \sum_{{\bf k}}
c_{{\bf k} \mu}^{(\alpha)}
e^{i{\bf k \cdot R}_{\alpha}} {\bf x_{k}}$,
with $\mu = x,y,z$ and $\alpha=1,2$.
In the present discussion we will stick to the diagonal coupling in
(\ref{hintpisces}) and keep only the direct coupling in
$\hat{\tau}^{z}_{1} \hat{\tau}^{z}_{2}$.
Our reasoning is as follows. Just as for the single spin-boson problem, we
expect diagonal couplings to dominate the low-energy dynamics of the
combined system, since the spins spend almost all their time in a diagonal
state (only a fraction $\Delta/\Omega_0$ of their time is in a non-diagonal
state, when the relevant system is tunneling under the barrier). In
certain cases the diagonal couplings can be zero (usually for symmetry
reasons), and then one must include non-diagonal
couplings
like $\frac{1}{2} \hat{\tau}^{\bot}_{\alpha}
\sum_{{\bf k}} (c_{{\bf k} \bot}^{\alpha}
e^{i {\bf k \cdot R}_{\alpha}} {\bf x_{k}} + H.c.)$.

Our reason for having a direct coupling in (\ref{direct}) of {\it longitudinal} form is 
then connected
   to our choice of diagonal couplings between the bath and the
    systems.
In fact, one may quite
	generally observe that in a field-theoretical context, any direct longitudinal
	 interaction $K_{zz} \hat{\tau}_{1}^{z} \hat{\tau}_{2}^{z}$
	  can be viewed as the result of a high-energy coupling of $\hat{\tau}^z_{1}$ and
	   $\hat{\tau}^z_{2}$ to the high-frequency or ``fast'' modes of some
	    dynamic field; ``fast'' in this context simply means much faster than the
	     low-energy scales of interest, so that the interaction may be treated as
	      quasi-instantaneous. In our case, integrating out these fast modes (those
	       with
		a frequency greater than $\Omega_{0}$) then produces the
longitudinal static interaction in $H_0$. In the next section we see how at the
frequencies of interest for the application of $H_{PISCES}$, a further longitudinal interaction
appears, mediated by those bath modes with frequencies $\omega$ such that $\Omega_0 > \omega \gg
\Delta_1, \Delta_2$.

One further term we have omitted from (\ref{hintpisces}) is an applied bias acting on one
		     or both of the spins, of the form $H_{ext} = \xi_{j} \hat{\tau}_{j}^{z}$, for
		       example. We will analyse its effects in a moment; its origin is
usually fairly obvious.

The crucial new ingredient, of course, in $H_{PISCES}$ is that we are now interested in the 
{\it correlations} between the behaviour of 2 systems (microscopic or macroscopic) whilst 
they are both influenced by the environment; even if $K_{zz}$ is zero, there will be 
correlations transmitted via the bath, of a dynamic and partly dissipative nature.
In the next section we will see how such physics comes 
out of the solution for the dynamics of $H_{PISCES}$.

In this section we shall stick firmly to the question of the 
{\it physical origin} of our models. It is then useful 
to mention 2 concrete examples of $H_{PISCES}$, where $\vec{\tau}_1$ and $\vec{\tau}_2$ represent 
macroscopic systems.  The first is a simple generalization of the ``Giant Kondo" 
model just discussed 
above - we imagine 2 nanomagnetic particles in a conducting background. 
The second involves the much more general "measurement problem", in which one
spin represents the low-energy Hilbert space of some quantum system, and the other the
low-energy Hilbert space of a measuring apparatus; both are coupled to a quantum environment of 
oscillators.

\vspace{4mm}

{\bf (i) \underline{Two Interacting Giant Kondo Spins}}:
We imagine starting 
from a high-energy Hamiltonian analogous to that for the single Giant Kondo system, now with
two nanomagnets coupled to the same conducting "substrate", ie., of form
\begin{equation}
H \ = \ H_o (\vec{S}_1, \vec{S}_2) \ + \ H_{bath} \ + \ H_{int}
\label{}
\end{equation}
where $H = (H_o(\vec{S}_1) + H_o (\vec{S}_2) + H_{dip}(\vec{S}_1, \vec{S}_2)$, 
describing 2 separate giant spins, coupled by a high-energy dipolar interaction; where 
$H_{bath} = \sum_{\vec{k}, \sigma} \epsilon_{k} c^{+}_{k, \sigma} c_{k, \sigma}$ as 
for the giant Kondo model; and finally
\begin{equation}
H_{int} \ = \ \frac{1}{2} \sum_{\vec{k}, \vec{q}} \hat{\vec{\sigma}}^{\alpha 
\beta} . \left(\bar{J}_1 \vec{S}_1 F_{\vec{q}}^{(1)} \ + \ \bar{J}_2 \vec{S}_2 F_{\vec{q}}^{(2)} 
\right) \ c^{+}_{k+q, \alpha} \ c_{k \beta}
\label{1.44}
\end{equation}
is the obvious generalization of the single giant Kondo coupling to the electron bath.

\vspace{3in}

FIG. 6: How the problem of coupled Kondo impurities might arise in practise. We imagine 2 conducting
nanomagnets (perhaps part of an array) embedded in a conducting or even semiconducting substrate. 
Electrons can pass across the boundary between the nanomagnets and the substrate. The nanomagnets are 
a distance $R$ apart, and have radius $R_o$.

\vspace{5mm}

The messy details of the truncation of this Hamiltonian to the PISCES form are given in the original 
paper.  The main point I wish to emphasize here is the way in which the truncation generates an 
{\it indirect} interaction between the 2 nanomagnets, of form 
${\epsilon} (R) \hat{\tau}^{(1)}_z \hat{\tau}^{(2)}_z$, 
where $R$ is the distance between the centres of the nanomagnets, and ${\epsilon}(R)$ is an 
indirect interaction which is calculated to be
\begin{equation}
{\epsilon}(R) \ \sim \ 2g_1g_2 S_1S_2 \int \frac{d^3 r}{V_1} \int \frac{d^3 r^{\prime}}{V_2}
\left(\frac{sin k_F \mid \underline{r} - \underline{r} \mid}{k_F \mid r-r\prime \mid} \right)^2
\label{1.45}
\end{equation}
which, not surprisingly, has the usual RKKY form; for 2 identical nanomagnets one finds \cite{PIS2}
\begin{equation}
{\epsilon}(R) \ \sim \ 144 \pi^{3} \varepsilon_F g^2 S^2 \ \frac{cos(2k_{F}R)}{(2k_F R)^2} \left[
\frac{cos (2k_{F}R_o) \ - \ sin(2k_{F}R_o)}{(2k_{F}R_o)^2} \right]
\label{1.46}
\end{equation}
where $\varepsilon_F$ is the Fermi energy of the substrate (ie., the bandwidth, in this
simple calculation). As remarked above, the generation of such a bath-mediated 
coupling is quite typical. In this case the relative strengths of the dipolar and RKKY couplings
depends very much on the distance between the particles, as well as the electron density- we will
discuss numbers in more detail in Chapter 3.

Another example of the PISCES model is provided 
by a pair of SQUID rings - in this case we generate an interaction (non-Ohmic) 
between the rings once we integrate 
out high-frequency EM field modes.  This is of course nothing but a fancy way of deriving the 
usual low-frequency inductive coupling between them.

\vspace{4mm}

{\bf (ii) \underline{Model for Quantum Measurements}}:

This brings us to a 2nd example, directly connected to the quantum theory of measurement.  The 
standard set-up in a quantum measurement involves some ``measuring coordinate" (necessarily 
macroscopic), and a quantum system which may be microscopic or very rarely macroscopic, which is 
being measured.  Both are in general coupled to their surrounding environment.  Now of course 
there are many different ways in which this set-up can be realised, but the essential features 
are already revealed in a model in which the quantum ``system" of interest exists in one of 
2 interesting quantum states, and the apparatus does the same.  In this case the apparatus 
works as a measuring device if (and only if) its final state is correlated with the initial 
state of the system.

Quite how this works has been discussed {\it qualitatively} for a few models, which tend 
to fall in one or another of 2 classes.  In the first class, the apparatus coordinate of interest 
can exist in one or other of 2 stable (or metastable) states, which are ``classically 
distinguishable".  The best-known example of this is the Stern-Gerlach experiment (where 
atomic beams are widely separated according to their microscopic spin state).  Similar examples 
are provided by SQUID magnetometers, and one may imagine a large number of ``Yes/No" 
experiments, in which the measuring system ends up with the relevant collective coordinate 
(the ``measuring coordinate") in one of the 2 quite different states.  By ``quite different", 
one usually means that (a) the overlap between the 2 quantum states is very small, and (b) that 
the classical counterparts of the quantum states are sufficiently different that one is justified 
in calling them ``macroscopically distinguishable".  Thus, in the case of Stern-Gerlach 
experiment, the beams are sufficiently widely separated that they no longer overlap, 
{\it and} they can be effectively separated by, eg., a microscope when they hit a screen 
or counter.  In the case of a SQUID magnetometer, the 2 relevant states differ by a single 
flux quantum through the SQUID ring - the overlap between the states is via tunneling, and 
hence exponentially small, and the currents flowing in the 2 states can differ by values $\sim$ mA.

A second class of models is typified by the Geiger counter.  In this case we again have a 
``Yes/No" measuring system, designed to discriminate between 2 states of some central system 
under investigation.  In this case the central system is typically subatomic, involving, eg., 
nuclear decay which has/has not occured.  However the measuring system clearly involves a 
measuring coordinate which is \underline{not} microscopic - the Geiger counter is metastable, 
and designed so that a large potential change (kV) occurs between the electrodes if a 
microscopic (subnuclear) decay product interacts with atoms inside the counter.  As has 
been repeatedly stressed in the literature \cite{}, this example well illustrates the 
massive {\it irreversibility} inherent in many measuring systems.  From this point of 
view any further coupling of the Geiger counter to another macroscopic system (eg., to 
Schrodinger's cat) seems almost superfluous as far as the measurement paradox is concerned - 
we are already at the classical level!

The important difference between these 2 sorts of measurement is that 
in the latter case there is a large energy difference 
between the initial and final states of the apparatus coordinate (the energy having been 
fed irreversibly into an ``environment" of other coordinates).

However, with our experience of the truncation procedure fresh in mind, we might legitimately 
ask whether it is possible to strip these coupled models down to a truncated form in which 
only 4 states are involved (2 for the central system, 2 for the apparatus).  All 
irreversible effects arise from a coupling of these states to an environmental bath - if this is 
modelled by oscillators then we are back to the PISCES model.

This is part of the rationale behind recent work by M. Dub\'{e}, E. Mueller, and myself 
\cite{PIS3}.  One aim of this work was to see how valid are some of the models used by 
quantum measurement theorists, by showing how these models may be derived from underlying 
``high energy" microscopic theory (and finding corrections to them).  We shall see that the 
results have interesting consequences for measurement theory.

Before describing the work, let us first see what sort of ``measurement model" we are aiming at.
One very commonly used model (known as the ``Bell/Coleman-Hepp" model \cite{hepp}) takes the form
\begin{equation}
H^{(A)}_{o} (\vec{\tau}) \ + \ H^{(s)}_{o} (\vec{\sigma}) \ + \ \frac{1}{2}
K \ \hat{\tau}_x (1-\hat{\sigma}_z)
\label{1.47}
\end{equation}
where $\vec{\tau}$ is a 2-level system referring to the apparatus, and $\vec{\sigma}$ a 
2-level system referring to the system.  The coupling is such that if $\vec{\sigma}$ is in 
initial state $|\uparrow >$, then there is no effect on $\vec{\tau}$, whereas if $\vec{\sigma}$
is initially $|\downarrow >$, then transitions can occur in the apparatus (with no change 
in the system state).  For this measurement scheme to function properly, one requires the 
initial state of $\vec{\tau}$ to be $|\Uparrow >$ or $|\Downarrow >$.  Typically $K$ will depend 
on time (ie., we turn the coupling on and then off), but this is not crucial for what follows.  
If we let $K = \pi$, then we have the classic ``ideal measurement" scheme \cite{}, for which
\begin{eqnarray}
|\Uparrow> |\uparrow> \longrightarrow |\Uparrow> |\uparrow> \nonumber \\
|\Uparrow> |\downarrow> \longrightarrow |\Downarrow> |\downarrow>  \;
\label{1.48}
\end{eqnarray}
ie., the final state of the apparatus is uniquely correlated with the initial state of the system.

Now in view of our earlier discussion of universality classes and fixed points for 
low-energy Hamiltonians, it is clear that such a scheme can be obtained starting with a 
large variety of initial high-energy Hamiltonians.  It is rather illuminating to consider 
one such starting point- we begin from a Hamiltonian
\begin{eqnarray}
H = \frac{1}{2}(M_A q^{2}_{A} + M_s q^{2}_{s}) + V_A(q_A) + V_s (q_s) - \xi_Aq_A
\nonumber \\  -  K_oq_sq_A + \frac{1}{2} \sum^{N}_{k=1} m_k(\dot{x}^{2}_{k} + 
\omega^{2}_{k}x^{2}_{k})
+ \sum^{N}_{k=1}(c^{(s)}_{k}q_s + c^{(A)}_k q_A)x_k ~\;
\label{1.49}
\end{eqnarray}
It is useful to visualize the ``2-dimensional" potential here, in the space of the coordinates 
$q_s, q_A$ (referring to system and apparatus); see below.  We assume potentials $V_s(q_s)$ and $V_A(q_A)$ 
which describe symmetric 2-well systems, with minima at $q_s = \pm 1, q_A = \pm 1$; however 
we also assume that the action required for the tunneling between the 2 apparatus states is 
considerably smaller than for the tunnelling between the 2 system states.  The coupling 
$-K_oq_Aq_s$ is a ``ferromagnetic" coupling, of quite generic form (produced by expanding the 
interaction in powers of $q_A$ and $q_s$).

When we truncate this down to low energies, we get a PISCES Hamiltonian of form
\begin{equation}
H \sim \Delta_A \hat{\tau}_x + \Delta_s \hat{\sigma}_x - \xi_A \hat{\tau}_z - {\cal K} 
\tau_z \sigma_z + \frac{1}{2} \sum^{N}_{k=1} \left[ m_k(\dot{x}^{2}_{k} + 
\omega^{2}_{k}x^{2}_{k}) + (c^{(A)}_{k} \hat{\tau}_z + c^{(s)}_{k} \hat{\sigma}^{z}_{k}) x_k\right]
\end{equation}
where ${\cal K}$ is a renormalized (but still high-energy) 
coupling, which includes the effect of oscillators above 
a UV cut-off frequency $\Omega_o$.  We will make 2 assumptions here, viz.,
\begin{equation}
(i) \; \Delta_A \gg \Delta_s
\end{equation}
\begin{equation}
(ii) \; {\cal K} > \xi_A \gg \Delta_n, kT
\end{equation}
These 2 assumptions are embodied in Fig.7 for the potential. 

\vspace{3in}

FIG. 7: The effective potential (shown as a contour map, with low energies shaded) 
for a system of 2 coupled two-well systems, with a mutual "ferromagnetic"
coupling and a bias applied to the apparatus coordinate $q_A$ (but not the system coordinate $q_s$).
On truncation to low energies this produces the biased PISCES model of eqtn. (61).

\vspace{5mm}

That $\Delta_A \gg \Delta_s$ 
implies that the apparatus reacts quickly to any change in the system state.  The purpose of 
$\xi_A$ is to hold the apparatus in state $|\Uparrow>$ when there is no coupling; when the 
renormalized coupling ${\cal K}$ acts, it must overcome this bias.  We also want the resulting net 
bias between initial and final states of the apparatus to be much greater than either 
$\Delta_A$ or $kT$, so that there is no chance the apparatus, once it has made a transition, 
can return back to its former state.

Without doing any calculations, we can immediately see from the Figure that this system will 
behave like an ideal measuring device.  The combined system-apparatus starts off either in 
$|\Uparrow \uparrow>$ (and stays there) or in $|\Uparrow \downarrow>$ (in which case it can 
tunnel inelastically to $|\Downarrow \downarrow>$; provided we wait long enough, it will 
always do this, no matter how weak is the coupling of $\vec{\tau}$ to the bath).  Typically 
the apparatus-bath coupling is strong, so that the apparatus will relax quickly.
Notice, however, that the relaxation rate must depend also on the size of the 
transition matrix element $\Delta_A$, and the strong coupling to the bath will renormalize 
$\Delta_A$ down to a considerably smaller value $\Delta^*_A$. Thus we must take care that 
$\Delta^*_A \gg \Delta^*_s$, after both have renormalised (and the weaker coupling of
the system to the bath means that $\Delta_s$ will decrease much less than $\Delta_A$).

In the next Chapter we will discuss the {\it dynamics} of models like these more 
thoroughly.  Notice already, however, one obvious remark that can be made about these models. 
This is that any interesting dynamics possessed by the {\it system} before 
the coupling to the apparatus is switched on, is {\it frozen} by this coupling.  
Thus, if the system is originally tunneling between $|\uparrow>$ and $|\downarrow>$ 
at a frequency $\Delta_s$, the large coupling ${\cal K}$ blocks this completely.  In this 
sense the measurement drastically interferes with the system dynamics.  We shall see in the 
next section that this is a quite general feature of the PISCES model, and most likely of 
measurements in general.  What is more, a good part of ${\cal K}$ comes from the oscillator bath 
itself (indeed $K_o$ already arises from truncation of even higher energy modes), and the 
{\it dissipative} effect of the remaining oscillators (having energy $< \Omega_o$) also 
strongly influences the dynamics of the system once ${\cal K}$ is switched on, in a way which 
is {\it quite different from that produced by the oscillators when} ${\cal K} = 0$.  This 
latter feature, discussed in detail in Dub\'{e} and Stamp \cite{PIS1} in the context of the 
PISCES model, is quite new as far as we know.  We thus see that models of this kind, which 
include both apparatus {\it and} system (as well as the environment), have interesting 
light to shed on measurement theory.  The PISCES model appears to be the first attempt 
to discuss all 3 partners in the measurement operation (system, apparatus, environment) 
on an equal and fully quantum-mechanical level.

\vspace{3in}

FIG. 8: The relationship between the measuring apparatus, the measured coordinate(s) $Q$ of interest,
and the environment (ie., all other relevant degrees of freedom), involved in a typical measurement. The
disciplines of mesoscopic and nanoscopic physics often involve similar situations, in which all variables
must be treated quantum-mechanically.

\vspace{5mm}

It is the experience of this author that many physicists are somewhat averse to the discussion
of abstract quantum measurement problems in the context of down-to-earth questions of solid-state
physics. This is unfortunate, since even if one is uninterested in the foundations of quantum
mechanics, there is no question that one learns a great deal about these down-to-earth problems
by simply placing them in the more general context that the measurement problem requires. This is
becoming really very obvious in the field of nanophysics, where both theorists and 
experimentalists must deal every day 
with the emergence of classical properties from quantum ones, and the 
relationship between the quantum and classical domains. In fact with a bit of hindsight 
one can see the steady evolution of condensed matter theory towards an ever-closer examination
of the interaction between observer (or probe), and a quantum system of interest, in the presence
of dissipation and/or decoherence (cf Fig.8). In modern mesoscopic or nanoscopic physics
there is often not a huge difference in the size of the probe and the quantum system of interest.
Condensed matter physics has come along way since the early days of statistical physics, when
only the {\it equilibrium} behaviour of a {\it macroscopic} system was of interest!

\vspace{7mm}

{\bf  2.4(c)  CANONICAL OSCILLATOR BATH MODELS}

Just as for the spin bath models, we have seen that it is possible to reduce the description 
of a large number of systems, at low energies, to that of a few effective oscillator bath models.

In this subsection I just briefly note some of the other ``canonical" oscillator bath models 
that have also been studied.  In contrast with the more recent spin bath models, many of the 
following are rather old, in some cases with a distinguished history - I will 
refer to reviews for historical details and references.

{\bf (i) Harmonic Oscillator coupled to Oscillators} : A very thorough review 
of this model was given a few years ago by Grabert et al \cite{grab}; we are interested in a 
low-energy Hamiltonian of the standard form in (50), where now
\begin{equation}
H_o (P,Q) = \frac{1}{2}[P^2/2M_o + M_o\Omega^{2}_{o}Q^2]
\end{equation}
and where in most cases one simply uses the bilinear coupling $H_{int} = \sum_{k} c_kx_k Q$.  
In this case the Hamiltonian is obviously exactly diagonalisable, and so the dynamics of 
the central oscillator is exactly solvable.

There are many obvious physical examples; some were already discussed by Feynman and Vernon, 
since many measuring systems are based on resonant absorption by one or more oscillators.  
An example which has stimulated a lot of recent work is that of a gravity wave detector - 
this is usually a very heavy bar of sphere, made from a conducting alley, in which $\Omega_o$ 
represents the principal oscillator (phases) mode.  The coupling to higher phonons, or to 
electron-hole pairs, is deliberately made very weak.  The purpose of these detectors is to 
absorb a single quantum of energy in this mode, from a gravitational wave. An enormous
literature exists on this kind of application; see in particular the books by Braginsky
and collaborators \cite{brag}, which are particularly interesting for the generality of their
coverage. It has indeed been remarked on many occasions that one can understand much of 
Quantum Mechanics with reference to 2 simple models (the 2-level system and the harmonic 
oscillator); in these lectures I am simply coupling these to an environment!

One can obviously also consider systems of coupled oscillators, in analogy with the PISCES model.

{\bf (ii) Tunneling/Nucleating System, coupled to Oscillators}:  Another canonical 
model which has been studied in hundreds if not thousands of papers is that of a particle 
tunneling from a potential well into an open domain - usually the problem is designed so that 
this domain is semi-infinite, and the relevant potential is 1-dimensional.  This restriction 
is actually rather weak, since many multi-dimensional tunneling or nucleation problems can be 
reduced to 1-dimensional one, as already noted in our discussion of domain wall tunneling 
in the spin bath.  In the context of tunneling of systems coupled to oscillators, this question 
has been discussed in great detail by A. Schmid \cite{schmid}.

The relevant Hamiltonian is exactly that in (2.1), with V(Q) representing the tunneling 
potential - this is the famous ``Caldeira-Leggett" model \cite{cal}.  So many discussions have been 
given of how models like this can be derived, that I do not propose adding to them here.

{\bf (iii) Free Particle coupled to Oscillators}.  If we let V(Q) = 0 in (2.1), we 
get a model which is of somewhat academic interest but which can be solved for a variety 
of couplings - it is particularly interesting for the study it allows of different initial 
conditions imposed on the environmental state, and an initial correlations between bath and 
particle \cite{grab}.

\vspace{4mm}

This concludes our survey of the various ways in which one may truncate high-energy 
Hamiltonians down to one or other of the universality classes mentioned in the introduction. 
We now proceed to a survey of the {\it dynamics} of these models.

\section{Tunneling Dynamics: Resonance, Relaxation, Decoherence}
\subsection{Generalities}

There are a number of features common to any problem in which some central quantum system 
tries to tunnel (or otherwise show coherent dynamics), whilst coupled to an environment.  Let 
us first note 2 important mathematical features. One is the division of coupling terms in 
$H_{eff}$ into ``diagonal terms" (operating when the system is {\it not} tunneling), 
and ``non-diagonal terms" (active {\it during} the tunneling).  The difference is 
very clear when the central system is a 2-level one, and easily shown diagramatically, 
for either spin or oscillator baths.  A second feature appears when one comes to average over 
the bath variables, to produce a reduced density matrix for the central system.  Mathematically 
this averaging supposes some distribution of probability over the possible bath states, which 
is assumed {\it invariant in time} - often it is simply a thermal average, assuming a 
thermal equilibrium distribution for the bath states.  For the time invariance to be physically 
reasonable, it is necessary that the energy transferred from the central system, while it is 
relaxing, be rapidly distributed over the bath states, via some kind of mixing.  This 
assumption is usually reasonable for the oscillator bath (although one has to be careful 
sometimes about non-linear effects).  For the spin bath, often containing a finite number N of 
spins (no thermodynamic limit), the assumption can be wrong, and then one has to think more 
carefully (see later).

All these problems also have obvious physical features in common.  The environmental 
wave-function ``entagles" with that of the central system, thereby destroying coherence in the 
latter; and if we start the central system in an excited state, it decays or relaxes 
incoherently by exciting bath modes.  In a large system this may occur via tunneling into a 
quasi-continuum of states (``Macroscopic Quantum Tunneling"), but at low energies it is often 
more common to have tunneling between discrete levels on either side of a barrier.  In this 
case, if the bias $\xi = (\varepsilon_L - \varepsilon_R)$ between these 2 levels is substantially 
larger than the tunneling matrix $\Delta$, resonance is lost and tunneling is suppressed; the 
overlap integral between $|L>$ and $|R>$ is now $\sim \Delta/(\Delta^2 + \xi^2)^{\frac{1}{2}}$, 
and so the new effective "tunneling amplitude" is $\Delta^2/(\Delta^2 + \xi^2)^{\frac{1}{2}}$
(compare the off-diagonal matrix elements in the density matrix of eqtn. (\ref{2wbiasdensityt}),
for the simple 2-level system).  
However at this point the bath can come to the rescue, by absorbing the energy difference 
$\xi$, and thereby mediate inelastic (and thus incoherent) tunneling. Other related processes of 
interest are quantum diffusion \cite{kag1} (in which the diffusion rate of a system in the quantum 
regime {\it increases} as one lowers the temperature), and quantum nucleation 
(which is essentially barrier tunneling).

Quite generally we may characterise the dynamics of the central system via the 
{\it reduced} density matrix; for the general Hamiltonian (1.2) this is
\begin{equation}
\rho(Q, Q^{\prime}; t) \ = \ Tr_{\{q_{k}\}} \left[ \rho_{tot} (Q, Q^{\prime}; \{q_k\}, 
\{q^{\prime}_{k}\};t) \right]\;
\label{3.1}
\end{equation}
where the trace is over the environmental variables (usually with a thermal weighting).  Our 
problem is to evaluate the {\it propagator} $K_{12}$ for $\rho$, defined by
\begin{equation}
\rho(Q, Q^{\prime}; t) \ = \ \int dQ_2 \int d Q^{\prime}_{2} K(1,2) \rho(Q_2, Q^{\prime}_{2}; 
t_2)\;
\label{3.2}
\end{equation}

In this equation, K(1,2) = $K(Q_1, Q^{\prime}_{1}; Q_2, Q^{\prime}_{2}$; $t_1$, $t_2$), and 
is the 2-particle Green function describing the propagation of the central system, first 
from $Q_1$ (at time $t_1$) out of $Q_2$ (at time $t_2$); and then back from $Q^{\prime}_2$ 
(at time $t_2$) to $Q^{\prime}_1$ (at time $t_1$).
K(1,2) simply describes the ``flow" of the reduced density matrix in ($Q, Q^{\prime}$) space 
(Fig.9).  ``Decoherence" in the "$Q$-space representation" then 
corresponds to the suppression of ``off-diagonal" parts of 
$\rho(Q, Q^{\prime})$, for which $Q$ and $Q^{\prime}$ differ significantly.

\vspace{3in}

FIG. 9: The flow of the reduced density matrix in the space of generalised coordinate variables
$(Q,Q')$. The propagator $K(1,2)$ then simply describes this flow. If there is strong decoherence
in the coordinate variables, the suppression of off-diagonal interference then tends to compress
the flow along the streamlines, as in the figure.

\vspace{5mm}

All of these features are true of tunneling systems, regardless of the bath.  However in what 
follows we shall see in detail how going between a spin and oscillator bath can change some 
very important details.

\vspace{7mm}

{\bf 3.1(a) OSCILLATOR BATHS}

As noted by Feynman and Vernon \cite{fey}, the oscillator bath environment is easily integrated 
out because the oscillator actions are {\it quadratic}.  Thus the central system 
density matrix has a propagator
\begin{equation}
K(1,2) \ = \displaystyle \int^{Q_2}_{Q_1} dQ \displaystyle \int^{Q^{\prime}_{2}}_{Q^{\prime}_{1}}
 dQ^{\prime} \ e^{-i/\hbar (S_o[Q] \ - \ S_o[Q^{\prime}])} {\cal F}(Q,Q')\;
\label{3.3}
\end{equation}
Here $S_o[Q]$ is the free central system action, and ${\cal F}(Q, Q^{\prime})$ is the famous 
``influence functional", defined in general by
\begin{equation}
{\cal F}(Q,Q') =  \prod_{k} \langle \hat{U}_k(Q,t)
 \hat{U}_k^{\dag}(Q',t) \rangle \;,
\label{neto.2}
\end{equation}
where $\hat{U}_k(Q,t)$ describes the evolution of the
$k$-th environmental oscillator, given that central system follows the path $Q(t)$ on its
"outward" voyage, and $Q'(t)$ on its "return" voyage. 
 Now we can always write this as
\begin{equation}
{\cal F}(Q,Q') = e^{-i\Phi (Q,Q') } = e^{-i \sum_{k=1}^N \phi_k (Q,Q') }\;,
\label{3.4}
\end{equation}
where $\Phi (Q,Q')$ is a {\it complex} phase, containing both real (reactive) contributions, 
and imaginary damping contributions. Thus ${\cal F}(Q,Q')$ acts as a weighting function, which 
weights different possible paths $(Q(t),Q'(t'))$ differently from what would happen if there 
was no environment (thus, if the oscillators couple directly to the coordinate via a coupling 
$F_k(Q) x_k$, then ${\cal F}_k (Q,Q')$ will typically 
tend to suppress ``off-diagonal" paths in which $Q$ 
and $Q'$ differ strongly).

The crucial result of weak coupling to each oscillator is that an expression  for each 
${\cal F}_k (Q,Q')$ can be written down (and thence for ${\cal F}(Q,Q') = \prod_k {\cal F}_k 
(Q,Q')$), in which $\phi_k (Q,Q')$ is expanded to 2nd order in the coupling only \cite{fey,cal}. Since 
the couplings are $O(N^{-1/2})$, summing over the $N$ oscillators then gives an expression for 
$\Phi (Q,Q')$ which is independent of $N$ and exact in the thermodynamic limit.

\vspace{3in}

FIG. 10: Graphical representation of the influence functional phase $\Phi(Q,Q')$. In (a) I show the 
4 contributions to 2nd order in the coupling $F_k(Q)$; the wavy line is the propagator $g_k(t-t')$
for the $k$-th bath oscillator. Exponentiation gives graphs to all orders; a typical example is 
shown in (b).

\vspace{5mm}

The second-order (in $F_{k}(Q)$) graphs contributing to $\Phi(Q,Q')$ are shown in Fig.10(a);
 the exponentiation of these gives graphs like those in Fig. 10(b).  The bosonic propagator for 
each oscillator, in equilibrium, is just
\begin{equation}
g_k(t) = {1 \over 2m_k\omega_k} \bigg[ e^{i\omega_kt } +2{\cos \omega_kt \over \exp \{ 
\beta \hbar \omega_k \} -1 } \bigg]\;~~~(\beta =1/k_BT )  \;.
\label{2.5}
\end{equation}
Summing over all bosonic modes allows us to subsume all environmental effects into a spectral
 function \cite{fey,cal}, whose form I quote here in the case where the coupling $F_k(Q)$ is 
 linear in $Q$, ie., $F_k(Q) = c_kQ$ (so that the system-oscillator coupling is bilinear):
\begin{equation}
J(\omega) \ = \ \frac{\pi}{2} \sum^{N}_{\kappa =1} \ \frac{\mid c_{\kappa} \mid^2}{M_{\kappa} 
\omega_{\kappa}} \ \delta(\omega \ - \ \omega_{\kappa})
\end{equation}
which of course just has a ``Fermi golden rule" form, characteristic of response functions.  As 
emphasized by Leggett, at low $\omega$ this function will often be dominated by an ``Ohmic" 
term $J(\omega) \sim \eta (\omega)$ (where in general $\eta$ depends on temperature). This is
partly because of course a linear form dominates over higher powers at low $\omega$. However the 
reason for the common occurence of the Ohmic form is that it is linked in the classical dissipative
dynamics to Ohmic friction. In this context it is important that even if one finds, in some
calculation, a non-Ohmic form for $J(\omega)$, it is almost certain that if one pursues 
the calculation to higher orders, an Ohmic coupling will be found. Often however this coupling
will usually be {\it temperature-dependent}, ie., we will have $J(\omega) 
\rightarrow J(\omega, T)$. For
a more detailed discussion of this point, see, eg., Dub\'e and Stamp \cite{dube}, and refs. therein.

The function $J(\omega)$ is usually called the "Caldeira-Leggett spectral function".

\subsection{Dynamics of the Spin-Boson Problem}

As our first example of the dynamics of a system coupled to an oscillator bath, we go to 
the famous spin-boson problem.
There are several detailed reviews of the spin-boson dynamics \cite{leg,AJL,weiss} (see also
Leggett's notes in this volume).  Here I summarize results relevant to the present topic.

Although one can imagine a whole variety of forms for the spectral function $J(\omega)$
in this problem, in many systems it is the Ohmic coupling which dominates at low energies, 
and I will only talk about this case here. Readers interested in the finer details can go to 
the reviews, particularly if they are interested in non-Ohmic spin-boson systems. 
For the Ohmic spin-boson problem one writes $J(\omega) = \pi \alpha \omega$, and 
enormous effort has been devoted to understanding the phase diagram and dynamics of the 
central spin, in contexts ranging from particle theory, the Kondo problem, 
defect tunneling and quantum diffusion, flux tunneling in SQUID's, and 
1-dimensional fermion problems. From these studies we recall several results.  First,
 the well known phase diagram for Ohmic coupling \cite{leg} (Fig.11). 
 
\vspace{3in}

FIG. 11: The "phase diagram" showing the different dynamic behaviours possible for an
{\it unbiased} 2-level system coupled to an Ohmic bath, as a function of the dimensionless
coupling $\alpha$ to the bath, and the temperature $T$. At $T = 0$ and $\alpha > 1$, the bath 
oscillators "quantum localise" the system- it freezes. For $\alpha \gg 1$, the system is 
quasi-localised for temperatures up to a crossover temperature $T_c \sim \Omega_o/2 \pi$. Elsewhere
the system shows overdamped relaxation (exactly exponential along the Toulouse line $\alpha = 1/2$),
except for low $T$ and low $\alpha$.

\vspace{5mm}
 
 The most important feature here is 
how delicate quantum coherence turns out to be, as a function of the dimensionless parameters 
$\alpha$ and $T/\Delta_R$, where $\Delta_R(\alpha)$ is a renormalized tunneling amplitude.
  This is of course because (i) the oscillators suppress $\Delta_o$ to $\Delta_R$, via a 
standard Franck-Condon effect, and (ii) the coupling $c_{\kappa}$ to $\hat{\tau}_z$ allows the 
oscillators to sense or ``measure" the state of $\hat{\tau}$, and thereby suppress coherent 
oscillations.

A second important result, which concerns many oscillator bath models (not just the spin-boson 
model), is the relationship between the classical and quantum behaviour of the system coupled 
to the oscillator bath.  It is of course trivially obvious that if the spectral function 
$J(\omega; T)$ contains all information about the system/bath coupling, then both the 
classical frictional dynamics and the low-T tunneling dynamics are determined by the same 
$J(\omega,T)$.  In this sense knowledge of the classical motion may help us in predicting 
the quantum dynamics (albeit at a different T).  However in those (not uncommon) cases where 
$J(\omega)$ = $\eta \omega$, with $\eta$ independent of $T$, Caldeira and Leggett showed that 
knowledge of $\eta$ obtained from the classical friction entirely determined the quantum dynamics.

One should, nevertheless, refrain from pushing this result too far.  In particular, it is a 
serious mistake to imagine that one can understand the {\it decoherence} in the 
quantum dynamics, solely from a knowledge of the classical dynamics.  A related mistake is 
the idea that there is some well-defined ``decoherence time", referring to exponential 
relaxation of off-diagonal matrix elements - as a rule this is incorrect, 
even in the oscillator bath models.

To better appreciate the above results, a brief word on the mathematical techniques is in 
order.  To solve for the dynamics of the spin-boson model, Leggett et al. \cite{leg} employed 
an approximation dubbed the ``NIBA" (``Non-Interacting Blip Approximation"), in order to 
evaluate the influence functional.  This approximation simply assumes that the system density 
matrix (with oscillators already averaged over) spends most of its time in a {\it diagonal} 
state (as opposed to an off-diagonal one).  It is useful to understand this approximation 
in 3 different ways.

\vspace{3in}

FIG. 12: A typical path $Q(t)$ for a 2-state system coupled to an environment. The 2 levels
result from truncating a higher-energy Hamiltonian containing many levels (here a giant
spin $\vec{S}$). The fast high-energy fluctuations are invisible, and only the rapid (on a
timescale $\Omega^{-1}$) transitions between the 2 classically stable states $\vec{S}_1$ and
$\vec{S}_2$ are seen. The system interacts diagonally (D) or non-diagonally (ND)
with the environment; in the latter case a non-linear coupling to an environmental triplet is
shown (as in the coupling to the spin bath).

\vspace{5mm}

(i)  diagramatically, the system is in a diagonal state (dubbed a ``sojourn" by Leggett et 
al. \cite{leg}), when both paths contributing to the density matrix evolution happen to be in the 
same state.  An off-diagonal (or ``blip") state, where the 2 paths at a given time are in 
opposite states, is then considered to be rare.

(ii)  mathematically, the suppression of the off-diagonal elements arises because the 
propagator $g_k(t)$, for the oscillators, is complex, and the imaginary part damps
out off-diagonal blip states. Notice that by "diagonal" we mean diagonal 
in $\hat{\tau}_z$; and the environment couples to $\hat{\tau}_z$.

Formally we can express this point in terms of the influence functional as follows. For an 
arbitrary path of the system, the complex phase $\Phi (Q,Q')$ in (\ref{3.4}) can be written
\begin{equation}
\Phi (Q,Q') =
\int_{0}^{t} d \tau \int_{0}^{\tau}ds
( i\Sigma (\tau -s)  \xi (\tau )\chi (s)
- \Gamma (\tau -s)   \xi (\tau )\xi (s)  )
\label{phase}
\end{equation}
where the double path is parametrised by sum and difference variables $\xi (s) = Q(s) - Q'(s)$
and $\chi(s) = Q(s) + Q'(s)$ (again we assume a coupling of the oscillators to $Q$). Now for 
the specific case of the spin-boson system system, the paths are very simple; they are nothing
but sums over outgoing or return paths of the form
\begin{equation}
Q_{(n)}(s)=1-\sum_{i=1}^{2n}\big[ sgn(s-t_{2i-1})+
sgn(t_{2i}-s) \big] \;,
\label{neto.1}
\end{equation}
where $sgn(x)$ is the sign-function, $n$ is the number of
transitions of the central system, occuring at
times $t_1,t_2, \dots ,t_{2n}$ (and here for definiteness we assume 
trajectories starting and ending in the same state). Notice that the paths have sudden jumps
because once we truncate our Hamiltonian to energies $ \ll \Omega_o$, we have no way of resolving
processes occurring over the time scale $\Omega_o^{-1}$ of the instanton jump- to all intents and
purposes they are instantaneous! I have shown in Fig.12 a typical such path (including the 
couplings to the environmental modes, which in a path integral formalism take place at particular
times). 

Notice that now the off-diagonal 
"blip" states are simply the ones having
finite $\xi(s)$ (and $\chi(s) = 0$). 
How and when are these suppressed by the oscillator bath? Well, we can see this by looking at the 
forms of the reactive and dissipative contributions to the phase $\Phi (Q,Q')$, which can easily
be calculated in terms of the spectral function $J(\omega)$ for the bilinear coupling model
(which just integrates over the complex propagator $g_k (t)$):
\begin{equation}
\Sigma (\tau -s) = \int^{\infty}_{0} d \omega \,
J (\omega ) \,
\sin\omega (\tau -s)
\end{equation}
\begin{equation}
\Gamma (\tau -s) = \int^{\infty}_{0} d \omega \,
J (\omega ) \,
\cos\omega (\tau -s) \coth(\omega /2T)
\end{equation}
>From this we see that whether blips are suppressed relative to sojourns depends very much on the 
form of $J(\omega)$ (for a detailed discussion see Leggett et al. \cite{leg}), but if they are, it is 
because of the imaginary part $\Gamma$ of the complex phase.

(iii) Finally and most physically, we can see that  
the diagonal state of the density matrix is favoured because the environment
is in effect continually ``measuring" $\hat{\tau}_z$ (the coupling
$\sum_{k} c_{k} x_{k} \hat{\tau}_z$ distinguishes between $|\uparrow>$ and $|\downarrow>$);
this suppresses quantum interference between the two states (ie., off-diagonal elements).
Notice that we can think of this as a ``dynamic localisation" of the system in one or
other state for long periods, and the opportunity for tunneling is reduced to periods when
resonance between the 2 wells persists long enough for tunneling to occur.
Note that this is not really the same as the "degeneracy blocking" we encountered before, 
which comes from a {\it static} bias field. Nevertheless it does bring us to perhaps the 
most important question about the spin-boson model, which is, what happens to the spin when
there is an {\it external} bias field $\xi$, and a finite temperature? 
I will simply quote the (practically very useful)
answer for you here \cite{}. Suppose the system starts up in a state $\vert \uparrow \rangle$ at $t=0$. 
Then at time $t$ later the probability $P_{11}(t)$ that it will still be
$\vert \uparrow \rangle$ is given by the simple {\it incoherent relaxation} expression 
$P_{11}(t) = e^{-t/\tau(T,\xi)}$, where for the most interesting case of Ohmic
coupling, the rate $\tau^{-1}$ is given by (assuming energy scale less than $\Omega_o$ as usual): 
\begin{equation}
\tau^{-1}(\xi,T)  = \frac{\Delta^{2}}{2\Omega_{0}}
\left[ \frac{2\pi T}{\Omega_{0}} \right]^{2\alpha -1}
\frac{\cosh(\xi /2T)}{\Gamma(2\alpha)}
| \Gamma(\alpha +i\xi /2\pi T) |^{2}
\label{decayratesover}
\end{equation}
where $\Gamma(x)$ is the Gamma function. This result is valid for all but the smallest biases,
and even when the bias is $\leq \Delta$, it is valid throughout the incoherent relaxation
region of the phase diagram in Fig. XXX. Thus even at very low bias we get relaxation at a rate
\begin{equation}
\tau^{-1}(T)  = \frac{\Delta^{2}}{2\Omega_{0}}
\frac{\Gamma^{2}(\alpha)}{\Gamma(2\alpha)}
\left[ \frac{2\pi T}{\Omega_{0}} \right]^{2\alpha-1}
 + O(\xi /T)^{2}
\; \; \; \; \; \; (\xi /T \ll 1)
\label{highT}
\end{equation}
whereas in the opposite limit where we apply a strong bias $\xi \gg T$ to the system, we 
get incoherent fluctuations between the 2 states:
\begin{equation}
\tau^{-1}(\xi)  = \frac{\Delta^{2}}{\Omega_{0}}
\frac{1}{\Gamma(2\alpha)}
\left[ \frac{\xi}{\Omega_{0}} \right]^{2\alpha -1}
\label{frozen}
\end{equation}
These results have been applied to a large variety of problems in physics, particularly those
involving conduction electron baths. Their application to SQUID dynamics has been particularly
emphasized by Leggett \cite{AJL,ajl84}.

\vspace{4mm}

{\bf \underline{Application to the Giant Kondo model}}: The application to the problem of a 
nanomagnet coupled to a conducting bath illustrates some of the dramatic features of the 
above results. Consider the example given previously of an $Fe$ particle with $S = 300$ and 
dimensionless coupling $\alpha_{\kappa} \sim 20$ to the electronic bath. Then at zero bias,
the incoherent fluctuation rate goes as the $40$-th power of temperature, whereas at low $T$,
the incoherent relaxation rate goes as the $40$-th power of bias! Consider now an array
of such particles, which we imagine to be functioning as a computer data storage bank. If we
are at low $T$, and small bias, the giant spins are "quantum localised"; they only relax over
eons of time (the crucial point is to stay below the crossover temperature $T_c = \Omega_o/2 \pi$; 
if $T = T_c/2$, the relaxation time is already $10^6$ years!). Thus even though these particles 
are no larger than small molecules, they can store information for astronomical time periods
below $T_c$ (and recall that $T_c$ can easily be 1 $K$ if we have a reasonably strong anisotropy!).
On the other hand suppose we keep $T$ well below $T_c$, and raise the bias. Then over the time-window
that might be of interest to data storage (ie., between $msec$ and years), we will see a very
sudden "switching-on" of the dynamics of the giant spins, in the small region of bias around 
$\xi \sim \Omega_o$. 

This result is obviously of potential practical importance (as already
discussed in the original paper \cite{prok2})! In this context an experimental test of these results
is highly desirable. Note that one cannot of course extend these results above $T, \xi > T_c$,
since our truncated Hamiltonian does not work there (and in fact when $T > T_c$, we go into the
thermally activated regime). Note also that these results also do not account for the effects
of the nuclear spin bath; for the combined effects of the 2 baths, see Prokof'ev and Stamp \cite{prok2}.

\subsection{Other Bosonic Models}

The spin-boson model reveals much about the low-energy dynamics of a system coupled to 
an oscillator bath - but much is obviously missing as well.  In what follows I quickly 
review the main results for a few other canonical models.
 
\vspace{7mm}

{\bf 3.3 (a) The PISCES MODEL}

This is the simplest non-trivial generalization of the spin-boson model.  As we saw in 
section 2.4, what is interesting here is the communication between the spins via the bath, 
and the way in which each affects the dynamics (including the coherence properties) of the 
other.

To characterize the dynamics of the model one calculates the 2-spin reduced density 
matrix $\rho(\tau_1 \tau_2; \tau_{1}^{\prime} \tau_{2}^{\prime}; t)$, which is a 4x4 time-
dependent density matrix.  Instead of discussing how these calculations were done (for which 
see the long paper of Stamp and Dub\'{e} \cite{PIS1}), I think it will be more useful here to 
give an intuitive feeling for the results, and some of their consequences for coupled 
mesoscopic or macroscopic systems.

The most interesting examples of the PISCES model have Ohmic coupling of each spin to the 
oscillator bath (cf. section 2.3).  In this case one can delineate the behaviour of each spin 
by a ``phase diagram" in the 4-dimensional space of couplings $\alpha_1, \alpha_2, {\cal J (R)}$, and 
$T$, where $\alpha_1, \alpha_2$ are the Ohmic coupling of each spin to the bath, and ${\cal J}$ the 
total static coupling between the spins, including the bath-mediated coupling $\epsilon (R)$ 
(section 2.4). The
other 2 energy scales in the problem are the $\Delta^*_{\beta}$, where $\beta = 1,2$ labels the 
2 spins, and the star superscripts indicate the {\it renormalised} tunneling matrix elements (after
coupling to the bath). Then there are 4 dimensionless parameters in the problem, viz., 
$\Delta^*_{\beta}/\alpha_{\beta}, {\cal J}/T$, and $T$ itself; this is our parameter space.

The simplest case to discuss is where the spins are identical (ie., 
$\Delta_1 = \Delta_2 = \Delta$, and $\alpha_1 = \alpha_2 = \alpha$).  The behaviour 
can then be displayed in a simple way as a function of ${\cal J}(\vec{R})/\alpha$ and $T$ (cf. Fig.13), 
and we can identify 4 different regions in this phase diagram, as follows:

\vspace{3in}

FIG. 13: The phase diagram for the symmetric PISCES model (both spins identical), for an Ohmic 
bath, as a function of the dimensionless parameters $ln(T/\Delta^*)$ and $ln({\cal J}/\Delta^*$, 
where ${\cal J}$ is the full renormalised coupling between them. We assume the 
dimensionless coupling $\alpha$ of each to the bath is $\ll 1$, so the mutual coherence phase 
appears.

\vspace{5mm}

(i) The ``Locked Phase'' (${\cal J} \gg T, \Delta^{*}_{\beta}/\alpha_{\beta}$):
In this regime, the effective coupling ${\cal J}$ is so strong that the 2
spins lock together, in either the states $| \uparrow \uparrow \rangle$ or
$| \downarrow \uparrow \rangle$ depending on the sign of ${\cal J}$.  
One finds that the combined ``locked spin'' oscillates between $|\uparrow \uparrow \rangle$ and
$| \downarrow \downarrow \rangle$ (for FM coupling), or between $|\uparrow \downarrow \rangle$ and
$| \downarrow \uparrow \rangle$ (for AFM coupling), at a renormalised frequency
$\Delta_{c} = \Delta_{1} \Delta_{2} / | {\cal J}|$. This result is not at all surprising-
in fact one would get the same result just by coupling 2 tunneling spins in the 
complete absence of the environment! However the oscillations
are now {\it damped} - in fact the locked spin now behaves like a single spin-boson
system, with a new coupling $\alpha_{c}=\alpha_{1}+\alpha_{2}
\pm 2 \alpha_{12}$ to
the oscillator bath, the $+$ ($-$) corresponding to ferromagnetic
(antiferromagnetic) coupling between the spins. For the dynamics of the locked
spin one simply then refers back to the spin-boson results.

(ii) The ``Mutual Coherence'' phase ($\Delta_{\beta}/\alpha_{\beta} \gg T \gg
{\cal J}$; ${\cal J} > \Delta_{\beta}$):
Here, the thermal energy overcomes the mutual coupling;
nevertheless if the dissipative couplings $\alpha_{\beta}$'s are sufficiently
small ($\alpha_{\beta} \ll 1$), it is possible for the energy scale
$\Delta_{\beta}^{*}/\alpha_{\beta}$ to dominate even if
$\Delta_{\beta} < {\cal J}$. In this case, even though we are dealing with a
strong coupling, and the bath dissipation is still important, some
coherence in the motion of each spin is maintained - moreover, the small
${\cal J}$ causes ``mutual coherence'' between the two spins, ie., their
damped oscillations are correlated to some extent. The analytic form of the
correlation functions is extremely complex (see the original paper \cite{PIS1}), and 
moreover is not given simply in terms of the single spin-boson results; the mutual
correlations are essential in this regime. However 
the physics is fairly easy to understand- we are seeing a kind of "beat" or
"breather" oscillation between the 2 spins, which the damping has not quite 
succeeded in destroying. In addition, each spin is exhibiting its own weakly damped
oscillations. The situation is reminiscent of 2 guitar strings, which couple through 
the air (and the guitar box), but where the air also causes weak damping.

\vspace{3in}

FIG. 14: The Probability $P_{\tau^z_1 \tau^z_2}(t)$ for the 2 PISCES spins to be in state
$\vert \tau^z_1 \tau^z_2 \rangle$ after a time $t$, if at $t = 0$ they started in state 
$\vert \uparrow \uparrow \rangle$. The system is in the "Correlated relaxation" regime, 
with time shown logarithmically, in
units of inverse temperature. We assume $\alpha_1 = 1.5$, $\alpha_2 = 2$, $T/\Delta_1 = 100$, 
$T/\Delta_2 = 300$, and a ferromagnetic coupling ${\cal J}/T = -0.04$.

\vspace{5mm}

(iii) The ``Correlated Relaxation'' or High-T phase ($T \gg \Delta_{\beta}/
\alpha_{\beta}, {\cal J}$). In this regime, the bath causes each spin to relax
incoherently; however, the relaxation of the two spins is still correlated
(indeed each spin relaxes in the time-dependent bias
generated by the other). This regime is, along with the locked regime, the most typical. 
The coupling between the 2 spins, via the bath oscillators, simply decoheres completely the 
spin dynamics of each, and they are both strongly damped. You can find the analytic
expressions for their damping in the original paper \cite{}; what is perhaps remarkable
is that the system exhibits 3 different damping rates, each of which is however a fairly simple
algebraic function of the damping rates for 2 different single spins, each 
coupled to an oscillator bath and in a {\it static} bias
field $\xi = {\cal J}$ (cf. equation (\ref{decayratesover})). In this sense the 
problem is not fundamentally different from the single spin-boson model. Here I just show a
typical relaxation of the 4 diagonal elements of the 2-spin density matrix, which 
illustrates the vastly different relaxation timescales which can result during the 
relaxation of 2 initially parallel spins (Fig.14).

(iv) Finally, and much less interesting, the ``perturbative regime''
(${\cal J} \ll \Delta_{\beta}^{*}$), in which the total coupling is so
weak that the 2 spins relax almost independently; all correlations can
be handled perturbatively, and only weakly affect the behaviour that one
calculates from the standard spin-boson model.

The remarkable thing is that {\it analytic} results for the dynamics of the 2 spins can
be found, although we shall not need these in this lightning survey.
Readers familiar with some particular case of the PISCES model will doubtless recognise
features of the phase diagram. Thus, eg., those who have looked at the problem of 2 coupled 
Kondo impurities will recognise the "locked regime" from the old studies of Wilkins
et al \cite{wilk}. The advantage here is that instead of just extracting a phase diagram
(which is easily done using renormalisation group methods), we have been able, using 
instanton methods, to extract the dynamic properties as well.

Perhaps the most important thing to take away from these results is just how easily the 
quantum dynamics of each spin can be decohered or even blocked by the other.  We should not be too 
surprised by this - recall how the environment causes a ``dynamical localisation" of the spin 
in the single spin-boson model.  However it is much more serious in the PISCES model, because 
to the weak random environmental bias we have added a much stronger dynamical bias coming from 
the other spin.  From this point of view it is actually interesting to compare the PISCES 
dynamics with that of a much simpler ``toy model", in which the environment is eliminated 
completely, and we just have 2 coupled spins (see Dub\'{e} and Stamp \cite{PIS1}).

Let us now fill out this picture by applying it to the 2 physical cases discussed in section 2.4.

\vspace{4mm}

{\bf (i) \underline{2 coupled Giant Kondo spins}}: As mentioned in the previous sub-sections, 
the dynamics of a Giant Kondo spin is of potential technological importance.  However it is also 
obvious that in a nanomagnetic array, interactions between the nanomagnets can alter the 
behaviour of a single nanomagnet.  Since there is obviously a large number of different cases 
that can be discussed \cite{prok2}, I will simply give you a taste of the results.  Consider again 
the example in which a Fe-based nanomagnet is embedded in a conducting film, whose conduction 
electron density can be varied.  We again assume $S = 300$, and now also assume 
$\Omega_o = 1K, \Delta = 1mK$, and 
a typical nuclear bias energy $E_o \sim 6mK$ (roughly 2 per cent of the $Fe$ nuclei have spins,
and the hyperfine coupling for each is of order $2-3 mK$).  
However we now assume a second such nanomagnet, 
at a distance $R$ from the first.  The effective bath-mediated interaction 
$\epsilon (R)$ between them depends 
strongly on $R$ (as $1/R^{3}$), but even more strongly on the conduction electron density - it goes 
as $(N(0))^4$, where $N(0)$ is the Fermi surface density of states!

Let us suppose that $R = 300 \AA$.  Then the dipolar interaction between the spins is 
$V_{dip} \sim 10^{-5}K$.  The conduction electron mediated interaction between the 2 
nanomagnets in a typical metallic film, with $g \sim 0.1$ (so that $\alpha \sim 20$) is 
then $\epsilon (R) \sim 30mK$, ie., 3000 times larger then this (and 30 times larger than $\Delta$).  
In this case it is clear that $\epsilon (R)$ has no real effect on the dynamics of the individual 
giant spins, which is completely controlled by the very large $\alpha$. As we saw previously,
the 2 spins relax independently at high $T$, but at low $T$, their dynamics are frozen into
a quantum localised state. As discussed by Prokof'ev and myself \cite{prok2}, this localisation could
be of great practical importance for future data storage at the molecular level.  
Notice however that if $R = 100 \AA$, $\epsilon (R)$ climbs to $1K$, ie., 
$\epsilon (R) \sim \Omega_o$, and the 
interaction will completely change the dynamics of the 2 spins - apart from tending to 
unlock the spin dynamics of each, it will also take them out of the Hilbert space of the 
2 lowest levels.  The dynamics of an array of such spins would be very complicated, 
even at $T = 0$.

Now instead suppose that we drop the conduction electron density by a factor of $\sim 10$, so 
that $ g \rightarrow 0.01$.  As previously, this means that $\alpha \rightarrow 0.2$, thereby 
unlocking the giant spin dynamics at all temperatures.  It also means that $\epsilon (R)$ is 
drastically reduced - it falls by a factor of $10^4$ (so that for $R = 300 \AA, \epsilon (R) $ is now 
$3 \times 10^{-6}K$), and is now smaller than $V_{dip}$!  It is also much smaller than $\Delta$, 
and in fact the dynamics of the spins, in zero applied field, will be controlled at low $T$ 
by the random nuclear bias, since $E_o $ is now the largest energy scale.  The spins will again have 
essentially independent dynamics, when $R = 300 \AA$; but now because we are in the perturbative 
regime.

Notice, however, that if instead $R = 100 \AA$, then $V_{dip}$ rises to $0.3 mK$ (and $\epsilon (R)$ 
to $0.1 mK$), ie., it is of similar size to $\Delta$ and $E_o$.  Obviously by playing with 
$\alpha$ and $R$ it is then possible to create a situation where mutually coherent 
oscillations can occur; otherwise, depending on the exact conditions, we will be either in the
correlated relaxation or the locked phases.  
A nanomagnetic array in the regime might be expected to show very 
interesting quantum diffusive properties in the $T = 0$ limit.

Finally one can imagine further reducing the electron density and/or $S$, to produce a situation 
where $\alpha \ll 1$, but $V_{dip}$ is of the same order as $\Delta$ (and both are larger 
than $E_o$). This is the "weak coupling" regime.
In this case one's first naive guess might be that a nanomagnetic array in this regime 
should behave coherently, with an energy band 
of width $\sim \Delta$.  It is clear that such an array is not beyond present technology to 
prepare.  This would be a remarkable display of macroscopic coherence behaviour, if true- but have 
we left anything out of the model? 

The answer is of course that we have. 
We have left out the nuclear dynamics, ie., the dynamics of the background spin bath.  This is 
not so important when $\alpha$ is large, but if $\alpha \ll 1$, it is essential, and it will destroy 
the coherence of our giant spins, as we shall see in the next subsection (as well as in section 4).

\vspace{4mm}

{\bf (ii) \underline{The Measurement Problem}}: Let us very quickly look at the analagous 
implications for our measurement model (which we have largely anticipated already, 
in introducing it above). Recall that the model discussed in section 2.3 
for a quantum measurement involved an overdamped but rapidly responding apparatus (ie., 
$\Delta_A$ large, $\alpha_A > 1$) and an underdamped system with slow dynamics ($\Delta_s$ small, 
$\alpha_s \ll 1$).  When a measurement is made, an interaction K(t) is switched on, which is 
sufficiently strong to allow the apparatus state to correlate with that of the system - however 
it is not strong enough to cause any transitions in the system state. What actually happens is 
that as soon as K(t) exceeds the small $\Delta_s$, the system dynamics are frozen, and the apparatus 
state subsequently evolves to correlate with the frozen system state.

The quantitative calculations of the ``Overdamped plus Underdamped" case \cite{PIS1} confirm this 
picture - I do not propose giving into the details here!  There is obvious practical application 
of these results to the problem of the observation of macroscopic quantum coherent behaviour (on 
the part of, eg., a SQUID system, or a nanomagnet) - the most obvious way of determining the 
dynamics of such systems is by coupling in another SQUID magnetometer, or a MFM (``Magnetic 
Force Microscope") to the tunneling system.  It is clear that the combined system/apparatus/
environment problem can be described by a PISCES model (provided we can ignore nuclear 
dynamics - see below)

\vspace{4mm}

{\bf (iii) \underline{Miscellaneous Models}}:
To complete this discussion of results for
the dynamics of various commercial oscillator bath models, a brief word on the dynamics
of the models noted in section 2.4(c).

The simplest such model is of a ``central" harmonic oscillator coupled to a bath of
oscillators.  If we assume the bilinear coupling $H_{int} = \sum_k c_kx_kQ$ between
the two, the problem is exactly solvable; the details were discussed at great length by
Grabert et al. \cite{grab}.  Just as in the standard problem of a classical dissipative
oscillator, the results are most easily discussed in the Laplace transform domain.
Thus suppose we have a central oscillator of mass $M_o$ and frequency $\Omega_o$.
The classical equation of motion then involves a dissipation coefficient $\eta (\omega)$,
in general frequency dependent; if $\eta (\omega) \rightarrow \eta $, we have Ohmic
dissipation.  In terms of the Caldeira and Leggett spectral function $J(\omega)$,
the Laplace-transformed function $\eta (s)$ is
\begin{equation}
\mu(s) = \int^{\infty}_{0} \frac{d \omega}{\pi} \frac{J(\omega)}{\omega} \frac{2s}{s^2 + \omega^2}
\end{equation}
Then the retarded response function $\chi(t)$ of the central oscillator is most conveniently
written as $\chi(t) = C(t) \Theta (t)$, where
\begin{eqnarray}
C(s) &= & \int^{\infty}_{0} dt C(t) e^{-st} \nonumber \\
&= & \frac{1}{M_o(\Omega^{2}_{o} + s^2) + s\eta (s)}
\end{eqnarray}
a result which demonstrates very nicely the correspondence, strongly emphasized by
Caldeira and Leggett \cite{cal}, and mentioned in section 3.2 above, between the classical
and quantum behaviour.

In the case of the tunneling problem, there is a huge literature on the dynamics.  This
work is so extensive that I will simply refer you to the literature here - it is not
central to the present discussion, and moreover is discussed in the chapter by Leggett
in this book.  The original long paper of Caldeira and Leggett \cite{cal} is a very good starting 
point. Hanggi et al. \cite{hang} discuss the connection with reaction rate theory, and a
very fine set of reviews of the whole field appear in the book edited by Kagan and Leggett
\cite{kagleg}.  Finally, a good introduction to the theoretical techniques is given by Weiss \cite{weiss}.

For work on the dynamics of a free particle, coupled to an oscillator bath, see the review of
Grabert et al. \cite{grab} again (as well as the paper of Hakim and Ambegaokar \cite{Hak}). 

So much for the dynamics of systems coupled to oscillator baths; now we turn to the very different
effect of the spin bath.

\subsection{Central Spin Model}

In our discussion of oscillator bath models I occasionally warned that the results would be 
seriously compromised by any spin bath that coupled to the central system.  Here we shall see 
how this happens, and review results for the dynamics of a central system coupled to a spin bath.  
It is crucial here to recall the essential difference between the 
oscillator and spin bath models- whereas the oscillators are only very weakly affected by 
the central system (so that their spectra and dynamics are hardly altered), the spins in the 
spin bath have their spectra and dynamics changed completely.

The discussion of the present section is based on papers by 
myself and Prokof'ev \cite{pro,prok,prok2,prok3} (see also ref. \cite{prokcm}), for which see the gory details - here I 
attempt an intuitive perspective on these, and briefly note their application to 
nanomagnets (discussed in detail in Chapter 4), and SQUIDs.

\vspace{7mm}

{\bf 3.4(a) THREE LIMITING CASES}:  

Let us first recall the low-energy effective 
Hamiltonian we derived for the central spin model, whose derivation was described in section 
2.2.  The Hamiltonian, in the absence of an external field, was
\begin{eqnarray}
H_{\mbox{\scriptsize eff}} & = & \left\{ 2 \Delta_o \hat{\tau }_ -  \cos
\bigg[ \pi S - i \sum_k ( \alpha_k \vec{n} \cdot \hat{\vec{\sigma }}_k  +
i \beta_o \vec{n} \cdot \vec{H}_o ) \bigg] + H.c. \right\} \nonumber \\
& + & {{\hat \tau }_z \over 2} \sum_{k=1}^N \omega_{k}^{\parallel} \:
{\vec l}_k \cdot {\hat {\vec \sigma }}_k  + {1 \over 2} \sum_{k=1}^N
\omega_{k}^{\perp}\: {\vec m}_k \cdot {\hat {\vec \sigma }}_k
+\sum_{k=1}^N \sum_{k'=1}^N V_{kk'}^{\alpha \beta } \hat{\sigma}_k^\alpha
\hat{\sigma}_{k'}^\beta ~ \;.
\label{1.24a}
\end{eqnarray}
where $\vec{\hat{\tau}}$ represents the central spin, and the $\{ \sigma_k \}$ the spin bath 
variables; the notation is that of section 2.2.

What we wish to do here is calculate the dynamics of $\vec{\tau}$ after integrating out the 
bath variables - our problem is similar to that addressed in section 3.2, only the 
bath has changed.  I shall describe the physics in a tutorial manner, by first introducing you 
to 3 limiting cases, and then showing how the complete solution is an amalgam of these cases.

\vspace{4mm}

{\bf (i) \underline{Topological Decoherence Limit}}:  This limiting case removes all of the 
static coupling between the central spin and the spin bath, and also removes the intrinsic 
spin bath dynamics, by supressing $V^{\alpha \beta}_{kk^{\prime}}$. The resulting Hamiltonian is
\begin{equation}
H_{\mbox{\scriptsize eff}}^{top} = 2 \Delta_o {\hat \tau }_x
\cos \big[ \pi S  -i \sum_{k=1}^N
 \alpha_k {\vec n}_k \cdot {\hat {\vec \sigma }}_k  \big] \;,
 \label{4.1a}
 \end{equation}
The essential feature of this model is that all of the spin bath dynamics is driven by the 
central spin - moreover, from the form of $H_{eff}$, we see that to the Berry phase $\pi S$ of 
the central spin is added a complex phase $-i \Sigma_k \alpha_k \vec{n}_k \cdot \vec{\sigma_k}$ 
coming from the bath.  Thus the main effect of the bath transitions is to add a {\it random phase} 
to the action incurred during each central spin transition - in effect the topological phase 
in the system dynamics is randomised. Both the physical and mathematical aspects of this case 
have been discussed in some detail in the original references , and I will not repeat 
this here.  The main ideas can be entirely understood in the special case where we choose
$\alpha_k$ to be entirely {\it imaginary}; and I will simply now refer to this imaginary
quantity as $\alpha_k$. The solution can then be written formally as
\begin{equation}
P_{11} (t) = {1 \over 2} \left\{ 1 + \langle \cos \big[ 4\Delta_o t
\cos \big( \Phi   + \sum_{k=1}^N
 \alpha_k {\vec n}_k \cdot {\hat {\vec \sigma }}_k \big)  \big]
  \rangle \right\}\;,
  \label{4.4a}
  \end{equation}
where $<...>$ is an average over the environmental states, and we have made the replacement 
$\pi S \rightarrow \Phi$, to take account of any high-frequency ($ > \Omega_o$) renormalization 
of the central spin Haldane phase, caused by the nuclear bath.  In the present model all bath 
states are degenerate, and this average is trivial - it can be written as a weighted integration 
over topological phase $\varphi$, as
\begin{eqnarray}
P_{11}(t) &= &
\sum_{m=-\infty }^{\infty} F_{\lambda '}(m)
\int_{0}^{2 \pi} {d\varphi \over 2 \pi } e^{i2m(\Phi -\varphi )}
\left\{ {1 \over 2}+{1 \over 2} \cos (2 \Delta_o ( \varphi ) t) \right\}
\label{4.8a} \\
&= &{1 \over 2} \left\{ 1+\sum_{m=-\infty }^{\infty} (-1)^{m}
F_{\lambda '}(m)
e^{i2m\Phi } J_{2m} (4\Delta_o t ) \right\} \;,
\label{4.8a}
\end{eqnarray}
where we define
\begin{equation}
\lambda = \frac{1}{2} \sum_{k} \mid \alpha_k \mid^2 \;; \;\;\;\;\;
F_{\lambda}(\nu) = e^{-4\lambda \nu^{2}}
\end{equation}
As mentioned in section 2.2, $\lambda$ is the mean number of bath spins flipped each time 
$\vec{S}$ flips.  If $\lambda > 1$, but not too large; I will call this the 
intermediate coupling limit. Then
\begin{equation}
F_\lambda( \nu ) = \delta_{\nu ,0 } + \; small\; corrections  \;\;\;\;\;
(intermediate)\;.
\label{4.12a}
\end{equation}
so that, very surprisingly, we get a {\it universal form} in the intermediate
coupling regime for $P_{11} (t) $  (here $\eta (x) $ is the step function):
\begin{equation}
P_{11}(t)  \longrightarrow  {1\over 2}
\biggl[ 1+J_0(4 \Delta_o t) \biggr] \equiv
\int {d\varphi \over 2 \pi } P_{11}^{(0)}(t,\Phi =\varphi)
\label{4.13a}
\end{equation}
(compare the angular average of the coherent series in (\ref{4.8a})),
>From this we can also compute the {\it absorption} of the 2-level giant spin, as a 
function of frequency, by Fourier transforming; we get 
\begin{equation}
\chi ^{\prime \prime } (\omega ) \longrightarrow
{2 \over (16 \Delta_o^2 -\omega^2)^{1/2}}\:
\eta (4 \Delta_o -\omega )\;\;\;\; (intermediate) \;.
\label{4.13b}
\end{equation}
We plot this universal form  in Fig.15. 

\vspace{3in}

FIG. 15: The universal form for the power absorption $\chi ^{\prime \prime } (\omega )$ in the 
case of pure topological decoherence.

\vspace{4mm}

The physics of this universal
form is simply one of phase cancellation.,
As explained in refs.\cite{pro,sta}, this phase cancellation
arises because successive flips of ${\vec S}$ cause,
in general, a {\it different}
topological phase to be accumulated by the spin environment, so that
when we sum over successive instantons for ${\vec S}$,
 we get phase randomisation
 and hence loss of coherence. 
The only paths that can then contribute to
coherent oscillations in  $P_{11} (t)$ are those for
  which the number of clockwise and anticlockwise flips in the spin bath are
  {\it equal} - in this case the topological phase
   "eaten up" by the environmental
   spins is zero. By looking at expression (\ref{4.13a}) one sees another way of describing this-
the universal behaviour comes from complete phase
randomisation \cite{sta},
so that all possible phases contribute equally to the
answer! The final form shows decaying oscillations , with an envelope
$\sim t^{-1/2}$ at long times. This decay can also be understood \cite{prok}
by noting that the "zero  phase" trajectories that contribute to
$P_{11}$
constitute a fraction $(2s)!/(2^{s}s!)^2 \sim s^{-1/2}$ of the total
number of possible trajectories, where  $s \sim  \Delta_o t$.
 Because of this decay,
 the $\delta$-function peak in the spectral function at
 $\omega = 4 \Delta_o \cos \Phi $ is now
 transformed to the spectral function of a 1-dimensional tight-binding model.

One should also note that in the the {\it strong} coupling limit, when 
$\mid \alpha_k \mid \rightarrow \pi/2$, so that each spin rotates adiabatically with $\vec{S}$,
we simply have 
\begin{equation}
P_{11} (t) = {1 \over 2} [1 + \cos (4\Delta_o t\cos \tilde{\Phi } ) ]
 \;,
 \label{abiab}
 \end{equation}
 where $\tilde{\Phi }=\pi S +N\pi /2$, i.e., the Haldane/Kramers
 phase is now $\tilde{\Phi }$, with the extra phase coming from the $N$ bath spins which
rotate rigidly with $\vec{S}$.

Recall that the above results are given for the special case where $\alpha_k$
is pure {\it imaginary}; for the general complex case see the original work.

\vspace{4mm}

{\bf (ii) \underline{Orthogonality Blocking Limit}}:  Let us now consider the case where again 
all bath dynamics is suppressed ($V^{\alpha \beta}_{kk^{\prime}} = 0$), but now we retain 
only the static diagonal part of $H_{eff}$, ie., we assume
\begin{equation}
H_{\mbox{\scriptsize eff}} = 2 \Delta_\Phi  \tau _x  +
{\hat \tau }_z {\omega_o^{\parallel} \over 2}
\sum_{k=1}^N  \:
 {{\vec l}_k \cdot {\hat {\vec \sigma }}_k } +  {\omega_o^{\perp} \over 2}
  \sum_{k=1}^N
  \: {{\vec m}_k \cdot {\hat {\vec \sigma }}_k } \;.
  \label{4.14h}
  \end{equation}
  \begin{equation}
  \Delta_\Phi = \Delta_o \cos \Phi  \;,
  \label{4.14g}
  \end{equation}
  where $ \omega_o^{\parallel} \gg \omega_o^{\perp}$; in (\ref{4.14h}) we
  use a "Kramers renormalised" matrix element $\Delta_\Phi$. The significance of the finite 
but small transverse term $\omega_o^{\perp}$, is that in any case where the initial and final 
fields $\vec{\gamma}_k$ acting on $\vec{\sigma}_k$ are not exactly parallel or antiparallel,
the diagonal coupling must have a transverse part (cf. section 2.2(b)).

Now in this case no transitions occur in the spin bath when the central spin $\vec{S}$ flips.  
However the bath spins still play a crucial role, because the transverse "field" term
$\omega^{\perp}_{o} \sum_{k} \hat{\sigma}^{x}_{k}$ changes the motion of $\vec{S}$. There is 
perhaps a temptation to treat this as nothing but the simple problem of the changed motion 
of $\vec{S}$ in this 
transverse field.   However this is quite wrong - if we do this we forget that the transverse 
term is coming from a set of spin variables (the $\{\vec{\hat{\sigma}}_k\}$) to each of which 
is associated a {\it wave-function} $\mid \Phi_k>$.  The crucial point here is 
that in the presence of the transverse coupling, there is a mismatch between the initial 
state $\mid \Phi^{in}_{k}>$ (before $\vec{S}$ flips) and the final state 
$\mid \Phi^{fin}_{k}>$ (after it flips); in general 
$\vert \langle \Phi^{in}_{k} \vert \Phi^{fin}_{k} \rangle \vert \; < 1$ 
if $\omega^{\perp}_{o} \neq 0$.  
When we average over all bath spins, this seriously alters the system dynamics.  
Since $\omega^{\perp}_{o} \ll \omega^{\parallel}_o$,
we define an angle $\beta_k$, describing this mismatch between the initial and final
states of $\vec{\sigma}_k$, according to
\begin{equation}
\cos 2\beta_k =  - \hat{\vec{\gamma} }_k^{(1)} \cdot
\hat{\vec{\gamma }}_k^{(2)}  \;,
\label{beta}
\end{equation}
where $\hat{ \vec{ \gamma} }_k^{(1)}$ and $\hat{\vec{ \gamma}}_k^{(2)}$
are unit vectors in the direction of the initial and
final state fields acting on $\vec{\sigma}_k$ (and here we assume these are almost
exactly antiparallel).
If we assume that  $\beta_k \ll 1$, then the {\it total} mismatch, of the usual
"Debye-Waller" or "Franck-Condon" form, is just $e^{-\kappa}$, where 
\begin{equation}
\kappa ={1 \over 2} \sum_k \beta_k ^2  \;,
\label{k.1}
\end{equation}
More generally, if the mismatch angles $\beta_k$ are not so small, we just have
\begin{equation}
e^{-\kappa } = \prod_{k=1}^N \cos \beta_k  \;.
\label{k.2}
\end{equation}

Because of the similarity to the famous ``orthogonality catastrophe" considered by Anderson 
(in which the electronic phase shift $\delta_k$ is analogous to the $\beta_k$ considered here), 
we have called 
this limit that of ``orthogonality blocking".  However the physics here is rather different 
because the bath spins behave quite differently from electrons (which, we recall, map to 
an oscillator bath).  As shown in the original work, \cite{} the easiest way to understand 
and to solve for the dynamics of $\vec{S}$ is to notice that $\mid \Phi^{fin}_{k}>$ is 
produced from $\mid \Phi^{in}_{k}>$ by the unitary transformation:
\begin{equation}
\mid {\vec \sigma }_k^{fin} \rangle = {\hat U}_k
\mid {\vec \sigma }_k^{in} \rangle =
 e^{ -i\beta_k {\hat \sigma }_k^x }  \mid {\vec \sigma }_k^{in} \rangle \;.
 \label{b.2}
 \end{equation}
Thus, {\it mathematically}, the problem is identical to one in which there is an amplitude 
$\beta_k$ for $\vec{\sigma}_k$ to flip each time $\vec{S}$ flips, and where typically $\beta_k \ll 1$.  
Physically, the mismatch between the 2 states arises because the fields 
$\vec{\gamma}^{in}_{k}$ and $\vec{\gamma}^{fin}_{k}$ acting on $\vec{\sigma}_k$ are 
neither parallel or exactly antiparallel.  Thus if $\sigma_k$ is initally aligned with 
$\vec{\gamma}^{in}_{k}$, it finds itself slightly misaligned with $\vec{\gamma}^{fin}_{k}$
after $\vec{S}$ flips
(semiclassically, $\vec{\sigma}_k$ must start precessing in the field $\vec{\gamma}^{fin}_{k}$).

This problem is thus similar to that of topological decoherence (with $\beta_k$ replacing 
$\alpha_k$, and $\kappa$ replacing $\lambda$), except that we now have the added complication 
that the spin bath states are not at all degenerate - in fact they are split by the large 
longitudinal coupling $\omega_{o}$.  Recall (end of section 2.2) that this 
spreads the nuclear bath states over a large energy range 
$E_o \sim \omega_{o} N^{\frac{1}{2}}$, so that in zero applied field, $S$ 
cannot make any transitions at all unless the internal spin bath field 
$\epsilon = \omega_{o} \sum_{k} \sigma^{z}_{k} \tau^{z}_{k}$ bias energy 
is the same before and after $\vec{S}$ flips.  Otherwise, since typically 
$\omega_{o} \gg \Delta$, the difference $2\epsilon$ between the intial and 
final energies of $\vec{S}$ simply blocks all tunneling.  This we immediately see that 
we require that either (i) the net polarisation $\tilde{M} = \sum_{k} \sigma^{z}_{k}$ of 
the spin bath is zero both before and after $\vec{S}$ flips (so $\epsilon = 0$); or (ii) $\tilde{M}$ 
changes from $M$ to $-M$ when $\vec{S}$ flips - meaning that {\it at least} $M$ spins flip 
(and $\epsilon = M\omega_{o}$ before and after $\vec{S}$ flips).

For details of the resulting calculations of $P_{\Uparrow \Uparrow}(t)$,and the resulting 
rather bizarre behaviour, I refer to the original papers \cite{prok,prok3}.

\vspace{4mm}

{\bf (iii) \underline{Degeneracy Blocking Limit}}:  Finally let us consider the case where 
again $V^{\alpha \beta}_{kk^{\prime}} = 0$ (no spin bath dynamics), and both the non-diagonal 
terms {\it and} the transverse diagonal term $\omega^{\perp}_{k}$ are zero.  
Thus we have
\begin{equation}
 H_{\mbox{\scriptsize eff}}=2 \Delta \tau _x +
 {1 \over 2} \tau _z \sum_{k=1}^N \omega_k^{\parallel} \:{\hat \sigma }_k^z \;;
 \label{4.27}
 \end{equation}
 with a spread of values of $\omega_k^{\parallel}$ of
 \begin{equation}
 \sqrt{\sum_{k} (\omega_k^{\parallel}-\omega_o)^2 }
 \equiv N^{1/2} \delta \omega_o \;.
 \label{4.28}
 \end{equation}
i.e., a distribution of width $\delta \omega_o$.

Clearly, this Hamiltonian is identical to the standard biased
 two-level system, with the bias energy $\epsilon $ depending on the particular
 environmental state; thus
  $\epsilon = \sum_{k=1}^N \omega_k^{\parallel} \sigma_k^z $.
  The introduction of this spread is to destroy the exact degeneracy between
   states in the same polarisation group. For coherence to take place,
   we require the initial and final states to be within roughly
   $\Delta$ of each other.

The crucial point is of course that $\omega^{\parallel}_{o} \gg \Delta$ in most cases, and in 
fact one often has $\delta\omega^{\parallel}_{o} > \Delta$ (thus, in a nanomagnet, 
$\delta\omega^{\parallel}_{o}$ comes not only from the variation of the local contact hyperfine 
couplings $\omega_j \vec{s}_j. \vec{I}_j$, but also from the large variety of weaker transfer 
hyperfine couplings between the $\vec{s}_j$ and nuclei on other non-magnetic sites).  A 
suitable dimensionless parameter which describes the extent of the spread in the 
$\omega^{\parallel}_{k}$, around the ``central" value $\omega_{o}$, is 
$\mu = N^{\frac{1}{2}} \delta\omega_o/\omega_o$.  In the limit where $\mu \rightarrow 0$, 
and then the 2$^N$ nuclear levels are organised into a set of sharp lines, each one 
corresponding to a particular polarisation group- however this case is rather academic.  
More generally we define a 
{\it density of states} function $W(\epsilon)$ for the nuclear levels in the 
presence of the hyperfine 
coupling (Fig.16). 

\vspace{3in}

FIG. 16: The density of states function $W(\epsilon)$ for the nuclear multiplet for 2 different values
of the parameter $\mu$ (see text). The case $\mu < 1$ is somewhat academic, and so $W(\epsilon)$ is 
Gaussian in practise.

\vspace{5mm}

If $\mu < 1$ the lines still can be seen, with however the 
different polarization groups overlapping.  If $\mu > 1$, however (and this is almost
invariably the case in any real system), this structure 
disappears and we are left with a Gaussian form \cite{pro,prok}:
\begin{equation}
W(\epsilon) = \left( \frac{2}{\pi E^{2}_{o}} \right)^{1/2} exp(-2\epsilon^2/E^{2}_{o})
\end{equation}
\begin{equation}
E_o = \omega^{2}_{o}N
\end{equation}
This form was already mentioned in section 2.2.  In nanomagnetic molecules $\mu$ is usually considerably
greater than unity, even for small magnetic molecules, because of the large number of protons in the 
molecule and its ligands. In nanomagnetic grains the nuclei of $O$ ions (as well as protons in
hydrated systems) as well as other magnetic species, have the same effect.

The dynamics of an {\it ensemble} of nanomagnets, each described by the above 
Hamiltonian, but averaged over the nuclear distribution $W(\epsilon)$ in a thermal ensemble, 
is then quite trivial to obtain - it is just the {\it weighted ensemble average} over the correlation 
function of a simple 2-level system in this internal bias field $\epsilon$.  In the usual case where 
$kT \gg \Delta$, we then have

\begin{eqnarray}
P_{11}(t) &= &\int d\epsilon W(\epsilon )
{e^{-\beta \epsilon } \over Z(\beta )}
\bigg[ 1-
{2 \Delta_\Phi ^2 \over \epsilon^2 + 4 \Delta_\Phi^2 } \left(
1 -\cos (2t\sqrt{\epsilon^2+4 \Delta_\Phi^2}  ) \right) \bigg] \\
\label{4.32}
&\longrightarrow & 1-2 A\sum_{k=0}^\infty J_{2k+1}(4 \Delta_\Phi t) \;.
\label{4.33}
\end{eqnarray}
where the reduction factor $A$ is just
\begin{equation}
A=2\pi  \Delta_\Phi /(\omega_o \sqrt{2\pi N})
\ll 1 \;.
\label{4.34}
\end{equation}
and the  spectral function is given by another universal form, viz:
\begin{equation}
\chi^{\prime \prime }(\omega ) =
A {8 \Delta_\Phi \over \omega \sqrt{\omega^2-16 \Delta_o^2} }
 \eta (\omega -4 \Delta_\Phi )\;,
 \label{4.35}
 \end{equation}

\vspace{3in}

FIG. 17: The universal form for $\chi^{\prime \prime }(\omega )$ in the case of pure degeneracy blocking.
Notice that absorption only occurs {\it above} the threshold defined by $\Delta_o$.

\vspace{5mm}

The result here is very easy to understand.  Only those nanomagnets near resonance (ie., 
with $\epsilon \sim \Delta$ or less) stand much of a chance of tunneling, since tunneling 
requires near degeneracy between intial and final states.  Those systems which do not have 
this degeneracy are ``degeneracy blocked"; and since $\Delta \ll E_o$, almost all of them 
are frozen.  This result of course arises because the only dynamics in this limiting model comes 
from the tunneling term itself. As soon as we give the spin bath some dynamics (either by 
reintroducing the non-diagonal terms, or even more importantly, by making 
$V^{\alpha \beta}_{kk^{\prime}}$ non-zero, so that the nuclear bath has independent dynamics), 
the blocking will be relieved, as we shall now see.

\vspace{7mm}

{\bf 3.4(b) GENERAL SOLUTION}  The discussion of the 3 limiting cases given above 
allows us to give a simple intuitive picture of the general solution to the problem (readers 
wanting details, proofs, etc., should go to the original papers \cite{prok2,prok3,prokcm}).

\vspace{4mm}

{\bf (i) \underline{Independent Bath Spins}}: Suppose we start with the (often rather academic)
case where the bath spins do not interact at all with each other (formally, this means
$V_{kk^{\prime}}^{\alpha \beta}$ is much less than all other energy scales, in particular
$\Delta$). The results are so pretty, and intuitively revealing, that I will quickly describe them 
to you.

Each of the 3 limiting cases above corresponds to one physical mechanism (already previewed
back in our initial discusion of this model, in section 2.2(b)). Now the magic is that the general 
solution is simply obtained by combining them! Mathematically, each mechanism corresponds to a
particular {\it statistical average}:
\begin{equation}
\hbox{ (a)  A "topological phase average" given by } \;\;\;\;
\sum_{\nu =-\infty }^{\infty} F_{\lambda '} (\nu )
\int {d\varphi \over 2 \pi } e^{i2\nu (\Phi -\varphi )} \;;
\label{q.5}
\end{equation}
\begin{equation}
\hbox{ (b) An "orthogonality average" given by  } \;\;\;\;
2 \int_0^\infty  dx x e^{-x^2} \;;
\label{q.6}
\end{equation}
\begin{equation}
\hbox{ (c) A "bias average" } \;\;\;\;
\int d\epsilon W(\epsilon ) { e^{-\beta \epsilon }
\over Z(\beta )}\;.
\label{q.7}
\end{equation}
Apart from the orthogonality average (the derivation of which is too
complicated to explain here), you have just seen these above. Now, consider a simple 2-level 
system in a bias $\epsilon$, but with a {\it renormalised} tunneling splitting 
\begin{equation}
\Delta_M(\varphi ,x) =2{\tilde \Delta}_o \vert \cos ( \varphi )
J_M(2x\sqrt{\gamma }) \vert  \;,
\label{5.8a}
\end{equation}
and associated energy splitting 
$E_M^2(\varphi ,x) = \Delta_M^2(\varphi ,x) +\epsilon^2$. This tunneling matrix element now depends 
explicitly on the topological phase, on the nuclear polarisation $M$, and on the orthogonality
variable $x$. The parameter $\gamma$ is just 
\begin{equation}
\gamma =  \left\{
\begin{array}{ll}
\lambda & \;\;\;\;\;\mbox{if
$\lambda  \gg \kappa$ (topological decoherence regime)} \\
\kappa &\;\;\;\;\;\mbox{if
$\kappa \gg \lambda $ (orthogonality blocking regime)}
\end{array} \right.  \;.
\label{q.5b}
\end{equation}
Recall now the very simple dynamics of an isolated biased 2-level system (eqtn. 
(\ref{2wbiasdensityt})); the probability that the system stays in state 
$\vert \uparrow \rangle$ after a time $t$ is $P_{11}^{(0)}(t) = 1 - (\Delta/E)^2 sin^2 (Et)$.
Then, perhaps amazingly, the same function for the central system, after all averaging over
the bath spins is carried out, is nothing but a generalisation of this; one gets 
\begin{equation}
 P_{11 }(t;T ) = 1-
 \int d\epsilon W(\epsilon ) { e^{-\beta \epsilon }
 \over Z(\beta )} \sum_{M=-N}^N
 \left( 1- P_M(t,\epsilon -M\omega_o ) \right)  \;;
 \label{6.25x}
 \end{equation}
in which the probability relaxation function for a given polarisation group $M$ is
\begin{equation}
P_M(t;\epsilon ) = 2 \int_0^\infty  dx x e^{-x^2}
\sum_{\nu =-\infty }^{\infty} F_{\lambda '} (\nu )
\int {d\varphi \over 2 \pi } e^{i2\nu (\Phi -\varphi )}
\bigg[ 1-{ \Delta_M^2(\varphi ,x) \over E_M^2(\varphi ,x) }
\sin ^2 ( E_M(\varphi ,x) t ) \bigg]               \;,
\label{q.11}
\end{equation}
That is all! Perhaps even more amazing, these integrals and sums can be done {\it analytically}
for almost all of the parameter regime (which, amusingly, is more than one can say for the 
spin-boson model, which just goes to show that a simple Hamiltonian does not always have simple
dynamics!). The physical interpretation of this result is of course obvious, just by looking
at our 3 limiting cases again; I will not go over it again. If you want to see analytic formulae
and pictures, see Prokof'ev and Stamp \cite{prok2,prok3,prokcm}; 
you will find that in some cases they do not look
anything like what you can get from a spin-boson model! 

This of course raises the question- can this "central spin" model ever behave like a spin-boson
model? The answer is yes. Lets first think about this physically- we wish the spins to somehow
behave like a set of oscillators, {\it weakly coupled} to the central spin. If you recall the 
form of the Caldeira-Leggett spectral function $J(\omega)$, involving a coupling constant
squared, divided by an energy denominator (cf. 2nd-order perturbation theory), then it is clear how
this can happen, for in our case the 2 relevant parameters are $\omega_k^{\|}$ and 
$\omega_k^{\parallel}$, and each is capable of playing the role of the coupling $c_k$ (with the 
other playing the role of the energy denominator). However {\it this only works if the coupling
is small} (compared to $\Delta$). An example is provided by the SQUID model we 
considered in section 2.3(a), summarized in the effective Hamiltonian (\ref{cases.3}); recall that 
there $\omega^{\parallel}_k/\omega^{\perp}_k \ll 1$, and both are very small. This problem (which
is related to that studied by Caldeira et al. \cite{castro}), is easily solved by
writing $P_{11}(t)$ as
\begin{eqnarray}
P_{11}(t) = \sum_{nm}^{\infty} (i\Delta_o)^{2(n+m)}
\int_0^t dt_1 \dots \int_{t_{2n-1}}^t dt_{2n}
\int_0^t dt'_1 \dots \int_{t'_{2m-1}}^t  dt'_{2m} {\cal F}(Q,Q')
\nonumber \\
{\cal F}(Q,Q') = \prod_{k} \langle \hat{U}_k(Q_{(n)},t)
\hat{U}_k^{\dag}(Q'_{(m)},t) \rangle \;,
\label{neto.3}
\end{eqnarray}
where as before $Q_{(n)}(t)$ refers to a path of the central system containing $n$ instanton jumps,
but now $ \hat{U}_k(Q_{(n)},t)$ refers to the time evolution of the $k$-th environmental {\it spin}
(not oscillator!), under the influence of the central spin (cf eqtn. (\ref{neto.2})). In the 
present case this is
\begin{equation}
\hat{U}_k(Q_{(n)},t) = T_{\tau} \exp \left\{ -i\int_0^t {ds \over 2}
\big[  \omega^{\perp}_k  \hat{\sigma}_k^z +
Q_{(n)}(s) \omega^{\parallel}_k \hat{\sigma}_k^x \big]
\right\} \;,
\label{neto.4}
\end{equation}
In general this formula for the "influence functional" is intractable (hence our alternative
approach above!). However in the present case, expanding to 2nd order in $\omega^{\parallel}_k$,
one gets at $T \gg \omega^{\perp}_k$ that 
\begin{equation}
{\cal F}(Q,Q')  = \exp \bigg\{ -\sum_k {(\omega^{\parallel}_k)^2 \over
8} \bigg\vert  \int_0^t ds
e^{2i\omega^{\perp}_k s }~[Q_{(n)}(s)-Q'_{(m)}(s)] \bigg\vert ^2
\bigg\} \;,
\label{neto.4a}
\end{equation}
Now in general this is still hard to solve; but for the SQUID it is simple, because even though
$\omega_k^{\|}$ is extremely small, there are so many nuclear spins that the parameter
$\gamma$ defined by $\gamma^2 = [(\sum_k \omega_k^{\|})^2 /2]^{1/2}$ is not (for our previous
model parameters $\gamma \sim 5 \times 10^{-4} K$, similar to the total longitudinal bias produced
by the nuclei), and so we expect $\gamma \gg \omega^{\perp}_k,~\Delta_o$ (unless $\Delta_o$ 
is very large indeed). In this case it is easy to see that 
${\cal F}(Q,Q') \sim  e^{-\gamma^2(t-t')^2}$, ie., the influence functional decays (and dephases
everything) before anything else can move (which, incidentally, is why we can ignore the nuclear
bath dynamics in studying this case). In path integral language, kink-antikink pairs are 
closely bound (justifying a "NIBA"); more prosaically, we get a fast "motional narrowing"
relaxation, with  $P_{11}(t)=1/2(1-e^{-t/\tau_R})$,
and with the relaxation time given just by the integral
\begin{equation}
\tau_R^{-1} = 2\Delta_o \int_0^{\infty}  e^{-\Gamma^2(t_2-t_1)^2}
\equiv {\sqrt{\pi} \Delta_o^2 \over  \Gamma } \;.
\label{neto.7}
\end{equation}
This result does not give out much hope for the experimental search for MQC in SQUIDs; it
indicates coherence will be destroyed by nuclear spins in a time scale $\ll \Delta_o^{-1}$.

\vspace{4mm}

{\bf (ii) \underline{Influence of Spin Bath Dynamics}}:

Suppose we now go back to the full spin bath problem. What will be the influence of the small
dynamical term $V_{kk^{\prime}}^{\alpha \beta}$ on the results. The answer is that it can be 
profound, because in a situation where the central system is more or less blocked from 
transitions by the strong degeneracy blocking field, the weak residual bath dynamics can "scan" 
the bias energy until the system finds a resonance. In general this can happen with the 
simultaneous flipping of some bath spins, so that the resonance condition must include the
change in energy of the nuclear bath as well; ie., we will end up summing over polarisation 
groups $M$. Thus, we have an {\it external} bias $\xi$, an internal longitudinal bias
$\epsilon (t) = \epsilon + \delta \epsilon (t)$ (where $\delta \epsilon (t)$ represents the 
time dependent fluctuations coming from the internal spin bath dynamics- or, at higher $T$, from 
external couplings which drive the spin bath, ie., $T_1$- processes in the case of a nuclear bath), 
and the condition for resonance is that $\xi + \epsilon + M\omega_o + \delta \epsilon (t) \sim 0$.
Moreover, the system should stay in resonance long enough to give the central spin a chance 
to tunnel!

This point is of very great practical importance; in fact it is obviously going to be 
entirely responsible for the existence of any dynamics in the low-$T$ limit 
(as was discussed in some detail by Prokof'ev and myself, a while ago, in the context of
nanomagnets \cite{prok2}), since every other relaxation mechanism is frozen! What is found  
is that for a {\it single isolated nanosystem} (ie., a nanomagnet), in the Quantum  
regime , the new correlation function should be given by the simple incoherent relaxation form
\begin{equation}
P_{11}(t) = \sum_M w(T,M) \int xdx e^{-x^2}
\sum_{\nu=-\infty}^{\infty} \int {\varphi \over 2\pi } F_{\lambda
'}(\nu) e^{i2\nu (\Phi -\varphi )} \big[ 1+e^{-t/\tau_M(x, \varphi ) } \big]
\;,
\label{diff.2}
\end{equation}
where the relaxation time $\tau_M(x, \varphi)$ for the $M$-th polarisation group is 
\begin{equation}
\tau_M^{-1}(\xi,\varphi ) = 2  \Delta_M^2(x, \varphi )/\pi^{1/2} \xi_o 
\label{diff.3}
\end{equation}
The derivation is simple but too long to be repeated here. The parameter 
$\xi_o \sim N^{1/2} \delta \omega_o$ is a measure of the total energy spread of each 
polarisation group in the density of states (recall Fig.16 ). In the absence of any such spread, 
we have $\xi_o \sim N^{1/2} T_2^{-1}$, where the residual couplings between the bath spins
give the total bath spectrum a linewidth $\xi_o$; here $T_2^{-1} \sim V_{kk^{\prime}}^{\alpha \beta}$.
We use this notation because in the case of a nuclear spin bath, this linewidth is 
parametrised (and in principle defined experimentally) by the transverse nuclear spin diffusion
time $T_2$. Typically $T_2^{-1} \sim 10-100 kHz$ in this case, although it can vary a lot. Notice
that these bath fluctuations are in the high-$T$ limit for anyone except a nanoKelvin experimentalist-
we cannot make them go away!
   This result is easiest to understand in the limit where both degeneracy blocking and 
topological decoherence are absent (no nuclei coflip with $\vec{S}$, and $H_o(\vec{S})$ has
axial symmetry). Then the averages collapse, and we get a unique relaxation rate given by 
$\tau_o = 2 \Delta_o^2/\pi^{1/2} \xi_o$. This is of course just the overlap between initial and final
states in a typical bias $\xi_o$, but be careful- it is a {\it relaxation rate}, involving the 
{\it irreversible} passage of the sytem from one state to the other (with an energy absorption 
$\sim \xi_o$ by the spin bath). The difference with a simple overlap integral becomes clear
if the calculation is redone in a finite bias $\xi$ (see Prokof'ev and Stamp \cite{prok2}); then we get 
instead that 
\begin{equation}
\tau (\xi) = \tau^{-1}_0 e^{-\vert \xi \vert /\xi_o}
\label{key}
\end{equation}
and we see that the rate falls of {\it exponentially} with the bias, as we take the system away from 
any possibility of reaching the resonant tunneling window, during its peregrinations over an 
energy range $\xi_o$. This formula is essential to understanding the low-$T$ dynamics of any
system described by a central spin model.

\vspace{4mm}

This concludes the very long discussion of canonical models and their dynamics. Enough, then, of 
Olympian theory- let's now get down to a merely Homeric epic, with some real physics (including
experiments)!

\section{Quantum Nanomagnetism}

Before plunging into a new field, it is certainly worth asking what fundamentally new ideas 
may come out of it.  So why study quantum nanomagnets?  Here are a number of reasons, each of 
which will appeal to different tastes:

(i)  They constitute a (very rich) testing ground for theorists interested in tunneling in 
complex systems; one can attack on many different fronts (numerical, analytical, experimental), 
and there is an almost inexhaustible variety of real examples to work with.  The spin 
Hamiltonians for magnetic nanomolecules lend themselves well to studies in quantum chaos.

(ii)  Much more than superconductors, they allow detailed study of the crossover between 
quantum and classical properties of physical systems, as one changes the system size, 
temperature, and coupling constants.  Understanding this crossover is not only important for 
the resolution of the notorious quantum measurement problem (and a better understanding of how 
decoherence works at this scale); it will also have enormous practical consequences in the 
next few decades, with the coming of quantum nanomagnetic devices.  Unravelling the relevant 
physics promises to be one of the most exciting challenges in condensed matter physics.

(iii)  In the field of magnetism itself, this crossover is of central interest; it has been 
highlighted by the discovery of high-temperature superconductors, heavy fermions, 1-dimensional 
magnets (spin chains), and a variety of ``frustrated magnets", many of which show tendencies 
towards ``spin liquid" behaviour (ie., quantum-disordered magnetic states), even at the 
macroscopic level.  More recently there has been the discovery, (in Mn-perovskite 
layered magnets, having the same structure as the high-Tc La-perovskite superconductors), 
of ``collosal magnetoresistance" phenomena; and the preparation of ``spin ladder" 
materials, which are essentially sets of coupled spin chains.  There is no question that 
the study of these quantum magnetic materials (and their remarkable superconducting 
properties when doped), is the central focus of most of the condensed matter research around 
the world at present.  It is quite extraordinary that almost all of these materials 
(including the magnetic molecules to be discussed presently) are {\it Transition Metal Oxides}, 
at or near a metal-insulator transition, with magnetism controlled by superexchange, and 
great sensitivity to doping.  Why all his should be, is one of the great mysteries in physics 
at the end of the 20th century.

(iv)  Some magnetic systems offer the prospect of observing genuinely macroscopic quantum 
phenomena, in the sense discussed by Leggett \cite{AJL}.  The most likely possibility here is 
in the tunneling of domain walls (or possibly their quantum nucleation), which at least 
theoretically should occur on a large scale \cite{staDW}, particularly in systems isotopically 
purified of nuclear spins \cite{dube}.  Some promising experiments in this direction have been 
done already \cite{gio,wer}.

>From this short list we see that nanomagnets appeal to physicists (and even philosophers 
of science) interested in the mysteries of quantum mechanics and the measurement problem; 
and at the same time to industrialists and venture capitalists, investing in the 
technology of the next century.  This curious mixture is not infrequent in subjects at 
the frontier of science.  

In the following I will give an overview of 2 important themes in this relatively new field, 
aimed at the non-specialist. 
These show (sometimes rather dramatically) how the 
concepts discussed in Chapters 2 and 3 can be  
applied to real physical systems.  Incidentally, although some of 
the experiments are understood theoretically, others are not, and there are some 
intriguing outstanding mysteries.  I will start with the now quite extensive 
results (theoretical and experimental) on quantum relaxation in magnetic micromolecules 
and grains; I will argue that these are fairly well understood.  
Then we look at ``macroscopic coherence" experiments in ferritin macromolecules, where the gap 
between theory and experiment is still controversial. I do not discuss other areas (eg., domain
wall tunneling), simply because of a lack of space.

\subsection{Resonant Tunneling Relaxation}

Let us start with a set of experiments in which no coherence exists at all.  Considerable 
press (some of it highly misleading) has been devoted recently to a series of experiments 
\cite{pau}-\cite{san} which have shown, with steadily increasing clarity, the role of tunneling in the 
relaxation of magnetisation in an initially polarized sample of magnetic molecules.  These 
experiments did not appear in a vacuum, but rather in the context of previous experimental 
efforts to see tunneling in magnetic particles and in the biological macromolecule ferritin 
(see section 4.3 below).  The possibility of doing tunneling relaxation experiments on 
molecule crystals of large-spin nanomolecules relies essentially on recent advances in the 
preparation of such molecules by molecular chemists \cite{}.

There have also been a large number of tunneling relaxation experiments on nanomagnetic 
particles.  I shall make no attempt to review all of these; however some of them are 
very interesting, specially as they indicate what remarkable possibilities exist for the future.

\vspace{7mm}

{\bf 4.1(a) MAGNETIC NANOMOLECULES- STRUCTURE and EXPERIMENTS}

I will only touch the surface here of a huge field, which was for many years the purview of
chemists and the occasional molecular physicist. Reviews of some of the chemistry are given
by, eg., Gatteschi et al \cite{Gatt}; and I would particularly recommend
the book of Kahn \cite{Kahn}. The low-$T$ experiments are too new to have been reviewed.

\vspace{4mm}

{\bf (i) \underline{Spin Structure at Low Energies}}:
The first experiments were done on the ``Mn-12" system, first in powder samples 
and then in large single crystals (refs.\cite{pau}-\cite{her}).  Early theoretical 
studies of this molecule, by the Grenoble-Firenze group, \cite{} led to the proposal of a 
``giant spin" Hamiltonians $H_o(\vec{S}) = (^{\|}H_{o}^{(2)} + ^{\perp}H_{o}^{(4)})$, where
\begin{equation}
\ ^{\|}H_{o}^{(2)} \ = \ -(D/S) S^{2}_{z}
\end{equation}
\begin{equation}
\ ^{\perp}H_{o}^{(4)} \ = \ -(K_4/S^3) \left[S^{4}_{+} \ + \ S^{4}_{-} \right]
\end{equation}
and in which S = 10, D $\sim$ 6-7K, and $K_4$ is hard to determine (present 
estimates for $K_4/D$ range from 0.005 to 0.03). 
Magnetisation and specific heat measurements indicate that, as far 
as the low-T dynamics is concerned, this is a good approximation for $T \leq 30K$.  The Mn-12 
molecule is of course very complicated, and early numerical attempts to establish the 
low-energy $H_{eff}$ did not succeed in predicting the S=10 ground state, partly because no 
magnetic anisotropy was included.  The difficulty of the problem can be seen by considering 
Fig.18(a) below, which shows the important exchange couplings, acting between the 12 different Mn ions 
in the molecule.  In reality these couplings are of course anisotropic, and there will also be 
single-ion anisotropy terms (cf eqn (10)).  Because the molecule lacks inversion symmetry, 
one also expects ``Dzyaloshinskii-Moriya" terms, of form $V_{DM} = \sum_{<ij>} \vec{D}_{ij} 
\cdot (\vec{s}_i \times \vec{s}_j)$, acting between spins. Thus any serious 
attempt to diagonalize 
the full spin Hamitonian (in a Hilbert space of dimension $5^8 4^4 = 10^8$), with a large 
number of couplings which are very difficult to determine, is rather pointless.  A different 
tactic was adopted by Tupitsyn et al, \cite{tup2}, following the early work of Sessoli et al. \cite{sess},
in an effort to explore the region $T \leq 
150K$.  It was assumed that the ``diagonal" exchange couplings in Fig.18(a) are so strong that at low 
T, the Mn$^{+3}$/Mn$^{+4}$ ionic pair is locked into a spin-1/2 state. This reduces the problem 
to an 8-spin one, with only 10$^4$ degrees of freedom (Fig.18(b)).  At this point one writes down 
an effective Hamiltonian for this 8-spin system, which is consistent with the known molecular 
symmetry, and which 
captures both exchange and anisotropy.  The simpliest such form, which ignores any time-
reversal symmetry breaking (which could come from Dzyaloshiniskii-Moriya interactions) can be 
written as follows:
\begin{equation}
\hat {H_o} = {\cal J}_1 \sum_{k=1}^{4}  {\vec S}_k \cdot {\vec \sigma}_{k} +
{\cal J}_2(\sum_{k=1}^{3} {\vec \sigma}_k \cdot {\vec \sigma}_{k+1} +
{\vec \sigma}_4\cdot {\vec \sigma}_{1}) +
K_{\|}\sum_{k=1}^{4} S_{k}^z \cdot \sigma_{k}^z, \;\;\;\;\;
S=2, \sigma={1 \over 2}  \;,
\end{equation}
where we couple 4 spin-2 spins $\vec{S}_k$ and 4 spin-1/2 spins $\vec{\sigma}_k$ (the 
latter should not 
be confused with nuclear spins!)

\vspace{3in}

FIG. 18: The structure of the exchange couplings in the Mn$_{12}$, after Sessoli et al. \cite{sess}.
In (a) is shown the couplings between the 12 Mn ions; whereas in (b) we see the couplings between the 
8 spins in the reduced model described in the text, where the very strongly coupled "diagonal" spin pairs 
(with estimated AFM coupling $J_1 \sim -215 K$ in (a)) are coupled to form single spin-1/2 at low energy. 
Sessoli et al. also estimate $J_2 \sim J_3 \sim -85 K$ and $\vert J_4 \vert \leq 45 K$. The 
renormalised couplings in (b) are discussed in the text.

\vspace{5mm}

We notice immediately that we have not {\it derived} this Hamiltonian.  In fact if the 
reader is getting used to the idea of low-energy effective Hamiltonians, 
he/she may well be asking - 
from what starting point {\it is} one supposed to derive $H_o$?  The answer is of course that the only 
reasonable starting point is one where we have some way of determing the input parameters 
experimentally.  One might of course attempt the very ambitious task of calculating the various 
couplings in the 12-spin Mn$_{12}$O$_{12}$-ac system, starting from an underlying Anderson 
lattice model for the molecule.  From a microscopic viewpoint this certainly makes sense - 
we are dealing with a hydrated transition metal oxide, to which the Anderson-Mott-Hubbard ideas 
apply rather directly.  But the problem is obvious - with no knowledge of the parameters 
in such a model, we can do no more than guess what superexchange couplings they might lead to, 
between the Mn ions.  In fact, the values generally assumed for these couplings are no more than 
educated guesses made by the chemists.

The natural result of this line of reasoning is to treat the above 8-spin Hamiltonian as our 
starting point and attempt to {\it deduce} the values of the parameters ${\cal J}_1$, ${\cal J}_2$, and 
$K_{\|}$ from experiments. The way this was done by Tupitsyn et al. was simple.  They first 
calculated the magnetisation $M_z$ in a longitudinal field $H_z$, including all 10$^4$ levels 
of the 8-spin model.  The results for $M_z(H_z,T)$ were compared with a 
large number of experimental 
results on a single oriented Mn$_{12}$ - ac crystal, taken at different 
$T$ and $H$.  The parameters 
${\cal J}_1, {\cal J}_2$, and $K_{\|}$ were then adjusted to give 
the best possible fit to these experiments; 
it was found that this gave ${\cal J}_1$ = -85K, ${\cal J}_2$ = 55K, and $K_{\|}$ = 7.5K.

To test this model, it was then used to predict the results of a {\it transverse} 
magnetisation measurement, ie., the magnetisation $M_x(H_x,T)$ was calculated over a wide range 
of $H_x$ and $T$ values.  These were then compared with transverse susceptibility measurements.  
This is a sensitive test of the theory, because the transverse field mixes together states 
which have little to do with each other in a longitudinal field. The results were very good for 
$T \leq$ 150K, undoubtedly because above this temperature, the ``diagonal couplings" are too 
weak to prevent the break-up of the 4 fake spin-1/2 pairs, into their spin-2 and spin-3/2 
constituents.
Thus up to 150K this sort of exercise is very useful, whereas the spin-10 ``giant 
spin" Hamiltonian breaks down seriously above 30K. With further thermodynamic measurements, 
eg., measurements of $C_v(T,H)$ (with the phonon contribution subtracted off) it 
should be possible to refine these models even more.

I should emphasize, however, that such calculations, although they can in principle give 
us a good understanding of the gross structure and origin of the eigenstates up to 
$\sim 100K$, are utterly useless when it comes to understanding the {\it tunneling}.  This 
is because, as we have already seen in Chapter 2, the tunneling matrix elements are not 
only {\it exponentially smaller} than the energy scale relevant to these calculations 
(ie., composed to anisotropy energies); they also depend exponentially on any change 
in these energy scales! Thus any uncertainty in our knowledge of the correct 
effective Hamiltonian (whether it be at the level of the giant spin $H_o(\vec{S})$, 
or at the more microscopic level just discussed) will be magnified exponentially when 
it is used to try and determine the tunneling matrix elements.  I shall return to this 
point again below, in discussing the interpretation of the experiments.

\vspace{4mm}

{\bf (ii) \underline{Quantum Relaxation Experiments}}:
Let us now proceed directly to the low-T experimental results on the Mn-12 and Fe-8 systems, 
(ie., for $T \leq 5K$). We begin with Mn-12, for which many experiments have been done in 
recent years.
The clearest of these are those by Thomas et al. \cite{tho} on a single crystal.  
The essential results 
found by the various authors were

(i)  When $T > 2K$, the relaxation of an initially polarized sample of molecules was found 
to decay roughly exponentially (although more recent measurements by these authors also 
find non-exponential relaxation around the resonances at high $T$).  The remarkable thing 
here was the series of resonant increases in the 
experimental relaxation rate $\tau^{-1}(H,T)$ around fields 
$nH_1$, where $n = 0, \pm 1, \pm 2, 
\ldots$, and $H_1 \sim 0.44T$. The crucial point is that if we add 
a longitudinal field bias term $-g\mu_B SH_z$ to (4.1), then the resonant fields correspond 
precisely to ``level crossings", ie., to fields at which eigenstates $|m\rangle$ of (4.1) become 
degenerate with eigenstates $|-m+n\rangle$ (here we define eigenstates $|m\rangle$ via 
the equation $^{\parallel}
H_{o}^{(2)} |m\rangle = \varepsilon^{o}_{m} |m\rangle$, where 
$\varepsilon^o_m = -D_{m}^{2}/S$).  The 
existence of resonant increases in $\tau^{-1}$ at these fields, and a proposed explanation in 
terms of thermally activated resonant tunneling, was first given by Novak and Sessoli \cite{nov} 
(see also Barbara et al. \cite{bar}).

(ii)  Below 2K, the relaxation appears to become rather $T$-dependent in the Mn-12; moreover, 
recent unpublished data shows it is not at all exponential, a fact 
emphasized in refs. \cite{pau,fri} (but not ref. \cite{tho}).

(iii)  The resonant peaks in $\tau^{-1} (H,T)$ appear to be Lorentzian around the $H_n$.  Their 
half-width is rather large (roughly 0.1$T$).

The Mn$_{12}$ experiments were followed by some very striking results \cite{san} obtained in the 
``Fe$_8$" molecule.  In many ways this molecule is very 
similar to Mn-12; a giant spin is again formed via -oxo bonds, in a typical example of 
antiferromagnetic superexchange in a transition metal oxide.  In this system, all eight $Fe$ ions 
are in a spin-5/2 valence state, which align again in a spin-10 ground state, which again is 
described to first approximation by an easy axis term $^{\parallel}H^{(2)}_{o}$ (eg. (4.1)); 
in this system $D \sim 2.7K$.  The lowest transverse term which breaks rotational symmetry 
about $S_z$ must take the form $^{\perp}H^{(2)}_{o} = -(E/S)[S^{2}_{x} - S^{2}_{y}]$ (a term 
disallowed in $Mn_{12}$, which crystallizes in tetragonal form); just as in $Mn_{12}$, 
$^{\perp}H_{o}/^{\parallel}H_{o} \ll 1$.

The relaxation results in $Fe_8$ molecular crystals share several features in common with those 
of $Mn_{12}$.  Again, one sees resonantly enhanced relaxation at fields $H_n$; but now the 
relaxation rate $\tau^{-1}(H,T)$ increases over 4 orders of magnitude around the zero field 
resonance (which, incidentally, is shifted from $H=0$ by several hundred Gauss)!  A particularly 
valuable feature of the $Fe_8$ system is that one may see reasonably rapid relaxation down to 
very low $T$; in fact, Sangregorio et al. \cite{san} followed it down to 70mK.  Very direct evidence 
appears in these experiments of a crossover, below $T \sim 360-400 mK$, to a regime of 
quantum relaxation.  The relaxation $M(t)$ of the magnetisation $M$ in an initially polarised 
sample is completely $T$-independent below 360 mK.  It is very noticeably non-exponential; in 
fact the authors attempted to fit it using a stretched exponential.

How we might understand these results? Before looking at the relaxation itself,
I think we need some cautionary remarks concerning the use of the giant spin Hamiltonian in
discussing this sort of experiment. These concern the relationship between the values of
the tunneling matrix elements $\Delta_{m}$, between levels $|m\rangle$ and $|-m\rangle$ of
the longitudinal part of the giant spin Hamiltonian. In what follows we will 
treat a matrix element like  $\Delta_{10}$,
 coupling the 2 lowest levels of $^{\parallel}{\cal H}_o^{(2)}$ in Mn$_{12}$ and
 Fe$_8$, as an {\it independent parameter}. One can of course calculate $\Delta_{10}$ if
 the weak anisotropy couplings are known. Thus
 a 2nd-order (in $S^+, S^-$) anisotropy
 term $^{\perp}{\cal H}_o^{(2)} = E(S_x^2-S_y^2)$ yields
 \begin{equation}
 \Delta_{10}={4S^2E^{S} (2S-1)!! \over (2D)^{S-1} (2S)!! } \approx
 {4S^{3/2} \over \sqrt{\pi } } E (E/2D)^{S-1} \;.
 \label{Delta-2}
 \end{equation}
 Using $E=-0.046~K$ for
 $Fe_8$ molecules \cite{san} then gives $\Delta_{10}
 \sim 10^{-9}~K$!  On the other hand the 4th-order anisotropy term
 $^{\perp}{\cal H}_o^{(4)} = B(S_+^4+S_-^4)$ yields
 \begin{equation}
 \Delta_{10}={(BS^2)^{S/2} S^2 (2S)! \over (D)^{S/2-1} 16^{[S/4]} [S!!]^2 }
 =BS^2 (BS^2/D)^{S/2-1} {(2S)! \over S^{S-2} 16^{[S/4]} [S!!]^2 } \;.
 \label{Delta-4}
 \end{equation}
 In the Mn$_{12}$ system the $^{\perp}{\cal H}_o^{(2)}$
 anisotropy term is absent, and the parameter $g_4=(BS^2)/D$ is
 rather small (estimates range from
 $g_4 \approx 0.03$  down to $g_4 \sim 0.005$. 
 For $g_4=0.03$ one gets $\Delta_{10} \approx 10^{-10}~K$.

 From these examples we learn that even rather small higher-order
 anisotropy terms may determine high-order tunneling matrix elements. This is very simply because,
 as already noted in section 2, a higher-order transverse coupling (eg., a coupling of 8-th order
 in $S^+, S^-$) appears raised to a lower power in $\Delta_{10}$ than a 2nd- or 4th- order
 coupling term, and so can play an important
 role even if it is much smaller than a 2nd- or 4th-order term.
Consequently, in, eg., Fe$_8$, $\Delta_{10}$ may be considerably larger
 than Eq.(\ref{Delta-2}), because of such higher-order
 terms (up to the twentieth order in $S^+, S^-$, in fact!).
 They are almost impossible to obtain
 experimentally, since their contribution to the transitions
 studied by EPR spectroscopy is likely to be negligible. All I am doing here, of course, is 
 repeating the warning made earlier about the exponential dependence of tunneling
 amplitudes on microscopic parameters, in a different way.

 Note that as a corollary to this argument, one
 has to be particularly cautious in predicting the effects of the
 applied weak transverse field $H_{\perp}$ on $\Delta_{10}$, since this field
has a contribution of order $H_{\perp}(H_{\perp}/D)^{2S-1}$, ie., to the $2S$-th power in
the field! Essentially this means that its direct effect at low $T$ is very small- thus,
the effect of any applied transverse field on $\Delta_{10}$ is negligible unless $H$
is of order {\it Tesla}. Likewise, the effect of {\it static} internal transverse fields (such
as transverse hyperfine or dipolar
fields when $kT \ll D$) on $\Delta_{10}$ is negligible.
\vspace{7mm}

{\bf 4.1(b) THEORETICAL INTERPRETATIONS}

To deal with relaxation we must take proper account of environmental effects.
Historically the first attempt was made by 
Villain and co-workers, even before resonant tunneling was 
established \cite{vil,pol}.  They considered 
the coupling between a giant spin, described by the Hamiltonian in equations (4.1) and (4.2), 
and a bath of phonons.  The physics of spin-phonon couplings is well understood at the 
microscopic level \cite{abr}, and Politi et al. \cite{pol} gave a treatment of 
the phonon-mediated tunneling relaxation of a giant spin $\vec{S}$, 
for the particular case of a single isolated $Mn_{12}$ molecule. However this theory disagreed
with the experiments. More recently they have extended it, and some other authors 
\cite{Garun, Bartol} have given similar treatments. 

In what follows I will lay considerable stress on the fact that none of these treatments is 
capable of explaining the low-$T$ relaxation in the {\it Quantum} Regime, for the simple reason
that at these temperatures the phonons can play no role. It can be then more or less 
{\it deduced} from the experiments that the relaxation must be mediated by nuclear spins,
which constitute the only dynamic environment remaining in the quantum regime. In this sense
the low-$T$ results are a kind of "smoking gun" for the role of the spin bath in the
quantum dynamics of the system. However it is clear that at higher $T$ both phonons and 
nuclear spins must be involved (as well as the rapidly fluctuating magnetic dipolar fields).

\vspace{4mm}

{\bf (i) \underline{Phonon-mediated Relaxation}}:
Let us briefly review this theory and its results.  This is useful because  
despite the fact that the calculations of Villain et al.\cite{vil,pol} are correct, 
their results disagree strongly with the 
experiments, particularly at low $T$ and low $H$
(ie., in the Quantum regime),  
and it is important to understand why.

Consider first the form of the coupling between \b{S} and phonons (parametrized, as 
in Chapter 2, by operators $b_{\vec{q}}, b^{+}_{\vec{q}}$, frequencies 
$\omega_q \sim c_sq$, and a Debye temperature $\Theta_D$).  For a strongly easy-axis 
``Ising-type" anistropy, there is the transverse term 
already given previously (cf. eqtn. (\ref{1.ph})).
That the coupling should be proportional to $\Omega_o$, the ``bounce frequency" 
introduced in section 2, is easily shown using instanton methods where applicable \cite{prok2}, 
but is also fairly obvious from dimensional arguments.  For strongly anisotropic 
``Ising-like" molecules like $Mn_{12}$ or $Fe_8$, $\Omega_o \rightarrow D$ (recall that 
$\Omega_o$ is roughly the energy separation between the 2 lowest levels of the 
symmetric van Hemmen/Sut\"{o} Hamiltonian, and the next levels).  The same result is 
found using standard spin-phonon theory \cite{abr}.

There are of course other spin-phonon coupling terms, but before asking why this one is the 
most important, let us see what it does to the dynamics of $\vec{S}$.  Consider 
2 possible transitions between eigenstates $\mid m>$ of the longitudinal part 
$^{\parallel}H^{(2)}_{o}$ of the giant spin Hamiltonian (here $\mid m> = \mid S>, 
\mid S-1>, \ldots \mid -S>$).  These are mediated simultaneously by the tunneling term 
$^{\perp}H_o(s)$, and by the coupling to the phonons.  Now the crucial point here is that 
a non-diagonal coupling like the one just given allows a transition 
$\mid m> \rightarrow \mid-(m \pm 1) >$, provided $\mid m>$ and $\mid -m>$ are in 
resonance (ie., provided $\mid \varepsilon_m - \varepsilon_{-m} \mid \leq \Delta_m$, 
where $\Delta_m$ is the tunneling matrix element between $\mid m>$ and $\mid -m >$, coming 
from $^{\perp}H_o$, and $\varepsilon_m$ is defined by 
$^{\parallel}H_o \mid m> = \varepsilon_m \mid m>$). This inelastic transition goes at 
a rate proportional to $\xi^3$,where $\xi = (\varepsilon _m - \varepsilon_{-m-1})$ 
is the relevant energy difference between initial and final states, for Debye phonons.  
This standard result \cite{abr} comes from the available phase space for phonons.  A 
simple calculation \cite{pol,prok2} gives a non-diagonal relaxation rate
\begin{equation}
\tau^{-1}(\varepsilon, T) \sim e^{-\varepsilon_{m}/kT} 
\frac{S^2\Delta^{2}_{m}}{\Theta_D} \left(\frac{\varepsilon}{\Theta_D}\right)^3
\end{equation}
at low T.  On the other hand a {\it diagonal coupling} (ie., one not involving 
operators $S_x, S_y$) will have a much smaller available phase space.  This is why 
the non-diagonal coupling dominates.

Suppose now that we are in the low-T limit; this arises precisely when 
$kT \leq \Omega_o/ 2 \pi$.  The crossover occurs experimentally for $T_c \sim 2K$ 
in $Mn_{12}$, and $T_c \sim 0.4 K$ in $Fe_8$.  Below $T_c$, only the 2 lowest levels 
are involved.  In this case a more accurate formula for the phonon relaxation 
rate is \cite{prok2}:
\begin{equation}
\tau^{-1} (H,T) \sim \frac{S^2\Delta^{2}_{10}}{\Theta_D} 
\left(\frac{\xi}{\Theta_D}\right)\coth (\xi/2kT)
\end{equation}
Now we immediately notice 2 things about this result.  First, $\tau(H,T)$ is 
{\it extremely long} for low fields (ie., where the bias 
$\xi = (\varepsilon_{10} - \varepsilon_{-10})$ is generated only by internal bias fields 
like nuclear hyperfine fields or intermolecular dipolar fields - in this case 
$\xi \sim 0.1 - 1K$ only).  Thus for $S = 10$, and for the values of $\Delta_{10}$ estimated 
for $Mn_{12}$ and $Fe_8$ (for which $\Delta_{10}$ is almost certainly less than 
10$^{-8}K$, perhaps much smaller), one finds that $\tau (H,T)$ at low $T$ and low $H$ becomes 
longer than the Hubble time!

The 2nd thing we notice is that around $H=0$, this sort of spin-phonon theory predicts 
a rapid increase ($\sim H^3$) of $\tau^{-1}(H)$ as $|H|$ is increased.

Unfortunately these 2 results are flatly contradicted by the experimental results, 
which show (for $Fe_8$ and $Mn_{12}$) a very sharp resonant {\it maximum} in 
$\tau^{-1}(H)$ around $H \sim 0$; in the middle of this the relaxation rates can be 
inverse seconds only (although, as already noted, the short time relaxation is far from 
exponential in the low T regime).

Thus we see that the low-T results are quite fatal to a theory which only involves a phonon 
environment.  However, this is by no means all that one can deduce from these low-T results.

\vspace{4mm}

{\bf (ii) \underline{Nuclear Hyperfine, and Molecular Dipole fields}}: 
Suppose we now include the 
coupling of each molecule to (i) nuclear spins, and (ii) the magnetic dipolar fields 
generated by its neighbours- these are the two important couplings left in the problem.  

Since, at low $T$, we have seen that phonons are irrelevant to the relaxation, let us 
consider the problem solely in the presence of these 2 couplings.  Below $T_c$ the 
problem now reduces to an effective Hamiltonian.
\begin{eqnarray}
H= &\: &  {1 \over 2} \sum_{ij} V_{ij}^{(d)} \tau_z^{(i)}\tau_z^{(j)}
+ \sum_i \Delta_{10} \tau_x^{(i)} 
+  \sum_{ik} V^{(N)}(\tau_z^{(i)}, \vec{I}_k) + H^{NN}\;,
\label{Hamiltonian}
\end{eqnarray}
where the first term describes the {\it static} dipolar-dipolar interactions
between molecules, the second describes tunneling, the third couples
magnetic molecules to nuclear spins $\{ \vec{I}_k \}$, and the last
term describes interactions between the nuclear spins.  This 
effective Hamiltonian operates in the subspace of the two lowest
levels of each molecule; we choose the basis set to be $\vert S_z=
\pm S \rangle $; $\tau_z$ and $\tau_x$ are Pauli matrices, and $\{i\}$,
$\{ j \}$ label molecular sites. Only the longitudinal dipolar interaction appears; transverse
"flip-flop" processes are exponentially small in the parameter $(\varepsilon_9 - 
\varepsilon_{10})/kT$.

Now it is essential to notice that if we also treat the hyperfine fields as static, we 
are faced with a really gross contradiction with experiment.  As already mentioned, the 
bias $\xi = (\varepsilon_{10} - \varepsilon_{-10})$ due to the combined nuclear and 
dipolar couplings, is not enough to allow phonons to act; moreover, the transverse fields 
coming from these couplings are far too small to make any noticeable change to $\Delta_{10}$.  
However the most crucial point is simply this: all but a very 
tiny fraction of molecules are blocked from any 
transitions at all, because $\xi \gg \Delta_{10}$ for almost all molecules. From this 
point of view it is quite incredible that there is any relaxation at all in the low-$T$ 
limit; recall that the energy barriers in these experiments are very large compared to $kT$ 
(at the lowest temperatures in the Fe-8 experiments, $kT$ was nearly 500 times less
than the barrier height, and some 100 times smaller than the energy $\sim 2{\cal D}$ 
required to excite 
any but the 2 lowest levels of each molecule!). 
{\bf Nevertheless one sees rapid relaxation - why?}

I have very recently argued, in a paper with N.V. Prokof'ev \cite{PRL}, that the explanation is 
more or less obvious, and can in fact be {\it deduced} from the experiments.  
The basic point is that no relaxation at all can proceed unless we give the environment 
some dynamics; as we have just seen, a static Hamiltonian won't work.  But at $70 mK$ there is only 
one source of such dynamics, as we have already seen in Chapters 2 and 3; it comes 
from the transverse ``spin diffusion" fluctuations in the hyperfine field, which have 
a typical frequency $T^{-1}_{2} \sim 10 -100 kHz$.  Unlike phonons, or intermolecular 
dipolar flip-flop transitions, or nuclear $T_1$ transitions (all of which disappear in 
the low-T limit, and are exponentially small in the ratio ${\cal D}/kT$), 
the nuclear $T_2$ dynamics persists down to temperatures in the $nK$ 
range, and is fairly independent of $T$ in the present experimental temperature range.

We have already seen in Chapter 3 what these rapid hyperfine fluctuations can do.  
For the present problem, the hyperfine coupling is sufficiently weak that no nuclear 
spins are likely to be flipped when $\vec{S}$ flips.  Thus a molecule near resonance 
(ie., where $\xi$ is small), will relax {\it incoherently} between states 
$\mid 10>$ and $\mid -10>$ (or vice versa) at a rate given by
\begin{equation}
\tau^{-1}_N(\xi) \approx
\tau^{-1}_0 e^{-\vert \xi \vert /\xi_o} \; .
\label{tauN2}
\end{equation}
\begin{equation}
\tau^{-1}_0 \approx {2 \Delta_{10}^2 \over
\pi^{1/2} \Gamma_2 }\; .
\label{tauN}
\end{equation}
where we have assumed a fluctuating bias field $\xi(t) = \xi + \delta\xi(t)$, for 
which the fluctuations $\delta \xi (t)$ 
extend over a range $\xi_o \sim T^{-1}_{2}$ (compare eqtn.(\ref{key})).  
The basic physics is simple; not only do the fluctuations greatly widen the 
``resonance window" width (from $\Delta_{10}$ to $\xi_o$, an increase of several orders 
of magnitude), they also allow the transitions to be {\it inelastic} 
(and therefore to cause irreversible relaxation). Thus, at very short times, molecules with 
a bias within $\xi_o$ of zero are sucked into the resonance window, where they undergo
irreversible inelastic relaxation.

At this point the long-range dipolar fields come in.  Once a fraction $\xi_o/W$ of 
molecules has relaxed via this mechanism (where $W \sim 0.1-1K$  is the total spread in 
longitudinal bias fields around 
the sample), the adjustment of the long-range dipolar fields caused by these transitions 
is quite enough to bring more molecules into the resonance window. At first glance
this seems like a very complex problem; however, to treat 
it is actually quite straightforward, using standard kinetic theory methods.

We begin by defining a normalised 1-molecule distribution function
$P_{\alpha } (\xi ,\vec{r}; t)$, with
$\sum_\alpha \int d\xi \int d\vec{r} P_{\alpha } (\xi ,\vec{r} ; t ) =1$.
It gives the
probability of finding a molecule at position $\vec{r}$, with
polarisation $\alpha =\pm 1$ (ie., in state $\vert S_z= \pm S
\rangle$), having a bias energy $\xi$, at time $t$. Molecules having
bias energy $\xi$ undergo transitions between  $\vert S_z= S\rangle$
and $\vert S_z= -S \rangle$ at a rate given by (\ref{tauN2}).
Flipping these molecules then changes the
dipolar fields acting on the whole ensemble, bringing
more molecules into near (or away from) resonance,
and leading to a self-consistent
evolution of $P_{\alpha } (\xi )$ in time. The general solution of
this problem requires a kinetic equation for $P_{\alpha}(\xi
,\vec{r}; t)$.

 To derive a kinetic equation for $P_{\alpha } (\xi ,\vec{r}
 ; t ) $, we again assume that dipolar and hyperfine fields are frozen
 (apart from the nuclear $T_2$ fluctuations just discussed), {\it unless} a
 molecule flips. All kinetics then come from these flips, along with
 the resulting adjustment of the dipolar field. We may then derive a
 kinetic equation in the usual way, by considering the
 change in $P_\alpha $ in a time $\delta t$, caused by molecular
 flips, at the rate $\tau_N^{-1}(\xi )$, around the sample. This yields
 \begin{eqnarray}
  \dot{P}_{\alpha } (\xi ,\vec{r} )= & &
   - \tau_N^{-1}(\xi ) [P_{\alpha } (\xi ,\vec{r} )
   -P_{-\alpha } (\xi ,\vec{r} ) ] \nonumber \\
   & -& \sum_{\alpha '} \int {d\vec{r}\: ' \over \Omega_0 } \int
   { d\xi ' \over  \tau_N (\xi ' )}
   \bigg[ P_{\alpha \alpha '}^{(2)} (\xi  , \xi ';\vec{r},\vec{r}\: ')
   - P_{\alpha \alpha '}^{(2)}
   (\xi -\alpha \alpha ' V_D(\vec{r} -\vec{r}\: ')
   , \xi ';\vec{r},\vec{r}\: ') \bigg] \;,
   \label{kinetic}
   \end{eqnarray}
where $P_{\alpha \alpha '}^{(2)} (\xi , \xi ';\vec{r},\vec{r}\: '; t)$
is the two-molecule distribution, giving the normalized joint
probability of finding a molecule at site $\vec{r}$, in state $\vert
\alpha \rangle $ and in a bias $\xi$, whilst another is at
$\vec{r}\: '$,  in state $\vert \alpha ' \rangle $, and in a bias $\xi
'$. $P^{(2)}$ is linked to higher multi-molecule distributions by a
BBGKY-like hierarchy of equations. The
first term on the right-hand
side of (\ref{kinetic}) describes the local tunneling relaxation, and
the second non-local term (analogous to a collision integral) comes
from the change in the dipolar field at $\vec{r}$, caused by a
molecular flip at $\vec{r}\: '$; the dipolar interaction $V_D(\vec{r} )
= E_D [1-3\cos^2\theta ] \Omega_0/r^3$, where $\Omega_0$ is the
volume of the unit molecular cell, and $\int d\vec{r}\: '$ integrates
over the sample volume.

As discussed in our paper, with this kinetic 
equation we can make a number of very clear predictions
concerning the low-T relaxation in these magnetic molecular crystals.  These include
(i)  The prediction that an initially polarized sample will, at short times, relax 
according to a universal $\sqrt{t}$ law (where $t$ is the elapsed time); the sample 
magnetisation will obey
\begin{equation}
M(t) = \left[1-(t/\tau_{short})^{1/2}\right] M_o
\label{universal}
\end{equation}
where $M_o$ is the saturated magnetisation, provided $1-(M/M_o) \ll 1$.
(ii)  The timescale $\tau_{short}$ depends radically on the sample shape.  For 
an ellipsoidal sample one finds
\begin{equation}
\tau_{short}^{-1} = {\xi_0 \over E_D \tau_0 }\:
{32\pi  \over 3^{5/2} (c^2+16\pi^2/3^5 ) } \;.
\label{tau-sh}
\end{equation}
where the constant $c$ is given by magnetostatic theory; for a prolate spheroid, one finds
\begin{equation}
c=(2 \pi /3)  [a^4+a^2-3a\sqrt{a^2-1}\ln (a+\sqrt{a^2-1} ) -2 ]/(a^2-1)^2
\end{equation}
where $a$ is the ratio of the longitudinal axis to its perpendicular;
analytic formulas can be found for any ellipsoid. 

The analytic simplicity of these results arises because in the short-time limit, 
$P^{(2)}(\xi, \xi^{\prime};\vec{r},\vec{r}^{\prime}; t)$ is factorizable, and because 
for an ellipsoid, the initial demagnetisation field is homogenous (a standard result 
of magnetostatic theory).  For a sample of arbitrary shape, we still get $\sqrt{t}$ 
relaxation, but in this case $\tau_{short}$ above is replaced by
\begin{equation}
(\tau_{short}^{(inh)})^{-1} \sim  \xi_0 N(0)
\tau_{short}^{-1} \;,
\label{tau-short-r}
\end{equation}
where $N(0) = \int d \vec{r} \sum_{\alpha} P_{\alpha}
(\xi=0 , \vec{r}; t = 0)$
is the {\it initial} "density of states" for the dipolar field distribution,
integrated over the whole sample, at bias $\xi = 0$; typically $N(0) \sim
 1 / E_{Dm}$, where $E_{Dm}$ is the average demagnetization field.

Notice that the prediction for these timescales shows that they depend essentially on both the
nuclear fluctuation timescale $T_2$ {\it and} on the sample shape. Once the condition
$1-(M/M_o) \ll 1$ is relaxed, we must worry about the
non-factorisability of $P^{(2)}$, and analytic calculations are no longer so easy.  
However it is easy to both verify the above results for short times, and to extend them 
to longer times, using Monte Carlo (MC) simulations of the relaxation in samples of 
various shapes \cite{PRL}.  
It is also interesting to see how the 1- molecule distribution function 
$P_{\alpha}(\xi, \vec{r}; t)$ itself relaxes; this is shown in Fig.19, at short times, 
for a homogeneous spherical sample.

Why is the short-time behaviour so simple (and {\it universal})? Essentially because we have 
a small parameter (the fraction of flipped spins). These spins behave like a set
of dilute dipoles, creating a solution to the kinetic equation which is in fact an old friend
to anyone who has worked in NMR. Consider first for simplicity the case
of an ellipsoidal sample, so that the initial field is homogeneous. 
Then at short times the solution to our kinetic equation,
if the normalised magnetisation at $t = 0$ is unity, just becomes
\begin{equation}
\dot{M}(t) = -M(t){ 2 \over \tau_0  }
\int d\xi e^{ - \vert \xi \vert /\xi_0 }
{\Gamma_d(t) / \pi \over [\xi - E(t) ]^2 +\Gamma_d^2 (t) } \;.
\label{dilute}
\end{equation}
where the field distribution in an ellipsoidal sample is nothing but a 
Lorentzian distribution (up to a high energy cut-off
$E_D$) related to that found by Anderson for dilute {\it static} dipoles:
\begin{eqnarray}
P_\alpha (\xi ) & =& {1+\alpha M(t) \over 2}
~ {\Gamma_d(t) / \pi \over [\xi - \alpha E(t) ]^2
+\Gamma_d^2 (t) } \;; \nonumber \\
\Gamma_d(t) & =& {4\pi^2 \over 3^{5/2} } E_D (1-M(t))  \; ;
\label{Lorentz1} \\
E(t) & = & cE_D (1-M(t)) \;,
\label{Lorentz2}
\end{eqnarray}
where $c$ is the sample shape dependent coefficient defined above, and $E(t)$
is the time-dependent internal field (assuming here that the external field is zero).

\vspace{3in}

FIG. 19: Monte-Carlo (MC) simulation of the time evolution of the time-dependent density of 
states $N_{\alpha}(\xi, t) = \int d \vec{r} P_{\alpha} (\vec{r},\xi ; t)$ of the 
unflipped spins in a spherical sample of diameter 50 sites, with spins arranged in a cubic lattice.
Energy and density of states use units where $\xi_0 = 1$ and $E_D = 20$, and results are shown for 
$\alpha = +1$ at time 
$t = 0$ (for an initially polarized sample) and for $\alpha = -1$ at 
$t = 0.1 \tau_0$. The Lorentzian shape is distorted by lattice effects at high energy (giving
the hump at $E = -15$). The noise, and the finite width of the 
$\delta$- function at $t = 0$, are finite size effects.

\vspace{5mm}

This explains the universality of the $\sqrt t$ behaviour for an ellipsoidal sample (and
note the distinctive and strong dependence of $\tau_{short}$ on the sample shape- this should
be testable experimentally). But what about samples of arbitrary shape? Then the 
problem becomes essentially inhomogeneous. We thus return to
the kinetic equation (\ref{kinetic}), and notice that
if the demagnetisation varies on a length scale much greater than
the average distance between flipped spins, then (\ref{dilute})
is simply modified to
\begin{equation}
\dot{M}(\vec{r} ,t) = -M(\vec{r} ,t){ 2 \over \tau_0  }
\int {d\xi \over \pi }
{\Gamma_d(\vec{r} ,t)  e^{ - \vert \xi \vert /\xi_0 }
\over [\xi - E(\vec{r} ,t) ]^2 + \Gamma_d^2 (\vec{r} ,t) } \;.
\label{dilute-r}
\end{equation}
where $\Gamma_d(\vec{r} ,t)$ and $E(\vec{r} ,t)$ are defined in terms
of $M(\vec{r} ,t)$ analogously to (\ref{Lorentz1})
and(\ref{Lorentz2}); the solution is then identical to
(\ref{universal}) except that $\tau_{short}$ is modified to
$\tau^{(inh)}_{short}$. You can think about this in a geometrical way. In an ellipsoidal
sample, with initially uniform field, the molecules are either initially in the "resonance
window" (of width $1/\xi_o$) or they are not. If they are, then the rapid decrease in the 
instantaneous relaxation rate, giving the square root form, comes about because the molecules
are being pushed away from resonance as the inhomogeneous Lorentzian fields develop. 
Only a Lorentzian shape gives a $\sqrt t$ relaxation; and when the flipped spins become
dense enough, the Lorentzian form must break down (in fact it probably becomes more gaussian, at
least for intermediate times).
Now turning to a sample of arbitrary shape, one should imagine the surface inside the sample
which is the locus of points where the total field (ie., the sum of external field $H_o$ and the 
internal field $E (\vec{r},t)$) is zero. When the sample starts relaxing, you can imagine that 
an expanding region around this surface becomes involved in the relaxation, with molecules
inside it being either brought into or forced out of resonance. In a sample of complicated shape
(particularly with edges and corners) this surface (and the related density of states) will be 
complicated, with van Hove singularities of various kinds in $N (\xi)$ (in principle classifiable using
Morse theory). However, this does not {\it in any way} affect the way in which the function 
$N_{-}(\xi, t)$, integrated over the sample, develops a Lorentzian shape around $\xi = 0$ at short
times.

I strongly emphasize that as soon as we have a reasonably large concentration of flipped 
molecules (roughly 10 to 15 per cent, according to the MC simulations), this analysis
must begin to break down, because the bias field distribution, due to the flipped spins,
will stop being Lorentzian. Moreover, we also expect the assumption of factorisation of the
2-molecule distribution function to break down, and the problem then apparently becomes
intractable- in fact we have a kind of {\it Quantum Spin Glass} problem. We expect a 
complicated crossover to a long time behaviour which shows "ageing" behaviour, and whose form
seems very difficult to predict.

However, one can also make analytical predictions for long times under a slightly different 
circumstance, viz., if the sample is first depolarized at {\it high} $T$, to a value 
$M/M_o \ll$ 1, and the cooled to the low-T quantum regime.  In this case we also have
factorizability of the 2-particle distribution function; and
in fact another analytic solution for the homogeneous
(ie., ellipsoidal) case can be found from (\ref{kinetic}), when $M \ll 1$ and
$P^{(2)}_{\alpha \alpha '}(\xi , \xi '; \vec{r},\vec{r}\: ') =
P_{\alpha }(\xi )P_{\alpha '}(\xi ')$. One finds {\it exponential}
relaxation, at a rate
\begin{equation}
\tau_{long}{-1}\approx { 2\xi_0 \over E_{max} \tau_0 [1+\kappa \ln
(E_{max} /\pi \xi_0 ) ] } \;,
\label{tau-long}
\end{equation}
where $\kappa \sim 1$ is a numerical coefficient, and $E_{max}$ is the
spread in dipolar fields in this nearly depolarized limit.

It will be very interesting to see if these low-T predictions are confirmed.  I 
emphasize that the low-$T$, low-$H$ limit is crucial here - in this case the predictions 
can more or less be deduced directly, along the lines just described.  In my personal 
opinion, confirmation of the $\sqrt{t}$ law, and the dependence of $\tau_{short}$ on 
$T_2$ and on sample shape, would give very powerful evidence for the nuclear relaxation 
mechanism discussed.  In any case, we see here a very nice example of the role a nuclear 
spin bath can play in tunneling dynamics.  The essential result which the 
spin bath theory gave us is the formula for $\tau^{-1}_{N}(\xi)$ in eqtn (\ref{tauN}) (and this result is 
just a special case of eqtn (\ref{diff.3})).

What more theoretical work can be done in this low-$T$ limit? The first obvious thing is to 
extend this kind of theory to much higher fields (both longitudinal, which will bring 
level $\mid 10>$ into resonance successively with levels $\mid 9>$, $\mid 8>$, etc; and 
transverse fields, which will slowly reduce the total barrier height, as well as slowly 
changing the tunneling matrix elements.  The effect of longitudinal fields will not be 
simple - even in the very low-T limit, the resonant tunneling between $\mid 10>$ and 
$\mid -9>$ will be quite different from that between $\mid 10>$ and $\mid -10>$, since 
both spontaneously-emitted phonons and dipolar processes will play a role.

Another unsolved problem of great interest at low $T$ is the effect of an AC field on the 
nanomolecular dynamics.  A number of papers have already been written on this, 
essentially treating the problem as a simple Landau-Zener problem, involving only the 
giant spin levels and the AC field. However I should emphasize strongly that  
the Landau-Zener transitions, when they take place, are doing so in the presence 
not only of the applied AC field, but also of the much more rapidly fluctuating nuclear 
fields, even at the very lowest temperatures (and of course at higher $T$, they have
to contend with the large and wildly fluctuating intermolecular dipolar fields!). 
The experiments have already shown a 
very rich behaviour (see particularly recent work of Novak et al., \cite{nov2}),
which depend strongly on $H$ and $T$.

\vspace{7mm}

{\bf (iii) \underline{Thermally-Activated Regime}}:
I concentrated on the low-$T$, low-$H$ regime above, in discussing the experiments, simply 
because their interpretation in this regime was straightforward (and provides clear evidence 
for nuclear spin effects).  The experiments at temperatures above $T_c$ are more complex to 
interpret (because of the large number of physical processes which are simultaneously 
involved), but are also of considerable interest.

It is always a good idea, in trying to understand a set of experiments, to see what happens 
in regimes where the theory is simple - and this often means looking for ``limiting cases", 
in which one or more parameters are very small (or very large), so that some processes 
dominate and others are irrelevant.  We have just seen this in the low-$T$ limit (where 
all dynamics except transverse nuclear spin fluctuations are frozen out).  Is there any 
analogous simplifying limit at higher $T$?

At first glance this is not obvious - at higher $T$ many more giant spin levels are 
involved, as are now the phonons, and quite a number of recent papers have dealt with 
processes involving phonon-mediated transitions between these levels.  For reasons 
quite obscure to me, most of these papers treat both the nuclear hyperfine and the 
molecular dipoles fields as {\it static}, when in fact they are both 
fluctuating rather violently and rapidly in time.  Thus the relaxation, at first, 
seems very messy.

In fact, however, things are simple for the following reasons:

(i)  The timescales involved in the various physical processes are much shorter than the 
experimental relaxation timescales (and moreover this is exactly what should have been 
expected once $T$ is considerably greater than $T_c$.

(ii)  The competition between the experimentally increasing tunneling matrix elements, 
and the experimentally decreasing thermal occupations, as one rises up through higher and 
higher giant spin levels, means that at given values of $H$ and $T$, one transition will 
dominate the tunneling relaxation.

There is nothing particularly radical about these circumstances - in fact they are 
perfectly normal.  Let us see how they arise.

We note that first that thermal equilibration between levels on the  {\it same side} 
of the energy barrier is very fast, over timescales $\sim \mu secs$: this is easily shown 
if we assume spin-phonon coupling strengths.  Such processes do {\it not} involve 
tunneling.  We also notice that, in contrast to the low-$T$ case, dipolar ``flip-flop" 
processes now occur very rapidly - in fact at a rate typically even faster than spin-phonon 
transitions.  Thus, for small $H$, transitions like 
$\vert 9 \rangle \vert -10 \rangle \rightarrow \vert 10 \rangle \vert -9 \rangle$ 
(or vice-versa) happen at a frequency 
exceeding $10^6 Hz$ once we are well out of the quantum regime.

These 2 processes are the only ones which can cause transitions between 2 giant spin 
levels on the same side of the barrier - they were both frozen at low $T$, but their 
rate increases exponentially fast as $T$ rises.  The now very rapid dipolar fluctuations 
also completely change the nuclear spin dynamics; the {\it longitudinal} 
nuclear relaxation rate $T^{-1}$ rises with the dipolar flip-flop rate, since 
the latter drives the former.  This is important, because it means that the bias energy 
$\xi = g\mu_B HS_z$ acting on each molecule is fluctuating very fast (because of the 
fluctuations of both $\vec{H}_{dip}$ and $\vec{H}_{hyp}$) over a fairly large 
range of energies (for $Fe_8, \vec{H}_{dip} \gg \vec{H}_{Hyp}$, and $\xi$ fluctuates 
over a range $\sim 0.5K$, whereas for for Mn$_{12}$, the hyperfine field is much larger,
indeed not much smaller than the dipolar fields, so that the fluctuations in $\xi$ 
may well be some 50 per cent larger). 
Notice that although we do not know how fast $T_1$ is (it has not 
been measured for any of these molecules at the 
relevant temperatures), we do expect it to have roughly the same $T$-dependence as the 
dipolar flip-flop rate; and we expect $T_1 \ll \tau_{exp}$ (where $\tau_{exp}$ is the
experimental relaxation rate), once we are well inside 
the thermally-activated regime.

Thus the picture that energies in the thermally activated regime is one where molecules 
cycle between giant spin states on the same side of the barrier, and between all the 
nuclear multiplets available for a given giant spin state $\vert m \rangle$.  
Inelastic tunneling transitions are much slower.  What then controls the 
tunneling rate in this regime?

The answer is startlingly simple.  The transitions on one side of the barrier occur so 
quickly that they {\it disappear completely from the physics} - their only role 
is to keep the molecules on one side of the barrier in a state of quasi-equilibrium, in 
which the probability of occupying a state of energy $\epsilon$ is just 
$Z^{-1} e^{-\epsilon /kT}$, where $Z$ is the total partition function.  Under these 
circumstances the relaxation rate is controlled only by the coupling to phonons, and 
we expect {\it exponential relaxation}!

Let me stress again that there is nothing in this conclusion that will surprise 
many - in fact a number of papers have already simply {\it assumed} the existence 
of quasi-equilibrium thermal populations on each side of the barrier, in attempting to 
calculate a relaxation rate \cite{}. However this assumption obviously needs to be 
justified, particularly as we know that the low-$T$ relaxation is predicted to be 
{\it non}-exponential! More generally, as emphasized some time ago \cite{prok2}, the crossover
to exponential relaxation will occur on a timescale roughly equal to $T_1$, and so in
the region of temperature where one sees crossover from quantum to thermally activated
behaviour (ie., for $T \sim {\cal D}/2 \pi$), one may even expect to see $\sqrt t$ relaxation
for $t \ll T_1$, with a crossover to exponential relaxation for $t \gg T_1$. Experimental 
measurements of $T_1$ would be very useful here!

However a cautionary note. It is not entirely obvious that 
one will always expect to see exponential relaxation in the experiments, even at high $T$.  
The reason for this is very simple.  In a typical relaxation experiment, as we have 
already seen at low $T$, the total field (internal field $E(\vec{r},t)$ plus external 
applied field $\vec{H}_o$) changes in time as the system relaxes; moreover, so does 
the distribution function $P_{\alpha}(s, \vec{r},;t)$, and hence the density of 
states $N(\xi; t)$ available for transitions.  Consequently, in
the course of the relaxation, the relaxation rate $\tau^{-1}(H)$ must become a function of 
time - essentially we are sweeping through regions of different relaxation rate. Thus the high-$T$
relaxation is probably going to be rather complex (and indeed the various experiments do not
obviously agree in this regime).

\subsection{Macroscopic Quantum Coherence?}

Ever since the well-known early discussions of Leggett et al. \cite{leg,AJL,ajl80}, on the 
possibility of superpositions of ``macroscopically distinguishable" quantum states in 
SQUIDs, there has been intense interest in its experimental realisation.  Discussions 
of this are usually framed in the context of ``Macroscopic Quantum Coherence" between 
2 low energy states; and following Leggett (see his chapter in this book), most of 
these discussions have been in the context of the spin-boson model.

Despite a number of attempts to find Macroscopic Quantum Coherence (MQC) in 
superconducting systems, no results of this kind have yet been reported. 
In section 3.4(b) above, I already discussed why I think that the observation of MQC
in superconductors is going to be very difficult, because of the nuclear bath effects.
However in the different field of magnetism, a rather dramatic claim for the discovery 
of MQC has been made by the group of Awschalom et al.\cite{aws,gid}, in a series of papers going 
back to 1992.  There has also been strong opposition to these claims from various 
quarters, most notably in the work of A. Garg \cite{awsgarg}.  

I don't think this issue is yet settled, but it is clearly of interest; the purpose of the present 
sub-section is simply to present the results, and to explain the role of the nuclear 
spin bath in the problem.

\vspace{7mm}

{\bf 4.2(a) EXPERIMENTAL RESULTS on FERRITIN MACROMOLECULES}

The series of experiments reported by Awschalom et al. actually began with results 
\cite{FeCo} on arrays of Iron pentacarbonyl (FeCo$_5$) ellipsoidal grains, of rather large 
size.  The AC susceptibility absorption was sharply peaked in these experiments, at 
very low frequencies (ranging from $\omega_o \sim$ 60 Hz for grains of size 
150 \AA $\times$ 700 \AA, up to $\omega_o \sim$ 400 Hz for grains of size 
150 \AA $\times$ 380 \AA).  Curiously, it was found that $\omega_o$ varied roughly 
exponentially with the volume of the particles (decreasing with larger volume) in a 
way that would be expected if the particle magnetisation was tunneling coherently 
between 2 orientations, or else being driven by the AC field between 2 eigenstates, 
which themselves are superpositions of 2 ``semiclassical" states 
$\vert \Uparrow$ and $\vert \Downarrow$ (the particles are uniaxial).

Nevertheless, as recognized by these authors, this interpretation is untenable, 
principally because the energy scale $\omega_o$ is so small (a frequency 400 Hz 
is equivalent to a temperature $\sim 8 \times 10^{-8}K$).  For all the grains to show
such a coherent response would require that any external bias field, acting on the grains,
would be so small that it would split the degenerate states by a bias 
energy $\xi$ considerably less than $\omega_o$.  This is of course 
impossible; apart from anything else the dipolar magnetic interaction between 
the grains is at least 5 orders of magnitude larger than $\omega_o$, for even 
the smallest grain size!  In fact, to this date I am aware of no explanation 
offered for these results.

However in 1992 experiments done on frozen solutions of ferritin molecules in the 
protein apoferritin were reported \cite{aws}. Ferritin is a very interesting molecule, 
found naturally in all eukaryotic cells; it is of roughly spherical shape, and 
diameter $\sim 80 \AA$.  The core contains some 4500 $Fe^{3+}$ ions (spin 5/2), and 
is protected from the outside world by a ``cage" of apoferritin protein molecules.

In bulk the ferrihydrite would order antiferromagnetically (giving a N\'{e}el vector 
for this system of magnitude $N = \vert \vec{N}\vert  \sim 18,000 \mu_B$).  In reality of course 
the molecules possess a surface moment, which must certainly vary from one molecule to 
another - in their first paper Awschalom et al. \cite{aws} estimated this moment by various 
means, yielding values ranging from 217 $\mu_B$ to 640 $\mu_B$.  The main result of 
this experiment, like the earlier FeCo$_5$ one, was a sharp resonance in 
$\chi ''(\omega)$ (the AC absorption). However this time the frequency was 
$\sim 940 kHz$ (ie., an energy $\sim 45\mu K$).  This was only seen in strongly 
diluted (1000:1) solutions.

More recently this group has done similar experiments on artificially engineered 
samples of ferritin molecules with smaller core sizes, and again they saw $\omega_o$ 
rise as a roughly experimental function of the inverse particle size \cite{}.  
One should also notice that no resonance was found in experiments on an apoferritin 
sample containing no ferritin, and that increased dilution of the sample 
tended to sharpen the absorption lines.

In contrast to the FeCo$_5$ results, Awschalom et al. believe that the ferritin results 
provide evidence for the resonant absorption of EM waves at a frequency $\omega_o$ 
corresponding to the tunneling matrix element $\Delta_o$, between N\'{e}el states 
$\vert \uparrow \rangle$ and $\vert \downarrow \rangle$.  This belief is partly based on the estimated 
value for the anisotropy field, which is not incompatible with a tunneling matrix 
element of this size.

A large number of objections have been raised to this straightforward interpretation. These 
include (i) the existence of randomly varying dipolar couplings between the molecules (the 
molecules are not oriented)\cite{Sta92} 
(ii) the field dependence of the resonance frequency and linewidth \cite{awsgarg}
(iii) the power absorption in the experiments; \cite{awsgarg} (iv) a claimed contradiction with high-$T$
blocking \cite{Tej}; and (v) nuclear spin effects \cite{pro,prok,gar}. 
Many of these queries will probably only 
be definitively settled if and when 
other experimental groups carry out an independent check on the experiments. I will make no attempt to 
address all of the issues, and refer the reader to the literature.
Instead, and in the context of this chapter, we will look at the single question of the the effect 
of nuclear spins on the coherence.

\vspace{7mm}

{\bf 4.2(b) NUCLEAR SPIN EFFECTS on MQC in FERRITIN}

To investigate the effects of the nuclear bath (and other environmental spins) on the 
dynamics of the ferritin molecules, we need a realistic effective Hamiltonian. The problem
is analogous to the SQUID problem we already looked at, and it is pedagogically useful because
it deals with an antiferromagnetic system. We begin with a single ferritin molecule, for which the
biaxial Hamiltonian (9) is usually used (with the understanding that $\vec{S}$ now 
represents the Ne\'el vector, and that the extra surface moment will in reality complicate things-
the following is not meant to be a complete study). Awschalom et al. \cite{aws,gid} give a value   
$1.72$ Tesla (in field units) for $(K_{\parallel}K_{\perp})^{1/2}$, so we will take a small 
oscillation frequency $\Omega_o \sim 40 GHz$ as a rough estimate. If we also suppose that the 
resonance at frequency $\omega_o$ represents resonant tunneling, then $\Delta_o \sim 1MHz$ for the
sample of naturally occurring ferritin.

>From these numbers we immediately see that topological decoherence effects will be very small; 
with roughly 100 $Fe^{57}$ nuclei in the sample (some 2 per cent of all the nuclei have a spin),
and a hyperfine coupling $\omega_k \sim 50-60 MHz$, we get $\alpha_k \sim 2 \times 10^{-3}$, and
so the mean number of nuclear spins co-flipping with the ferritin is $\lambda \sim 4 \times 10^{-5}$.
However as noted in Prokof'ev and Stamp \cite{pro,prok}, 
and discussed at some length later by Garg \cite{gar},
the degeneracy blocking effects are very severe; the spread $E_o$ in the nuclear multiplet around
each ferritin level is roughly $500-600 MHz$, and so the mean bias is some 600 times greater than 
$\Delta$ (the contrary claim of Levine and Howard \cite{lev1} was later retracted \cite{lev2}). There will 
also be an effect of orthogonality blocking, whose effect is very hard to quantify; it will 
come from the surface moments on the ferritin, and even more from the surface interactions with the 
apoferritin (particularly between surface moments and any paramagnetic spins in the apoferritin;
this was described as an effect of "loose spins" in the early papers \cite{pro}). The ensuing random
fields acting on the ferritin moment mean that the initial and final states of the Ne\'el vector will
not be exactly antiparallel.

If we ignore degeneracy blocking, and also ignore the nuclear spin dynamics, then the situation does
not in fact look too bad for the experiments, at least at first glance. This is because the lineshape
in a situation of pure degeneracy blocking (already shown in Fig.XXX), is rahter sharply peaked,
and so resembles the experiments. There is certainly a problem of power absorption, but as noted by
Awschalom et al., this is very hard to quantify \cite{awsgarg, Tej}. 
I ignore here the other difficulties mentioned 
above (points (i)-(iv)).

However there is also the problem of the nuclear spin dynamics at low $T$; moreover, as we have 
seen already in the context of the magnetic molecules, this will drive a time-dependent change 
in the dipolar field distribution in the sample (which presumably starts off being pretty much
Lorentzian in the sample of dilute and randomly-ordered molecules). However there is one big 
difference here from the case of Mn$_{12}$ and Fe$_8$ discusssed earlier. This is that 
in all of the experiments done on ferritin so far, $\Delta_o$ is {\it bigger} than the inverse
timescale $T_2^{-1}$ of the nuclear fluctuations. This changes things a great deal- now the bath 
fluctuations are slow and thus have a much smaller effect on coherent tunneling.

One could go on and discuss this problem in quantitative detail. However I will not for the 
following simple reason. This is that {\it none} of the calculations described here (or those 
appearing in the literature) have yet addressed what is the key theoretical problem in the 
interpretation of the ferritin experiments, which is the AC response of a central spin,
coupled to a spin bath, when the driving field is {\it fast}. This is nothing but the 
{\bf Landau-Zener} problem for a system coupled to a spin bath, and it is not yet solved! I 
would stress that the solution to this problem is not at al obvious- indeed with a 
{\it finite} spin bath it is clear that under many circumstances a response function (in the 
strict sense of linear response) does not even exist! Thus, since whereof one cannot speak..

\subsection{Acknowledgements}:

Much of the work of my own described herein was done with Drs. M. Dube' and N.V. Prokof'ev,
whom I thank for innumerable discussions. At the Seattle workshop I also enjoyed very 
useful discussions with Drs. P. Ao, G. Bertsch, O. Bohigas, A. Bulgac, H. Grabert, P. Hanggi,
P. Leboeuf, A.J. Leggett, A. Lopez-Martens, 
C. Sa de Melo, B. Spivak, D.J. Thouless, S. Tomsovic, and J. Treiner,
although not all of the subjects we discussed appear here! I thank the Institute for Nuclear Theory
(Seattle), the Canadian Institute for Advanced Research, NSERC Canada, and the Laboratoire de 
Champs Magnetiques Intenses (Grenoble), for support during the time this chapter was written.

\end{document}